\documentclass[journal]{IEEEtran}

\usepackage{cite}
\usepackage[dvips]{graphicx}
\usepackage[cmex10]{amsmath}
\usepackage{color}
\usepackage{bm}
\usepackage{theorem}
\usepackage{amscd}
\usepackage{amssymb}

\newcommand{\ca}{\widehat}

{\theorembodyfont{\normalfont}
\theoremheaderfont{\normalfont\it}
\newtheorem{thm}{ Theorem}
\newtheorem{dfn}[thm]{ Definition}
\newtheorem{lmm}[thm]{ Lemma}

\newtheorem{cjt}[thm]{ Conjecture}}

{\theorembodyfont{\normalfont}
\theoremheaderfont{\normalfont\it}
\newtheorem{prf}{ Proof:}}

{\theorembodyfont{\normalfont}
\theoremheaderfont{\normalfont\it}
}

{\theorembodyfont{\normalfont}
\theoremheaderfont{\normalfont\it}
}

{\theorembodyfont{\normalfont}
\theoremheaderfont{\normalfont\it}
}

{\theorembodyfont{\normalfont}
\theoremheaderfont{\normalfont\it}
}

{\theorembodyfont{\normalfont}
\theoremheaderfont{\normalfont\it}
\newtheorem{rmk}{ Remark.}}

\newcommand{\bra}[1]{\mbox{$\left\langle#1\right|$}}
\newcommand{\ket}[1]{\mbox{$\left|#1\right\rangle$}}

\newcommand{\outpro}[2]{\mbox{$\ket{#1}\!\bra{#2}$}}
\newcommand{\proj}[1]{\mbox{$\ket{#1}\!\bra{#1}$}}

\newcommand{\nn}{\nonumber}
\newcommand{\beq}{\begin{eqnarray}}
\newcommand{\eeq}{\end{eqnarray}}

\begin{document}

\title{A Coding Theorem for Bipartite Unitaries\\in Distributed Quantum Computation}

\author{Eyuri Wakakuwa, Akihito Soeda and Mio Murao
 
\thanks{This work is supported by the Project for Developing Innovation Systems of MEXT, Japan and JSPS KAKENHI (Grant No.~23540463, No.~23240001, No.~26330006, No.~15H01677 and No.~17H01694).  We also gratefully acknowledge to the ELC project (Grant-in-Aid for Scientific Research on Innovative Areas MEXT KAKENHI (Grant No.~24106009)) for encouraging the research presented in this paper. This paper was presented in part at ISIT 2015.}
\thanks{E. Wakakuwa is with the Department of Complexity Science and Engineering, Graduate School of Frontier Sciences, The University of Tokyo, Chiba 277-8561, Japan (email: wakakuwa@edu.k.u-tokyo.ac.jp).}
\thanks{A. Soeda is with the Department of Physics, Graduate School of Science, The University of Tokyo, Tokyo 113-0033, Japan.}
\thanks{M. Murao is with the Department of Physics, Graduate School of Science, The University of Tokyo, Tokyo 113-0033, Japan,  and is with Institute for Nano Quantum Information Electronics, The University of Tokyo, Tokyo 113-0033, Japan.}
%\thanks{Copyright (c) 2017 IEEE. Personal use of this material is permitted.  However, permission to use this material for any other purposes must be obtained from the IEEE by sending a request to pubs-permissions@ieee.org.}
}

\maketitle

\begin{abstract}
We analyze implementations of bipartite unitaries by means of local operations and classical communication (LOCC) assisted by shared entanglement. We employ concepts and techniques developed in quantum Shannon theory to study an asymptotic scenario, in which two distant parties perform the same bipartite unitary on infinitely many pairs of inputs. We analyze  minimum cost of entanglement and classical communication per copy. For two-round LOCC protocols, we derive a single-letter formula for the minimum cost of entanglement and classical communication, under an additional requirement that the error converges to zero faster than $1/n^4$, where $n$ is the number of input pairs. The formula is given by the ``Markovianizing cost'' of a tripartite state associated with the unitary, which can be computed by a finite-step algorithm. We also derive a lower bound on the minimum cost of resources, which applies for protocols with arbitrary number of rounds.
\end{abstract}

\begin{IEEEkeywords}
Entanglement cost, classical communication cost, quantum Shannon theory, approximate recoverability.
\end{IEEEkeywords}

\section{Introduction}

Distributed quantum computation is a task in which a group of distant parties collaborates to perform a large quantum computation, by using classical communication, quantum communication and shared entanglement as resources. One of the most extensively investigated tasks is implementation of bipartite unitaries by local operations and classical communication (LOCC) assisted by shared entanglement. Here, two distant parties, say Alice and Bob, have quantum systems $A$ and $B$ in an unknown state $|\varphi\rangle^{AB}$, and aim to perform a known unitary $U^{AB}$ by LOCC, using some resource entanglement shared in advance. Although this task can be implemented simply by using quantum teleportation, it was shown that the cost of entanglement and classical communication can be reduced by constructing a more efficient protocol, depending on the unitary to be implemented \cite{eisert00}.

The following two questions then naturally arise: (i) How can we find efficient protocols which consume less resources for a given bipartite unitary? and (ii) What are the minimum costs of resources required for implementing that unitary? Although these questions have been addressed, e.g. in \cite{eisert00, cirac01, groisman05, chen05, ye06, berry07, zhao08, yu10, cohen10,soeda11, stahlke11}, most of the studies so far assume particular forms of the resource entanglement or of the bipartite unitary to be implemented. A general method to address these problems is yet discovered.

In the present paper, we address the above questions in an information theoretical scenario for the first time, by applying the concept of ``block coding''. Here, the two parties perform the same bipartite unitary  on a sequence of input pairs at once. We consider an asymptotic limit of infinite pairs and vanishingly small error, and analyze the minimum cost of entanglement and classical communication per copy required for the task. We mainly focus on protocols consisting of two-round LOCC as the first nontrivial case,  for simplicity. Our approach is different from previous  ones which treated single-shot cases\cite{eisert00, cirac01, groisman05, chen05, ye06, berry07, zhao08, yu10, cohen10,soeda11, stahlke11}.

The main result of this paper is that we derive a single-letter formula for the minimum cost of entanglement, forward and backward classical communication in two-round protocols, under an additional assumption that the error converges to zero faster than $1/n^4$, where $n$ is the number of input pairs. The formula is represented in terms of the ``Markovianizing cost'' (\!\cite{waka15_markov,waka15_rec}) of a state associated with the unitary, which can be computed by a finite-step algorithm. The result is applicable for any bipartite unitary.

It is left open, however, whether the same converse bound holds when we drop the requirement on the convergence speed of the error. We relate this problem to another open problem regarding an ``asymptotic symmetry'' of approximate recoverability, that is, whether a tripartite quantum state $\rho^{ABC}$ is approximately recoverable from $\rho^{BC}$ if it is approximately recoverable from $\rho^{AB}$, up to a dimension-independent rescaling of error of recovery. We prove that an affirmative answer to the  asymptotic symmetry implies an affirmative one to the  converse bound.

We also derive a lower bound on the minimum cost of entanglement and classical communication, which is applicable for any protocol with arbitrary number of rounds, in terms of a parameter called the {\it Schmidt strength} of the unitary. It turns out that the lower bound is achievable for a class of bipartite unitaries called generalized Clifford operators. 

The structure of this paper is as follows. In Section \ref{sec:formulation}, we introduce the formal definition of the problem. The results are summarized in Section \ref{sec:results}. In Section \ref{sec:preliminaries}, we review results on Markovianization and state merging. Section \ref{sec:singletworound} analyzes single-shot two-round protocols for implementing a bipartite unitary. Outlines of the proofs of the main result are presented in Section \ref{sec:achievabletworound}. In Section \ref{sec:algorithm}, we discuss general properties of the Markovianizing cost of  unitaries. The Markovianizing cost for two classes of bipartite unitaries is computed in Section \ref{sec:exa} as examples. In Section \ref{sec:open}, we investigate an open problem regarding the convergence speed of the error from a viewpoint of approximate recoverability. Section \ref{sec:ccpowerlocc} analyzes the power of a LOCC protocol for transmitting classical information. In Section \ref{sec:lowerbound}, we provide a lower bound on the cost of resources for an arbitrary LOCC protocol. Conclusions are given in Section \ref{sec:conclusion}. The detailed proofs of lemmas and theorems in the main part are presented in Appendices.\\

{\it Notations.} $|\Phi_d\rangle$, $|\Phi_{K_n}\rangle$ and $|\Phi_{L_n}\rangle$ represent the maximally entangled state with the Schmidt rank $d$, $K_n$ and $L_n$, respectively. $\pi_d$ is the maximally mixed state of rank $d$. The fidelity and the trace distance between two quantum states $\rho$ and $\sigma$ are denoted by $F(\rho,\sigma)$ and $\|\rho-\sigma\|_1$, respectively. We abbreviate $F(\rho,|\psi\rangle\!\langle\psi|)$ as $F(\rho,|\psi\rangle)$. For a quantum operation $\mathcal E$, we abbreviate ${\mathcal E}(|\psi\rangle\!\langle\psi|)$ as ${\mathcal E}(|\psi\rangle)$. Otherwise we follow the notations introduced in \cite{waka15_markov}.

\section{Formulations}
\label{sec:formulation}

\begin{figure}[t]
\begin{center}
\includegraphics[bb={0 0 550 205}, scale=0.43]{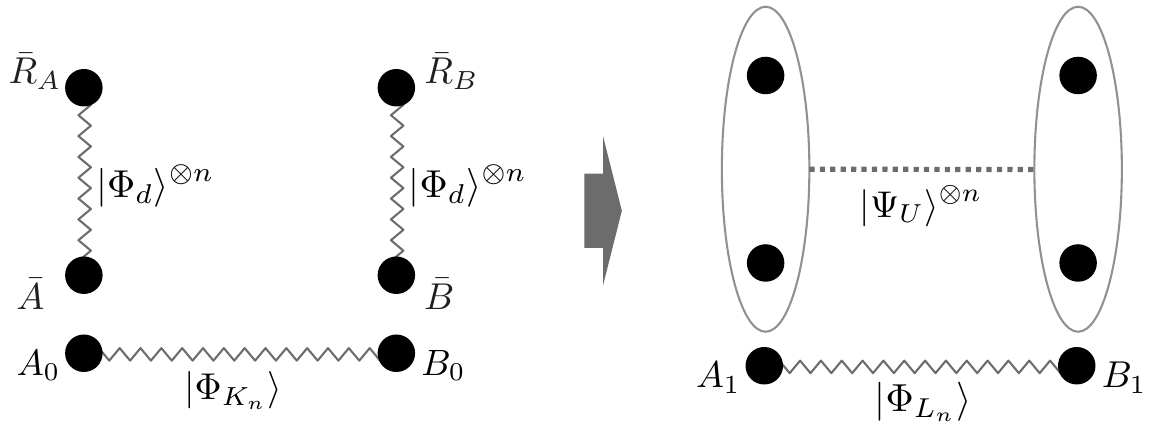}
\end{center}
\caption{The task is to apply $(U^{AB})^{\otimes n}$ on $(\ket{\Phi_d}^{AR_A}\ket{\Phi_d}^{BR_B})^{\otimes n}$ by using resource entanglement $|\Phi_{K_n}\rangle^{A_0B_0}$. $R_A$ and $R_B$ are reference systems that Alice and Bob cannot access.  We denote the composite systems $R_A^n$ and $R_B^n$ by ${\bar R}_A$ and ${\bar R}_B$, respectively. The entanglement cost is defined as the difference between the amount of initial entanglement and that of final entanglement shared by Alice and Bob, i.e., $K_n$ and $L_n$.}
\label{fig:dfntask}
\end{figure}

Suppose that Alice and Bob are given an unknown pure quantum state $\psi$ on a composite system $A^nB^n$, which is paired off as $A_1B_1\cdots A_nB_n$.  We consider a task in which they perform the same bipartite unitary $U^{AB}$ on each pair of systems $A_1B_1,\cdots ,A_nB_n$, or equivalently, perform $(U^{AB})^{\otimes n}$ on $A^nB^n$. For accomplishing this task, Alice and Bob apply a quantum operation consisting of local operations and classical communication (LOCC), in which case they need to consume an entangled state shared in advance as a resource. It is required that the error vanishes in the limit of $n$ to infinity. Our interest is to find the minimum cost of entanglement, forward and backward classical communication per copy for accomplishing this task, in an asymptotic limit of $n$ to infinity. Following the notations of \cite{waka15_markov}, we denote composite systems $A^n$ and $B^n$ by ${\bar A}$ and ${\bar B}$, respectively. 

We assume that the entangled state shared in advance is in the form of the maximally entangled state $\Phi_{K_n}^{A_0B_0}$. In general, an entangled state $\Phi_{L_n}^{A_1B_1}$ may be retrieved at the end of the operation.  Thus we define a {\it protocol} for accomplishing this task by a triplet of $\Phi_{K_n}$, $\Phi_{L_n}$ and a quantum operation ${\mathcal M}_n$ that consists of local operations and classical communication. The entanglement cost of a protocol is quantified by the number of copies of Bell pairs in the state $\Phi_{K_n}^{A_0B_0}$, subtracted by that in $\Phi_{L_n}^{A_1B_1}$ (i.e. $\log{K_n}-\log{L_n}$). 
 
For evaluating the error of a  protocol for this task, we adopt the average fidelity over all pure input state $|\psi\rangle$ on ${\bar A}{\bar B}$. That is, defining 
\begin{align}
\rho({\mathcal M}_n,\psi):={\mathcal M}_n(|\psi\rangle^{{\bar A}{\bar B}}|\Phi_{K_n}\rangle^{A_0B_0}),\nn
\end{align}
we will use the following function to evaluate the error of ${\mathcal M}_n$:
\begin{align}
&\!\!\!F_{\rm av}({\mathcal M}_n,K_n,L_n;U^{\otimes n})\nn\\
&\!\!\!:=\int_{\rm Haar} p(d\psi)\:F(\rho({\mathcal M}_n,\psi),(U^{\otimes n}|\psi\rangle)^{{\bar A}{\bar B}}|\Phi_{L_n}\rangle^{A_0B_0}).\label{eq:ensfide}
\end{align}
Here, the average is taken with respect to the Haar measure on ${\mathcal H}^{\bar A}\otimes{\mathcal H}^{\bar B}$. 

 Another task is one in which Alice and Bob apply $(U^{AB})^{\otimes n}$ on $(|\Phi_d\rangle^{AR_A}|\Phi_d\rangle^{BR_B})^{\otimes n}$ by LOCC, using an entangled state $\Phi_{K_n}^{A_0B_0}$ and retrieving $\Phi_{L_n}^{A_1B_1}$  as well at the end. Here, $R_A$ and $R_B$ are imaginary reference systems that are inaccessible to Alice and Bob (see Figure \ref{fig:dfntask}). We introduce notations
\begin{eqnarray}
|\Psi_U\rangle:=U^{AB}|\Phi_d\rangle^{AR_A}|\Phi_d\rangle^{BR_B}\nn
\end{eqnarray}
and
\begin{eqnarray}
\rho({\mathcal M}_n):={\mathcal M}_n((|\Phi_d\rangle^{AR_A}|\Phi_d\rangle^{BR_B})^{\otimes n}|\Phi_{K_n}\rangle^{A_0B_0}).\label{eq:deffinst}
\end{eqnarray}
The error of a protocol ${\mathcal M}_n$ is quantified by the following function:
\begin{align}
&F_{\rm en}({\mathcal M}_n,K_n,L_n;U^{\otimes n}):=F(\rho({\mathcal M_n}), |\Psi_U\rangle^{\otimes n}|\Phi_{L_n}\rangle^{A_1B_1}).\nn\\
&\label{eq:entfide}
\end{align}
 These tasks are equivalent since
\begin{align}
F_{\rm av}({\mathcal M}_n,K_n,L_n;U^{\otimes n})\approx1\label{eq:bFapprox}
\end{align}
if and only if
\begin{align}
F_{\rm en}({\mathcal M}_n,K_n,L_n;U^{\otimes n})\approx1,\label{eq:Feapprox}
\end{align}
which is proven in Appendix \ref{app:equivbe} using the relation between the average fidelity and entanglement fidelity\cite{nielsen2002simple}. Note that, since $R_A$ and $R_B$ are inaccessible systems, we  cannot apply the result of \cite{lo2001concentrating}, which analyzed convertibility of bipartite pure entangled states by LOCC protocols.

Let us introduce a rigorous definition.
\begin{dfn}\label{dfn:theprotocol}
Let $U$ be a bipartite unitary acting on two $d$-dimensional quantum systems $A$ and $B$. Let Alice and Bob have quantum registers $\{A_0,A_1\}$ and $\{B_0,B_1\}$, respectively, and let ${\mathcal M}_n$ be a quantum operation from ${\bar A}A_0\otimes {\bar B}B_0$ to ${\bar A}A_1\otimes{\bar B}B_1$. ${\mathcal M}_n$ is called an {\it $(r,n,\epsilon)$-protocol for implementing $U$ with the entanglement cost $E_n$}, if ${\mathcal M}_n$ is an $r$-round LOCC  and there exist natural numbers $K_n$, $L_n$ such that 
\begin{eqnarray}
F_{\rm en}({\mathcal M}_n,K_n,L_n;U^{\otimes n})\geq1-\epsilon
\label{eq:fidelityn}
\end{eqnarray}
and
\begin{align}
E_n:=\log{K_n}-\log{L_n}.\nn
\end{align}
\end{dfn}

Let $r_{A\rightarrow B}$ and $r_{B\rightarrow A}$ be numbers of times of classical communication  rounds from Alice to Bob and from Bob to Alice, respectively, in an $r$-round protocol ${\mathcal M}_n$. By definition, we have $r=r_{A\rightarrow B}+r_{B\rightarrow A}$. We denote by $C_{f,(\gamma)}$ the bit length of  the classical message transmitted in the $\gamma$-th communication from Alice to Bob, and by $C_{b,(\gamma)}$ the one from Bob to Alice. The forward classical communication cost $C_{f,n}$ of ${\mathcal M}_n$ is defined as the sum of numbers of classical bits transmitted from Alice to Bob in ${\mathcal M}_n$, that is,
\begin{align}
C_{f,n}:=\sum_{\gamma=1}^{r_{A\rightarrow B}}C_{f,(\gamma)}.\nn
\end{align}
In the same way, the backward classical communication cost $C_{b,n}$ of ${\mathcal M}_n$ is defined as
\begin{align}
C_{b,n}:=\sum_{\gamma=1}^{r_{B\rightarrow A}}C_{b,(\gamma)}.\nn
\end{align}

As mentioned above, our interest is to find the minimum cost of entanglement, forward and backward classical communication per copy for accomplishing this task, in an asymptotic limit of $n\rightarrow\infty$ and $\epsilon\rightarrow0$. This leads us to define  the achievability of the cost of resources.

\begin{dfn}\label{dfn:achievablerate}
A rate triplet $(E,C_f,C_b)$ is said to be achievable by an $r$-round protocol if  there exists a sequence of $(r,n,\epsilon_n)$-protocols for implementing $U$ ($n=1,2,\cdots$), with the entanglement cost $E_n=nE$, forward classical communication cost $C_{f,n}=nC_f$ and backward classical communication cost $C_{b,n}=nC_b$ for each $n$, such that $\lim_{n\rightarrow\infty}\epsilon_n=0$.  
\end{dfn}

\section{Results}\label{sec:results}

The results of this paper are summarized as follows.

\subsection{Result 1: Achievable Rate Region for Two-Round Protocols}\label{sec:result1}

In this paper, we mainly consider two-round protocols (i.e. $r=2$) starting with Alice's operation. In general, such a protocol proceeds as follows:  Alice first performs a measurement and communicates the outcome to Bob; Bob then performs a measurement and communicates the outcome to Alice; and, finally, Alice performs an operation. The first main result of this paper is that the optimal  rate of the cost of entanglement and of classical communication in a two-round protocol are given by a   quantity called the {\it Markovianizing cost of the unitary}, under an additional requirement that the error vanishes faster than $1/n^4$. 

A tripartite quantum state $\Upsilon^{ABC}$ is called a {\it Markov state conditioned by $B$} if it satisfies $I(A:C|B)_\Upsilon=0$\cite{hayden04}. {\it Markovianization} as formulated in \cite{waka15_markov} is a task in which $n$ copies of a tripartite state $\rho^{ABC}$ is transformed by a randomizing operation on ${\bar A}$ to a Markov state conditioned by ${\bar B}$. The {\it Markovianizing cost of $\rho^{ABC}$} is defined as the minimum  cost of randomness per copy required for the task, in an asymptotic limit of infinite copies and vanishingly small error. A rigorous definition is as follows.

\begin{dfn}\label{dfn:mofpsi}
A tripartite state $\rho^{ABC}$ is {\it Markovianized} with the randomness cost $R$ on $A$, conditioned by $B$, if the following statement holds. That is, for any $\epsilon>0$, there exists $n_\epsilon$ such that for any $n\geq n_\epsilon$, we find a random unitary operation ${\mathcal V}_n:\tau\mapsto2^{-nR}\sum_{k=1}^{2^{nR}}V_k\tau V_k^{\dagger}$ on ${\bar A}$ and a Markov state $\Upsilon^{{\bar A}{\bar B}{\bar C}}$ conditioned by ${\bar B}$ that satisfy
\begin{eqnarray}
\left\|{\mathcal V}_n^{{\bar A}}(\rho^{\otimes n})-\Upsilon^{{\bar A}{\bar B}{\bar C}}\right\|_1\leq\epsilon.
\label{eq:defmarkovianizing}
\end{eqnarray}
The {\it Markovianizing cost} of $\rho^{ABC}$ is defined as  $M_{A|B}(\rho^{ABC}):=\inf\{R\:|\:\rho^{ABC}$ is Markovianized with the randomness cost $R$ on $A$, conditioned by $B\}$.
\end{dfn}
We extend the notion of Markovianizing cost to a bipartite unitary as follows.
\begin{dfn}\label{dfn:markofu}
Let $U$ be a bipartite unitary acting on two $d$-level systems $A$ and $B$,  and consider a ``tripartite'' state
\begin{eqnarray}
|\Psi_U\rangle^{AR_A(\!BR_B\!)}:=(U^{AB}\otimes I^{R_AR_B})|\Phi_d\rangle^{AR_A}|\Phi_d\rangle^{BR_B}\label{eq:tripartiteu}
\end{eqnarray}
by regarding $B$ and $R_B$ as a single system. The Markovianizing cost of $U$ is defined as $M(U):=M_{A|R_A}(\Psi_U^{AR_A(\!BR_B\!)})$.
\label{dfn:markovianizingcostofu}
\end{dfn}

The main result of this paper is presented by the following theorem.  The proofs are given in Section \ref{sec:achievabletworound} and the corresponding appendices, after preparatory arguments in Section \ref{sec:preliminaries} and \ref{sec:singletworound}.

\begin{thm}\label{thm:misoptach}\hfill
\begin{itemize}
\item {\it Direct:} A rate triplet $(E,C_f,C_b)$ is achievable by  a two-round protocol for implementing $U$ if $E,C_f,C_b>M(U^\dagger)$.
\item {\it Converse:} A rate triplet $(E,C_f,C_b)$ is achievable by  a two-round protocol for implementing $U$ only if $E,C_f,C_b\geq M(U^\dagger)$, if we additionally require in Definition \ref{dfn:achievablerate} that
\begin{align}
\lim_{n\rightarrow\infty}n^4\cdot\epsilon_n=0.\label{eq:condconv}
\end{align}
\end{itemize}
\end{thm}

It is left open whether the same converse bound holds when we  remove Condition (\ref{eq:condconv}). As we will discuss in Section \ref{sec:open} in detail, this question is directly related to another question  of whether Equality (\ref{eq:mreqmrm}) holds without Condition (\ref{eq:condconv}). At the core of these questions lies an open problem regarding an ``asymptotic symmetry'' of approximate recoverability.

\subsection{Result 2: General Lower Bound on the Cost of Resources}\label{sec:result2}

 It  is proven in \cite{nielsen03,oppenheim04} that any bipartite unitary $U$ on $AB$ is decomposed as
\begin{eqnarray}
U^{AB}=\sum_{s=0}^{d^2-1}c_sE_s^A\otimes F_s^B,\label{eq:opsch}
\end{eqnarray}
where $c_s\:(s=0,\cdots,d^2-1)$ are nonnegative real numbers that satisfy
\begin{eqnarray}
\sum_{s=0}^{d^2-1}c_s^2=1,\;c_s\geq0\:(\forall s),\nn
\end{eqnarray}
and $E_s\in{\mathcal L}({\mathcal H}^A),\;F_s\in{\mathcal L}({\mathcal H}^B)$ are linear operators which are orthonormal with respect to the Hilbert-Schmidt inner product, i.e.,
\begin{eqnarray}
\frac{1}{d}{\rm Tr}[E_{s}^\dagger E_{s'}]=\frac{1}{d}{\rm Tr}[F_s^\dagger F_{s'}]=\delta_{ss'}.\label{eq:orthoef}
\end{eqnarray}
The Shannon entropy of $\{c_s^2\}_s$ is called the {\it Schmidt strength} of $U$. We denote it by $K(U)$, that is,
\begin{eqnarray}
K(U):=H(\{c_s^2\}_s)=-\sum_sc_s^2\log{c_s^2}.\label{dfn:kofu}
\end{eqnarray}

The following theorem states that a lower bound on the minimum  cost of entanglement and classical communication, in a protocol with arbitrary number of rounds of communication, is given by the Schmidt strength of the unitary. Proofs are given in Section \ref{sec:lowerbound} and the corresponding appendices.
\begin{thm}\label{thm:lowerbound}
A rate triplet $(E,C_f,C_b)$ is achievable only if $E,C_f,C_b\geq K(U)$.
\end{thm}

\begin{rmk}
Instead of the average fidelity (\ref{eq:ensfide}) and the entanglement fidelity (\ref{eq:entfide}), one could use the worst-case fidelity as a figure of merit of error. Let $\psi'$ be a pure state on a system ${\bar A}{\bar B}R$, with $R$ being a reference system, and define
\begin{align}
\rho({\mathcal M}_n,\psi'):={\mathcal M}_n(|\psi'\rangle^{{\bar A}{\bar B}R}|\Phi_{K_n}\rangle^{A_0B_0}).\nn
\end{align}
The worst-case fidelity is defined as
\begin{align}
& F_{\rm w.c.}({\mathcal M}_n,K_n,L_n;U^{\otimes n})\nn\\
&\quad:=\inf_R\inf_{\psi'} F(\rho({\mathcal M}_n,\psi'),(U^{\otimes n}|\psi'\rangle)^{{\bar A}{\bar B}R}|\Phi_{L_n}\rangle^{A_0B_0}),\nn
\end{align}
where the infimum is taken over all reference systems $R$ and all pure states $\psi'$ on ${\bar A}{\bar B}R$. Since we have $F_{\rm w.c.}\leq F_{\rm en}$, the converse bounds in Theorem \ref{thm:misoptach} and \ref{thm:lowerbound} remain to hold under this choice of the fidelity. However, we do not know whether the direct part in Theorem \ref{thm:misoptach} holds as well. It  is proven in  \cite{dur2015deterministic} and \cite{chiribella2015universal} that the ``superreplication'' of an unknown unitary gate is possible with vanishingly small average error, while it is {\it not} possible with vanishingly small worst-case error. In analogy to these results, it could be natural to expect that there is a gap between the minimum cost of resources for implementing a bipartite unitary with vanishing average error and the one for implementing it with vanishing worst-case error.
\end{rmk}

\section{Preliminaries} \label{sec:preliminaries}

We review  an alternative definition of the Markovianizing cost \cite{waka15_rec}, in addition to state merging \cite{horo05,horo07}. The results reviewed here are used in the following sections  to prove Theorem \ref{thm:misoptach}.

\subsection{Markovianization  in terms of Recoverability}\label{subsec:markov}

It is proved in \cite{hayden04} that the following conditions are equivalent:
\begin{enumerate}
\item {\it Vanishing QCMI:} $\rho^{ABC}$ is a Markov state conditioned by $B$, i.e., it satisfies
\begin{eqnarray}
I(A:C|B)_\rho=0.\nn
\end{eqnarray}
\item {\it Recoverability:} $\rho^{ABC}$ is recoverable from its bipartite reduced state on $AB$ and $BC$, that is, there exist quantum operations ${\mathcal R}:B\rightarrow AB$ and ${\mathcal R}':B\rightarrow BC$ such that 
\begin{eqnarray}
\rho^{ABC}={\mathcal R}(\rho^{BC})={\mathcal R}'(\rho^{AB}).\label{eq:exrecov}
\end{eqnarray}
\end{enumerate}
Based on this fact, the {\it Markovianizing cost in terms of recoverability} is introduced in \cite{waka15_rec}. In the same way as Definition \ref{dfn:mofpsi}, we consider a task in which $n$ copies of a tripartite quantum state $\rho^{ABC}$ is transformed by a random unitary operation on ${\bar A}$. Instead of requiring that the state after the operation satisfies Condition (\ref{eq:defmarkovianizing}), however, we now require that the state satisfies Condition (\ref{eq:exrecov}) up to a small error $\epsilon$. Rigorous definitions are as follows.

\begin{dfn}
A tripartite state $\rho^{ABC}$ is  said to be {\it$\epsilon$-recoverable from $BC$} if there exists a quantum operation ${\mathcal R}:B\rightarrow AB$ such that
\begin{eqnarray}
\left\|\rho^{ABC}-{\mathcal R}(\rho^{BC})\right\|_1\leq\epsilon.\nn
\end{eqnarray}
$\rho^{ABC}$ is {\it$\epsilon$-recoverable from $AB$} if there exists a quantum operation ${\mathcal R}':B\rightarrow  BC$ such that
\begin{eqnarray}
\left\|\rho^{ABC}-{\mathcal R}'(\rho^{AB})\right\|_1\leq\epsilon.\nn
\end{eqnarray}
\end{dfn}

\begin{figure}[t]
\begin{center}
\includegraphics[bb={0 0 722 348}, scale=0.33]{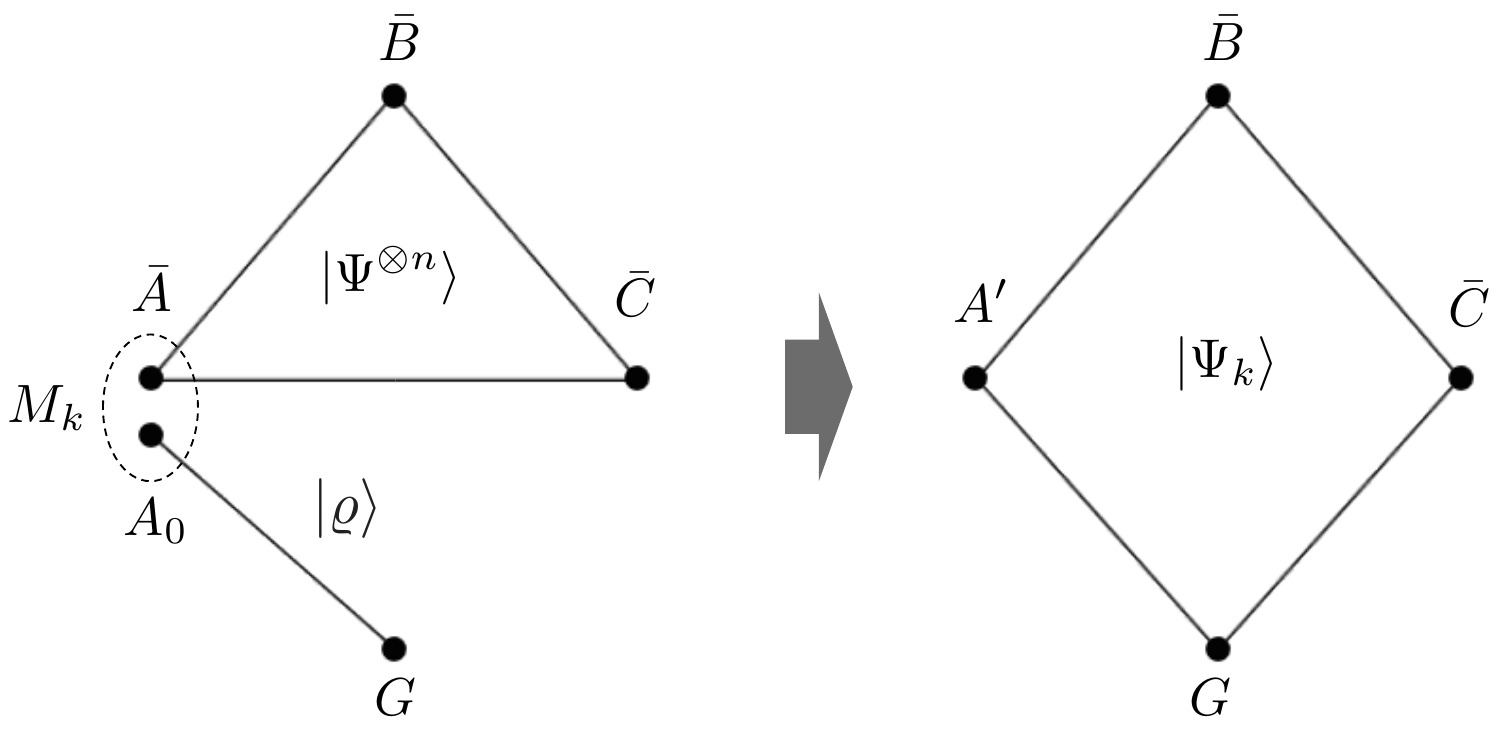}
\end{center}\caption{A graphical representation of Markovianization of a pure state by a measurement with an auxiliary entangled resource. After the measurement, the reduced state on $A'{\bar B}{\bar C}$ should be an approximately recoverable state.}
\label{fig:markovmeasure}
\end{figure}

\begin{dfn}\label{dfn:markfrombc}
A tripartite state $\rho^{ABC}$ is {\it Markovianized with the randomness cost $R$ on $A$, in terms of recoverability from $BC$},  if the following statement holds. That is, there exists a sequence of sets of unitaries $\{\{V_{n,k}\}_{k=1}^{2^{nR}}\}_{n=1}^\infty$, with each $V_{n,k}$ acting on $({\mathcal H}^A)^{\otimes n}$, such that ${\mathcal V}_n((\rho^{ABC})^{\otimes n})$ is $\epsilon_n$-recoverable from ${\bar B}{\bar C}$ for
\begin{align}
{\mathcal V}_n:\tau\mapsto2^{-nR}\sum_{k=1}^{2^{nR}}V_k\tau V_k^{\dagger}\nn
\end{align}
and $\lim_{n\rightarrow\infty}\epsilon_n=0$.

This allows us to define the {\it Markovianizing cost of $\rho^{ABC}$ in terms of recoverability from $BC$}  as $M^R_{A|BC}(\rho^{ABC}):=\inf\{R\:|\:\rho^{ABC}$ is Markovianized with the randomness cost $R$ on $A$, in terms of recoverability from $BC\}$.
\end{dfn}

We also consider  a Markovianization induced by a measurement, supplemented by auxiliary entanglement resource (Figure \ref{fig:markovmeasure}).

\begin{dfn}\label{dfn:rofpsi}
 Consider a tripartite pure state $|\Psi\rangle^{ABC}$, and let $A_0$ and $G$ be additional quantum systems.  A pair of a pure state $\ket{\varrho}^{A_0G}$ and a measurement on ${\bar A}A_0$, which is described by a set of measurement operators $\{M_k^{{\bar A}A_0\rightarrow A'}\}_{k\in{\mathbb K}}$, is called an $(n,R,\epsilon)$-Markovianization pair for a state $|\Psi\rangle^{ABC}$ if it satisfies the following conditions:
\begin{enumerate}
\item The measurement does not significantly change the reduced state on ${\bar B}{\bar C}$ on average, i.e.,
\begin{eqnarray}
\sum_{k\in{\mathbb K}}p_{k}\left\|(\Psi^{\otimes n})^{{\bar B}{\bar C}}-\Psi_{k}^{{\bar B}{\bar C}}\right\|_1\leq\epsilon,\label{eq:iamyourfather}
\end{eqnarray}
where $p_{k}$ is the probability of obtaining the outcome $k$, and ${\Psi_{k}}$ is the post-measurement state corresponding to the outcome $k$.
\item The post-measurement state is approximately recoverable on average, that is, there exist linear CPTP maps ${\mathcal R}_k:{\bar B}\rightarrow{\bar B}{\bar C}\:(k\in{\mathbb K})$ satisfying
\begin{eqnarray}
\sum_{k\in{\mathbb K}}p_{k}\left\|\Psi_{k}^{A'{\bar B}{\bar C}}-{\mathcal R}_k(\Psi_{k}^{A'{\bar B}})\right\|_1\leq\epsilon.\label{eq:recoverabilize}
\end{eqnarray}
\item The correlation between ${\bar B}{\bar C}$ and $G$ produced by the measurement is at most $nR$ bits in QMI, that is,
\begin{eqnarray}
I({\bar B}{\bar C}:G)_{av}:=\sum_{k\in{\mathbb K}}p_{k}I({\bar B}{\bar C}:G)_{\Psi_{k}}\leq nR.
\end{eqnarray}
\end{enumerate}  
 A state $|\Psi\rangle^{ABC}$ is  said to be {\it Markovianized with the correlation production $R$ by a measurement on $A$, in terms of recoverability from $AB$}, if there exists a sequence of $(n,R,\epsilon_n)$-Markovianization pairs ($n=1,2,\cdots$) such that $\lim_{n\rightarrow\infty}\epsilon_n=0$.

 Correspondingly, the {\it measurement-induced Markovianizing cost of $|\Psi\rangle^{ABC}$ in terms of recoverability from $AB$} is defined as $M^{R,m}_{A|AB}(\Psi^{ABC}):=\inf\{R\;|\;|\Psi\rangle^{ABC}$ is Markovianized with the correlation production $R$ by a measurement on $A$, in terms of recoverability from $AB\}$.
\end{dfn}

 The two types of Markovianizing costs defined above are equal  to that in Definition \ref{dfn:mofpsi} for pure states, if we impose an additional requirement on the convergence speed of the error in Definition \ref{dfn:rofpsi}\cite{waka15_rec}.

\begin{thm}\label{thm:mreqmrm}(Theorem 11 and  15 in \cite{waka15_rec})
For any tripartite pure state $|\Psi\rangle^{ABC}$, we have
\beq
 M_{A|B}(\Psi^{ABC})=M^{R}_{A|BC}(\Psi^{ABC})=M^{R,m}_{A|AB}(\Psi^{ABC}),\label{eq:mreqmrm}
\eeq
if we additionally require in Definition \ref{dfn:rofpsi} that
\begin{align}
 \lim_{n\rightarrow\infty}n\cdot \epsilon_n=0.\label{eq:condconvrec}
\end{align}
\end{thm}
 The following lemma relates the Markovianizing cost to other entropic quantities characterizing the state transformation induced by a Markovianizing measurement.

\begin{lmm}\label{thm:recov}(See Appendix \ref{app:prfrecov} for a proof.)
 There exists a nonnegative function ${\tilde\xi}(x)$ such that $\lim_{x\rightarrow0}{\tilde\xi}(x)=0$, and the following inequalities hold for any $n\in{\mathbb N}$, $\epsilon>0$ and $\{M_k^{{\bar A}A_0\rightarrow A'}\}_k$ satisfying $16\epsilon<n\leq1/4\epsilon$ and Inequalities (\ref{eq:iamyourfather}) and (\ref{eq:recoverabilize}):
\begin{eqnarray}
H(\{p_k\}_{k\in{\mathbb K}})&\geq&\Delta S(A')_{av}\;\geq\;\Delta S(A')_{av}\!-\!\Delta{S}(G)_{av}\nn\\
&\geq&nM_{A|B}(\Psi^{ABC})-n{\tilde\xi}( n\epsilon).\label{eq:markovmeasure3}
\end{eqnarray}
 Here, we defined
\begin{eqnarray}
\Delta S(A')_{av}&:=&nS(A)+S(A_0)_{\varrho}-\sum_{k\in{\mathbb K}}{p_{k}}S(A')_{\Psi_{k}},\nonumber\\
\Delta{S}(G)_{av}&:=&S(G)_{\varrho}-\sum_{k\in{\mathbb K}}{p_{k}}S(G)_{\Psi_{k}}.\nonumber
\end{eqnarray}

\end{lmm}

\begin{rmk}
It has been left open whether Equality (\ref{eq:mreqmrm}) holds when we  remove Condition (\ref{eq:condconvrec}). See  Proposition 13 (unproven) of \cite{waka15_rec}.
\end{rmk}

\subsection{State Merging}
 
\begin{figure}[t]
\begin{center}
\includegraphics[bb={0 0 512 212}, scale=0.43]{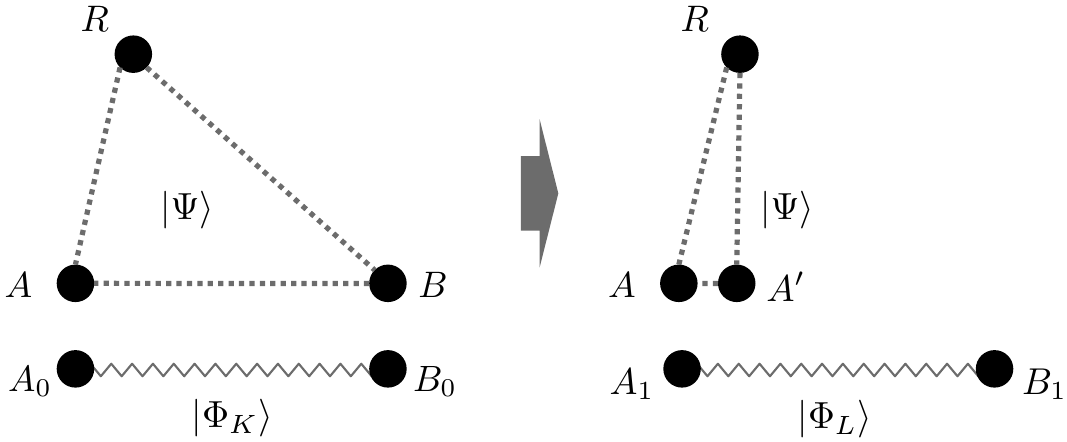}
\end{center}\caption{State merging is a task in which Bob transfers his share of $|\Psi\rangle^{ABR}$ to Alice. $R$ is an inaccessible reference system. For  the sake of presentation, we consider Bob as the sender and Alice as the receiver.}
\label{fig:statemerging}
\end{figure}

Suppose Alice and Bob share a tripartite pure state $|\Psi\rangle^{ABR}$ with an inaccessible reference system $R$. State merging (\!\!\cite{horo05,horo07}) is a task in which Bob sends his share of $\Psi$ to Alice so that Alice has both $A$ and $B$ parts of $\Psi$, or equivalently, so that Alice has the whole part of the purification of $\Psi^R$. (See Figure \ref{fig:statemerging}. For later convenience, we exchange roles of Alice and Bob in the  standard formulation.)  Our concern is the  cost of entanglement and classical communication required for  state merging. A rigorous definition is given as follows.

\begin{dfn}\label{dfn:sm}
Consider a tripartite pure state $\ket{\Psi}^{ABR}$. Let Alice and Bob have quantum systems  $\{A_{\!B},A_0,A_1\}$ and $\{B_0,B_1\}$, respectively, where $A_{\!B}$ is assumed to be identical to $B$. The following protocol ${\mathcal N}$ consisting of a sequence of quantum operations is called state merging of $\Psi$ with error $\epsilon$, entanglement cost $\log{K}-\log{L}$ and classical communication cost $C$. Here, ${\mathcal N}:AA_0BB_0\rightarrow AA_{\!B}A_1B_1$ is a one-way LOCC  from Bob to Alice, such that
\begin{eqnarray}
F(\rho({\mathcal N}), |\Psi\rangle^{AA_{\!B}R}|\Phi_{L}\rangle^{A_1B_1})\geq1-\epsilon
\label{eq:fidelityqsm}
\end{eqnarray}
for
\begin{align}
\rho({\mathcal N}):={\mathcal N}(|\Psi\rangle^{ABR}|\Phi_{K}\rangle^{A_0B_0}),\nn
\end{align}
 $K$ and $L$ are natural numbers, and $\Phi_{K}$ and $\Phi_{L}$ are the maximally entangled states with the Schmidt rank $K$ and $L$, respectively. $C$ is  the total amount of classical communication transmitted from Bob to Alice in ${\mathcal M}$, measured  in bits. 
\end{dfn}
There always exists  a state merging with an error determined by the initial state. The following theorem is obtained as a corollary of Proposition 3 and 4 in \cite{horo07} by letting $L=1$.
\begin{thm}\label{thm:achieveqsm}
Let $D_A:=({\rm Tr}[(\Psi^A)^2])^{-1}$ and $r_{\!{}_B}:={\rm rank}[\Psi^B]$. There exists  a state merging of $\Psi$ with entanglement cost $0$, classical communication cost $C=\log{r_{\!{}_B}}$ and error
\begin{align}
\epsilon\leq2\sqrt{2}(\sqrt{d_R/{D_A}}+1/{r_{\!{}_B}})^{1/2}.\label{eq:mergerrbound} 
\end{align}
\end{thm}
It is also proved in \cite{horo07} that the  cost of entanglement and classical communication in  a state merging are bounded from below as
\begin{eqnarray}
\log{K}-\log{L}\gtrsim nS(B|A)_{\Psi},\;\;C\gtrsim nI(B:R)_{\Psi},\label{eq:qsmentccopt}
\end{eqnarray}
when $\epsilon$ is sufficiently small. See Appendix \ref{app:statemerging} for details.

\section{Single-shot  Two-Round Protocols}\label{sec:singletworound}

\begin{figure}[t]
\begin{center}
\includegraphics[bb={0 0 532 202}, scale=0.43]{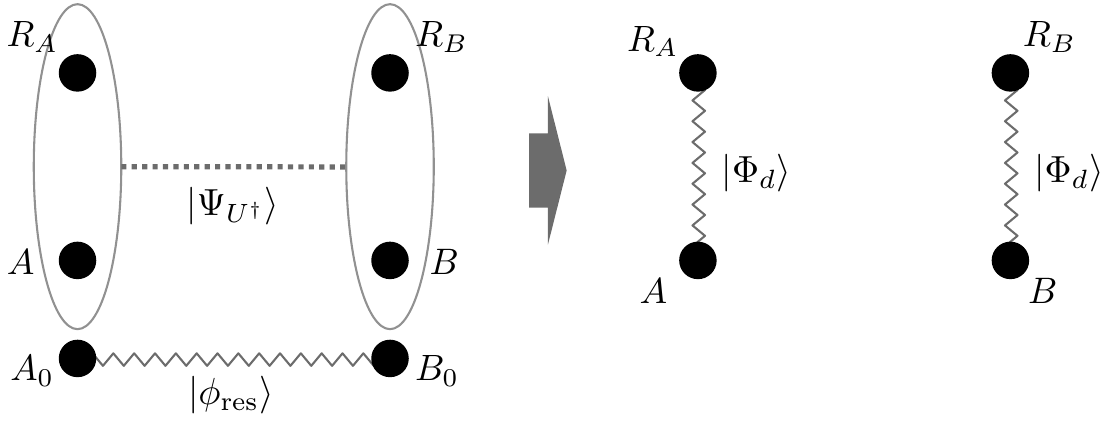}
\end{center}\caption{A graphical representation of a task that we analyze in Section \ref{sec:singletworound}. The task is to destroy correlation between $AR_A$ and $BR_B$, while preserving the maximal entanglement between $AB$ and $R_AR_B$. A certain amount of shared entanglement must be consumed to accomplish this task, since reference systems $R_A$ and $R_B$ are inaccessible.}
\label{fig:singleshottask}
\end{figure}

In this section, we consider $n=1$ (single-shot) case, and analyze a single-shot protocol ${\mathcal M}$ for implementing $U$ by two-round LOCC assisted by shared entanglement. The results obtained here are then applied to the asymptotic situation in Section \ref{sec:achievabletworound}.

Let ${\mathcal M}:AA_0\otimes BB_0\rightarrow AA_1\otimes BB_1$ be a two-round LOCC protocol for implementing $U$. ${\mathcal M}$ succeeds in implementing $U$  with high fidelity,  if
\begin{eqnarray}
F({\rho}({\mathcal M})^{AR_ABR_B},|\Psi_U\rangle)\geq1-\epsilon
\label{eq:1shoterror}
\end{eqnarray}
for  some small $\epsilon$, where
\beq
{\rho}({\mathcal M}):={\mathcal M}(|\Phi_d\rangle^{AR_A}|\Phi_d\rangle^{BR_B}|\phi_{\rm res}\rangle^{A_0B_0})\nn
\eeq
and $\phi_{\rm res}$ is a pure resource state shared in advance.  Since we have $\Phi_d^{A}\otimes\Phi_d^{B}=\Psi_U^{AB}$, and all purifications are equivalent up to local unitary transformations, there exists a unitary $\hat U$ on $R_AR_B$ such that
\beq
|\Phi_d\rangle^{AR_A}|\Phi_d\rangle^{BR_B}={\hat U}^{R_AR_B}|\Psi_U\rangle^{AR_ABR_B}.\nn
\eeq
Applying $U^{\dagger AB}$ on both sides yields
\beq
|\Psi_{U^\dagger}\rangle^{AR_ABR_B}&:=&U^{\dagger AB}|\Phi_d\rangle^{AR_A}|\Phi_d\rangle^{BR_B}\nn\\
&=&{\hat U}^{R_AR_B}|\Phi_d\rangle^{AR_A}|\Phi_d\rangle^{BR_B},\label{eq:sommisama}
\eeq
which leads to
\beq
{\rho}({\mathcal M},U^{\dagger})&:=&{\mathcal M}(|\Psi_{U^{\dagger}}\rangle^{AR_ABR_B}|\phi_{\rm res}\rangle^{A_0B_0})\nn\\
&=&{\hat U}^{R_AR_B}{\rho}({\mathcal M})^{AR_ABR_B}{\hat U}^{\dagger R_AR_B}.\nn
\eeq 
Note that $\mathcal M$ does not act on $R_AR_B$. Therefore, due to the unitary invariance of the fidelity, Condition (\ref{eq:1shoterror}) is equivalent to
\begin{eqnarray}
F({\rho}({\mathcal M},U^{\dagger})^{ABR_AR_B},\ket{\Phi_d}^{AR_A}\ket{\Phi_d}^{BR_B})\geq1-\epsilon.
\label{eq:1shoterror2}\end{eqnarray}

While $|\Phi_d\rangle|\Phi_d\rangle$ obviously has no correlation between $AR_A$ and $BR_B$, $|\Psi_{U^{\dagger}}\rangle$ has  a certain amount of entanglement depending on $U^{\dagger}$. Thus, for a given initial state $|\Psi_{U^{\dagger}}\rangle$ and a resource state $|\phi_{\rm res}\rangle$,  a successful protocol $\mathcal M$ decouples $AR_A$ and $BR_B$ while preserving the maximal entanglement between $AB$ and $R_AR_B$ (Figure \ref{fig:singleshottask}). Observe that both $|\Psi_{U^{\dagger}}\rangle$ and $|\Phi_d\rangle|\Phi_d\rangle$ are  maximally entangled states between $AB$ and $R_AR_B$ with Schmidt rank $d^2$. 

The main goal of this section is to derive conditions on operations that comprise $\mathcal M$, for the protocol to succeed with high fidelity.  It turns out that any successful protocol can be described as a combination of Markovianization and  a subsequent state merging. Consequently, as we describe in detail in Section \ref{sec:achievabletworound}, the minimum  cost of resources is derived by combining results on Markovianization and state merging presented in Section \ref{sec:preliminaries}. 

\begin{figure}[t]
\begin{center}
\includegraphics[bb={0 0 393 769}, scale=0.3]{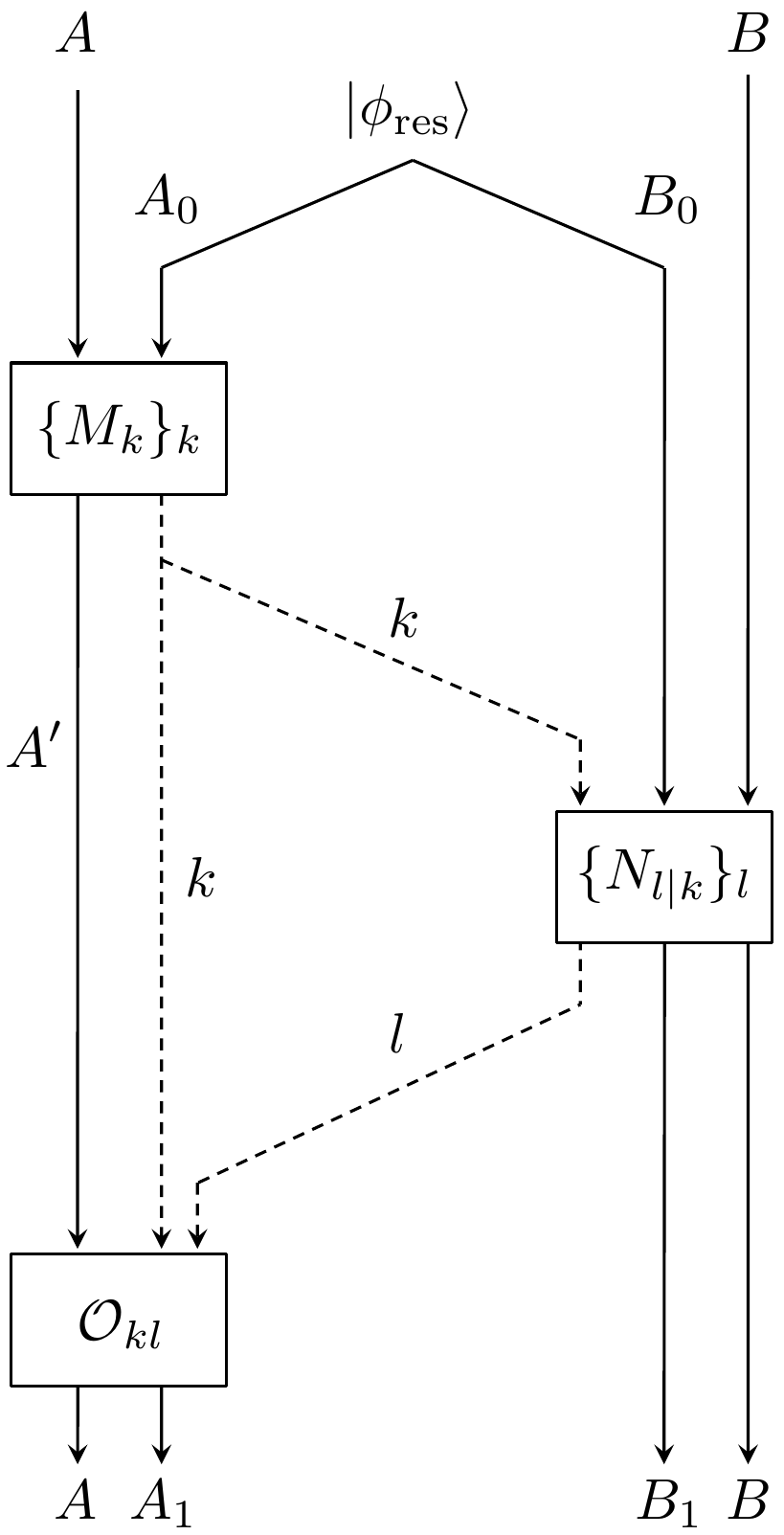}
\end{center}\caption{A graphical description of how the two-round protocol $\mathcal M$ proceeds. The solid lines represent quantum systems, the two dotted lines indicate classical communication, and the three boxes represent quantum measurements and operations.}
\label{fig:singleshotformmm}
\end{figure}

In the following, we fix a unitary $U$ acting on $AB$, and denote $\Psi_{U^\dagger}$ simply by $\Psi$. Without loss of generality, we assume that the two-round protocol $\mathcal M$ proceeds as follows (Figure \ref{fig:singleshotformmm}):
\begin{itemize}\setlength{\leftskip}{0.1cm}
\item[1.] Alice performs a measurement on $AA_0$, which is described by a set of measurement operators ${\mathbb M}=\{{ M}_{k}^{AA_0\rightarrow A'}\}_k$, and obtains an outcome $k$.
\item[2.] Alice communicates $k$ to Bob.
\item[3.] Bob performs a measurement on $BB_0$, described by ${\mathbb N}_k=\{{ N}_{l|k}^{BB_0\rightarrow BB_1}\}_{l}$, and obtains an outcome $l$.
\item[4.] Bob communicates $l$ to Alice.
\item[5.] Alice performs an operation which is described by a linear CPTP map ${\mathcal O}_{kl}:A'\rightarrow AA_1$.
\end{itemize}
Here, $A'$ is the output system of Alice's measurement such that
\begin{align}
{\rm dim}{\mathcal H}^{A'}\leq{\rm dim}{\mathcal H}^{A}\times{\rm dim}{\mathcal H}^{A_0}.\label{eq:dleqdd}
\end{align}
We denote the set of outcomes of Alice's measurement by $\mathbb K$.  (See Remark at the end of Appendix \ref{app:setting} for a treatment of protocols in which not all information about the measurement outcome $k$ is communicated to Bob.)

\subsection{Conditions on Alice's Measurement}
Let us first discuss general conditions regarding state transformations by Alice's measurement.  For a fixed $\phi_{\rm res}$, and for any linear operator $M:{\mathcal H}^A\otimes{\mathcal H}^{A_0}\rightarrow{\mathcal H}^{A'}$ such that $M^\dagger M\leq I$, we define a map ${\mathcal E}_{M}$ by
\begin{eqnarray}
&&{\mathcal E}_{M}(\tau):=p_{M}^{-1}M(\tau^{A}\otimes\phi_{res}^{A_0})M^\dagger,\nn\\
&&p_M:={\rm Tr}[M(\tau^{A}\otimes\phi_{res}^{A_0})M^\dagger].\nn
\end{eqnarray} 
We call ${\mathcal E}_{M}$ as an $M$-{\it induced map}. $M$ is supposed to be an element of $\mathbb M$, in which case $p_M$ describes the probability of obtaining a certain measurement outcome corresponding to $M$. Note that $p_M$ depends on the input state $\tau$ in general. Consequently, the $M$-induced map is not necessarily a linear map. The linearity of ${\mathcal E}_{M}$ is equivalent to the independence of $p_M$ from $\tau$, which indicates that the measurement is {\it oblivious} to the input, in the sense that it does not extract any information about the input state. This obliviousness condition plays an important role in proofs of most of the lemmas in this section (as well as in \cite{soeda11}). Thus we introduce an equivalent definition of approximate obliviousness as follows. 

\begin{dfn}
An $M$-induced map is {\it$\varsigma$-oblivious} if it satisfies
\begin{eqnarray}
\left\|\Phi_M^{R_A}-\pi_d^{R_A}\right\|_1\leq\varsigma,\nn
\end{eqnarray}
where $\Phi_M^{A'R_A}:={\mathcal E}_{M}^A(\Phi_d^{AR_A})$. A measurement ${\mathbb M}$ is {\it$\varsigma$-oblivious} if an $M_k$-induced map is $\varsigma_k$-oblivious for each $k$ and $\sum_{k\in{\mathbb K}}p_k\varsigma_k\leq\varsigma$.
\end{dfn}
We introduce some other conditions on Alice's measurement. In the following, we denote ${\mathcal E}_{M}(\Psi)$ as $\Psi_{M}$.

\begin{dfn}\label{def:epsilondecouple}
An $M$-induced map is {\it$\mu$-decoupling between $A'R_A$ and $R_B$} if it satisfies
\begin{eqnarray}
\left\|\Psi_{M}^{A'R_AR_B}-\Psi_{M}^{A'R_A}\otimes\Psi_{M}^{R_B}\right\|_1\leq\mu.\label{eq:dfnedec}
\end{eqnarray}
A measurement ${\mathbb M}$ is {\it$\mu$-decoupling between $A'R_A$ and $R_B$} if an $M_k$-induced map is $\mu_k$-decoupling between $A'R_A$ and $R_B$ for each $k$ and $\sum_{k\in{\mathbb K}}p_k\mu_k\leq\mu$.
\end{dfn}
 As in Definition \ref{dfn:markofu}, we now consider $B$ and $R_B$  as a single system and regard $\Psi_{M}^{A'R_ABR_B}$ as a ``tripartite'' state on $A'$, $R_A$ and $BR_B$.

\begin{dfn}\label{dfn:markrecrabra}
An $M$-induced map is {\it$\nu$- Markovianizing from $R_ABR_B$} if $\Psi_{M}^{A'R_ABR_B}$ is $\nu$-recoverable from $R_ABR_B$, that is, if there exists a linear CPTP map ${\mathcal R}:R_A\rightarrow A'R_A$ such that
\begin{eqnarray}
\left\|\Psi_{M}^{A'R_ABR_B}-{\mathcal R}(\Psi_{M}^{R_ABR_B})\right\|_1\leq\nu.\nn
\end{eqnarray} 
A measurement $\mathbb M$ is {\it$\nu$-Markovianizing from $R_ABR_B$} if  an $M_k$-induced map is $\nu_k$-Markovianizing from $R_ABR_B$ for each $k$ and $\sum_{k\in{\mathbb K}}p_k\nu_k\leq\nu$.
\end{dfn}

\begin{dfn}\label{dfn:markrecara}
An $M$-induced map is {\it$\iota$-Markovianizing from $A'R_A$} if $\Psi_{M}^{A'R_ABR_B}$ is $\iota$-recoverable from $A'R_A$, that is, if there exists a linear CPTP map ${\mathcal R}':R_A\rightarrow R_ABR_B$ such that
\begin{eqnarray}
\left\|\Psi_{M}^{A'R_ABR_B}-{\mathcal R}'(\Psi_{M}^{A'R_A})\right\|_1\leq\iota.\nn
\end{eqnarray}
A measurement $\mathbb M$ is {\it$\iota$-Markovianizing from $A'R_A$} if  an $M_k$-induced map is $\iota_k$-Markovianizing from $A'R_A$ for each $k$ and $\sum_{k\in{\mathbb K}}p_k\iota_k\leq\iota$.
\end{dfn}

The following two lemmas are at the core of the proofs of the main result, which translates the problem of finding the optimal costs  for implementing a bipartite unitary to that of computing the Markovianizing cost of a particular state.

\begin{lmm}\label{lmm:decoupleifmarkov}
A measurement $\mathbb M$ is $(3\varsigma\!+\!2\nu)$-decoupling between $A'R_A$ and $R_B$ if it is $\varsigma$-oblivious and $\nu$-Markovianizing from $R_ABR_B$. 
\end{lmm}

\begin{lmm}\label{lmm:markovifdecouple}
A measurement $\mathbb M$ is $(\varsigma+\mu)$-Markovianizing from $A'R_A$ if it is $\varsigma$-oblivious and $\mu$-decoupling between $A'R_A$ and $R_B$. 
\end{lmm}

Let us describe a simplified version of the proof of the above two lemmas in the case of $\mu=\nu=\varsigma=0$. The conditions of $\epsilon$-Markovianizing in Definition \ref{dfn:markrecrabra} and \ref{dfn:markrecara} are then equivalent to the condition that $\Psi_{M}^{A'R_ABR_B}$ is a Markov state conditioned by $R_A$. Suppose an $M$-induced map is $0$-oblivious, which implies $\Phi_{M}^{R_A}=\pi_d^{R_A}$. Using (\ref{eq:sommisama}), we see that
\beq
\Psi_{M}^{A'R_ABR_B}={\hat U}^{R_AR_B}(\Phi_M^{A'R_A}\otimes\Phi_d^{BR_B}){\hat U}^{\dagger R_AR_B},\label{eq:notfound}
\eeq
and consequently,
\beq
\Psi_{M}^{R_AR_B}&=&{\hat U}^{R_AR_B}(\pi_d^{R_A}\otimes\pi_d^{R_B}){\hat U}^{\dagger R_AR_B}\nn\\
&=&\pi_d^{R_A}\otimes\pi_d^{R_B}.\label{eq:hatupipi}
\eeq
Therefore, for the state $\Psi_{M}^{A'R_ABR_B}$, we have $I(A':B|R_AR_B)=0$ due to the local unitary invariance of QCMI, as well as $I(R_A:R_B)=0$. It follows that
\begin{eqnarray}
I(A':BR_B|R_A)&=&I(A':R_B|R_A)+I(A':B|R_AR_B)\nn\\
&=&I(A'R_A:R_B)-I(R_A:R_B)\nn\\
&=&I(A'R_A:R_B),\nn
\end{eqnarray}
which implies the equivalence between the conditions of decoupling and Markovianizing under the condition of obliviousness. 

For rigorous proofs, we need to relax the ``exact'' condition ($\mu=\nu=\varsigma=0$) to the ``approximate'' condition ($\mu,\nu,\varsigma>0$). See Appendices \ref{app:prfdecmar} and \ref{app:prfmardec} for details.

\subsection{Conditions for Achievability}\label{sec:condforach}

For the proof of the direct part of Theorem \ref{thm:misoptach}, let us consider how to construct a successful protocol. Let $\tilde B$ be a register on Bob's side which has a sufficiently large dimension. The following lemma states that Markovianization by Alice's measurement is a sufficient condition for the success of the first half of $\mathcal M$, in which $\Phi_d^{BR_B}$ is obtained from $|\Psi\rangle$.

\begin{lmm}\label{lmm:decoupleisometry}(See Appendix \ref{app:decoupleisometry} for a proof.) 
Suppose that a measurement $\mathbb M$ is $0$-oblivious and $\mu$-decoupling between $A'R_A$ and $R_B$, $\mu\in(0,1]$, and that $\Psi_{M_k}^{A'R_AR_B}$ does not depend on $k$. Then, there exist pure states $|{\tilde\Psi}^p\rangle^{A'R_A{\tilde B}}$, $|\Psi'\rangle^{A'R_A{\tilde B}BR_B}$ and isometries $W_k^{BB_0\rightarrow B{\tilde B}}\:(k\in{\mathbb K})$ such that
\begin{eqnarray}
\left\||\Psi'\rangle\!\langle\Psi'|-({\tilde\Psi}^p)^{A'R_A{\tilde B}}\otimes\Phi_d^{BR_B}\right\|_1\leq5\sqrt[4]{\mu}\label{eq:ohlonesomeme}
\end{eqnarray}
and $|\Psi'\rangle=W_k|\Psi_{M_k}\rangle$ for any $k\in{\mathbb K}$. In addition, $|{\tilde\Psi}^p\rangle$ satisfies
\begin{align}
\left\|({\tilde\Psi}^p)^{R_A}-\pi_d^{R_A}\right\|_1\leq3\sqrt[4]{\mu},\;{\rm rank}[({\tilde\Psi}^p)^{\tilde B}]\leq{\rm dim}{\mathcal H}^{B_0}.\nn
\end{align}
\end{lmm}
The following lemma immediately follows from Lemma \ref{lmm:decoupleifmarkov} and \ref{lmm:decoupleisometry}.

\begin{lmm}\label{lmm:decoupleisometryyy}
 Suppose that a measurement $\mathbb M$ is $0$-oblivious and $\nu$-Markovianizing from $R_ABR_B$, $\nu\in(0,1/2]$, and that $\Psi_{M_k}^{A'R_AR_B}$ does not depend on $k$.  Then, there exist pure states $|{\tilde\Psi}^p\rangle^{A'R_A{\tilde B}}$, $|\Psi'\rangle^{A'R_A{\tilde B}BR_B}$ and isometries $W_k^{BB_0\rightarrow B{\tilde B}}\:(k\in{\mathbb K})$ that satisfy
\begin{eqnarray}
&&\left\||\Psi'\rangle\!\langle\Psi'|-({\tilde\Psi}^p)^{A'R_A{\tilde B}}\otimes\Phi_d^{BR_B}\right\|_1\leq5\sqrt[4]{2\nu},\nn\\
&&|\Psi'\rangle=W_k|\Psi_{M_k}\rangle\quad(\forall k\in{\mathbb K})\nn
\end{eqnarray}
and
\begin{align}
\!\!\left\|({\tilde\Psi}^p)^{R_A}-\pi_d^{R_A}\right\|_1\leq3\sqrt[4]{2\nu},\;{\rm rank}[({\tilde\Psi}^p)^{\tilde B}]\leq{\rm dim}{\mathcal H}^{B_0}.\label{eq:rankineq}
\end{align}
\end{lmm}

The task remaining after obtaining $\Phi_d^{BR_B}$ is to obtain $\Phi_d^{AR_A}$ from tripartite pure states $|{\tilde\Psi}^p\rangle^{A'R_AB'}$, which is equivalent to performing state merging from Bob to Alice. Consequently, we can construct a successful protocol by combining Markovianization of $|\Psi\rangle$ and the subsequent state merging of $|{\tilde\Psi}^p\rangle$ from Bob to Alice.

\subsection{Conditions for Optimality}\label{sec:condforopt}

For the proof of the converse part of Theorem \ref{thm:misoptach}, let us analyze conditions on Alice's measurement imposed by (\ref{eq:1shoterror2}). Let ${\mathbb M}=\{{ M}_{k}^{AA_0\rightarrow A'}\}_{k\in{\mathbb K}}$ be Alice's measurement in protocol $\mathcal M$ that satisfies (\ref{eq:1shoterror2}). First, conservation of the maximal entanglement between systems $AB$ and $R_AR_B$ immediately implies that Alice's measurement must be oblivious. Second, since the final state is close to $|\Phi_d\rangle^{AR_A}|\Phi_d\rangle^{BR_B}$, correlation between $AR_A$ and $R_B$ is destroyed by $\mathcal M$. This part of decoupling must be accomplished by Alice's measurement alone, which implies that Alice's measurement must be Markovianizing due to Lemma \ref{lmm:markovifdecouple}. Hence we obtain the following lemma.

\begin{lmm}\label{lmm:markovifepsilon}(Appendices \ref{app:setting} and \ref{app:prfmarkovifepsilon})
The measurement $\mathbb M$ is $4\sqrt[4]{\epsilon}$-oblivious and $12\sqrt[4]{\epsilon}$-Markovianizing from $A'R_A$.
\end{lmm}

Let us continue to analyze conditions on Bob's measurement imposed by (\ref{eq:1shoterror2}). Let $B_E$ be an ancillary system, and let $W_{k}:BB_0\rightarrow BB_1B_E$ be an isometry such that the Naimark extension of Bob's measurement ${\mathbb N}_k$ is given by ${N}_{l|k}=\bra{l}^{B_E}W_{k}$. Define $|\Psi_{k}'\rangle:=W_k|\Psi_{k}\rangle$. The following lemma states that Bob's measurement is decomposed into two parts: (i) performing an isometry to obtain $\Phi_d^{BR_B}$, and (ii) performing a measurement on his share of a ``tripartite'' pure state on $A'$, $R_A$ and $B_1B_E$.

\begin{lmm}\label{thm:abcaabcc}(Appendix \ref{app:prfoblivi})
There exist pure states $|\Psi_{k}^p\rangle^{A'R_AB_1B_E}\:(k\in{\mathbb K})$ such that
\begin{eqnarray}
\sum_{k\in{\mathbb K}}p_{k}\left\|\Psi_{k}'-(\Psi_{k}^p)^{A'{R}_AB_1B_E}\otimes\Phi_d^{B{R}_B}\right\|_1\leq4\sqrt[4]{\epsilon}.\nn
\end{eqnarray}
\end{lmm}

\begin{figure}[t]
\begin{center}
\includegraphics[bb={0 0 787 950}, scale=0.3]{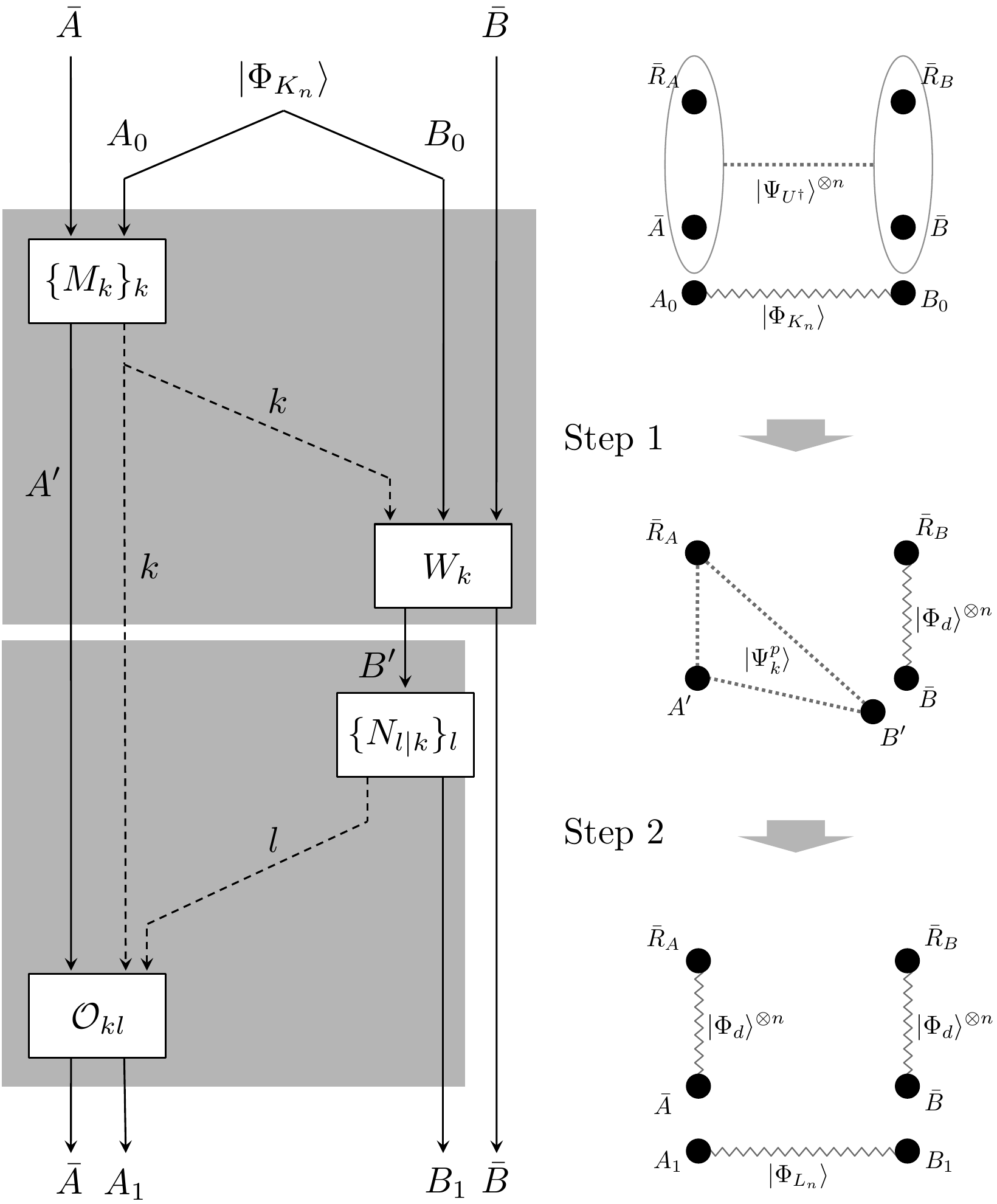}
\end{center}\caption{The two-step protocol is depicted. Step 1: Markovianization by Alice's measurement and an isometric operation by Bob, aiming at obtaining $|\Phi_d^{\otimes n}\rangle^{{\bar B}{\bar R}_B}$, and Step 2: state merging from Bob to Alice, for obtaining $|\Phi_d^{\otimes n}\rangle^{{\bar A}{\bar R}_A}$. In the proof of achievability, we consider a case where $A_1=B_1=\emptyset$ and $\Psi_k^p={\tilde\Psi}^p$. In the proof of optimality, $B'$ denotes a composite system $B_1B_E$, and $\{N_{l|k}\}_l$ is the projective measurement on $B_E$ in the basis $\{|l\rangle\}_l$.}
\label{fig:singleshotform}
\end{figure}

By the measurement on $B_E$ described by $\{|l\rangle\!\langle l|\}_l$ and an operation on $A'$ depending on $k$ and $l$, the maximal entanglement $\Phi_d^{AR_A}$ must be obtained from $|\Psi_{k}^p\rangle^{A'{R}_AB_1B_E}$. This transformation is equivalent to state merging from Bob to Alice. Consequently, any successful protocol is described as a combination of Markovianization of $|\Psi\rangle$ and the subsequent state merging of $|\Psi_{k}^p\rangle$ from Bob to Alice.

\section{Proof of Theorem \ref{thm:misoptach}}\label{sec:achievabletworound}
Let us return to the asymptotic scenario and prove Theorem \ref{thm:misoptach}. We consider protocols that transforms a state $|\Psi^{\otimes n}\rangle^{\bar{A}\bar{B}\bar{R}_A\bar{R}_B}|{\Phi_{K_n}}\rangle^{A_0B_0}$ into $|\Phi_d^{\otimes n}\rangle^{\bar{A}\bar{R}_A}|\Phi_d^{\otimes n}\rangle^{\bar{B}\bar{R}_B}$ $|{\Phi_{L_n}}\rangle^{A_0B_0}$, as depicted in the right side of Figure \ref{fig:singleshotform}. Conditions obtained in Section \ref{sec:singletworound} directly apply by the following correspondence:
\begin{eqnarray}
&&\!\!\!\!\!\!\!\!\!\!\!A,B,R_A,R_B\rightarrow{\bar A},{\bar B},{\bar R}_A,{\bar R}_B,\;\;\;U\rightarrow U^{\otimes n},\;\;\;\Phi_d\rightarrow\Phi_d^{\otimes n},\nn\\
&&\!\!\phi_{\rm res}\rightarrow\Phi_{K_n},\;\;\;\Psi\rightarrow\Psi^{\otimes n},\;\;\;\pi_d\rightarrow\pi_d^{\otimes n}.\nn
\end{eqnarray}

As presented in Section \ref{sec:singletworound}, two-round protocols for this task is decomposed into two steps (see the left side of Figure \ref{fig:singleshotform}). The first step is composed of Alice's measurement, forward classical communication and Bob's isometry. Markovianization by Alice's measurement satisfying the obliviousness condition is necessary and sufficient in order that Bob is able to obtain $(\Phi_d^{BR_B})^{\otimes n}$. The second step is composed of Bob's  measurement, backward classical communication and Alice's local operation. To obtain $(\Phi_d^{AR_A})^{\otimes n}$, it is necessary and sufficient that the second step implements state merging of a particular tripartite state.

\subsection{Direct Part}\label{sec:direct}
We prove the direct part of Theorem \ref{thm:misoptach}.  We assume $L_n=1$ in Definition \ref{dfn:theprotocol}, i.e., we consider a case where no entanglement is left after the protocol. The proof is by construction. Take arbitrary $R>M(U^\dagger)$, small $\epsilon,r>0$, choose sufficiently large $n$ and let $K_n=2^{n(R+r)}$. Divide the resource state $\Phi_{K_n}^{A_0B_0}$ as
\begin{eqnarray}
\Phi_{K_n}^{A_0B_0}\rightarrow\Phi_{2^{nR}}^{A_0B_0}\otimes\Phi_{2^{nr}}^{{\tilde A}_0{\tilde B}_0}.\label{eq:initialent}
\end{eqnarray}
Consider a protocol consisting of the following steps.
\begin{enumerate}\setlength{\leftskip}{-0.15cm}
\item {\bf Alice's measurement}: By  Definition \ref{dfn:markofu}, \ref{dfn:markfrombc} and Theorem \ref{thm:mreqmrm}, there exists a random unitary operation ${\mathcal V}_n:\tau\mapsto2^{-nR}\sum_{j=1}^{2^{nR}}V_j\tau V_j^{\dagger}$ on $\bar A$ such that ${\mathcal V}_n(\Psi^{\otimes n})$ is $\epsilon$-recoverable from ${\bar R}_A{\bar B}{\bar R}_B$. Using $V_j$ in ${\mathcal V}_n$, construct Alice's measurement ${\mathbb M}=\{M_k^{{\bar A}A_0\rightarrow{\bar A}}\}_{k=1}^{2^{nR}}$ as
\begin{eqnarray}
M_k^{{\bar A}A_0\rightarrow{\bar A}}=\frac{1}{\sqrt{2^{nR}}}\sum_{j=1}^{2^{nR}}\exp{\left(i\frac{2\pi jk}{2^{nR}}\right)}\langle j|^{A_0}\otimes V_j^{\bar A}.\nn
\end{eqnarray}
${\mathbb M}$ is $0$-oblivious and $\epsilon$-Markovianizing from ${\bar R}_A{\bar B}{\bar R}_B$. In addition, the reduced state of the post-measurement state on ${\bar A}{\bar R}_A{\bar R}_B$ does not depend on $k$. Indeed, we have $p_k=2^{-nR}$ and
\begin{eqnarray}
p_k^{-1}M_k(\tau^{\bar A}\otimes\Phi_{2^{nR}}^{A_0})M_k^\dagger={\mathcal V}_n(\tau)\nn
\end{eqnarray}
for  $k=1,\cdots,2^{nR}$. Alice performs the measurement defined above. 
\item {\bf Forward classical communication}: Alice sends the measurement outcome $k$ to Bob.
\item {\bf Bob's isometry}: Due to Lemma \ref{lmm:decoupleisometry}, there exist pure states $|{\tilde\Psi}^p\rangle^{{\bar A}{\bar R}_A{\tilde B}}$, $|\Psi'\rangle^{{\bar A}{\bar R}_A{\tilde B}BR_B}$ and isometries $\{W_k^{{\bar B}B_0\rightarrow {\bar B}{\tilde B}}\}_{k=1}^{2^{nR}}$ that satisfy
\begin{eqnarray}
|\Psi'\rangle=p_k^{-1/2}(M_k\otimes W_k)|\Psi^{\otimes n}\rangle\ket{\Phi_{K_n}}\nn
\end{eqnarray}
for any $k$, and satisfy
\begin{eqnarray}
|\Psi'\rangle&\approx&|{\tilde\Psi}^p\rangle^{{\bar A}{\bar R}_A{\tilde B}}|\Phi_d^{\otimes n}\rangle^{{\bar B}{\bar R}_B},\label{eq:nene}\\
({\tilde\Psi}^p)^{{\bar R}_A}&\approx&\pi_d^{{\bar R}_A}\label{eq:jean}
\end{eqnarray}
with a small error. Bob performs $W_k$.
\item {\bf State merging}: Alice and Bob perform state merging of
\begin{eqnarray}
|{\check\Psi}^p\rangle^{A'\bar{R}_AB'}:=|{\tilde\Psi}^p\rangle^{\bar{A}\bar{R}_A{\tilde B}}|\Phi_{2^{nr}}\rangle^{{\tilde A}_0{\tilde B}_0},\label{eq:defpsicheck}
\end{eqnarray}
where $A'=\bar{A}{\tilde A}_0$ and $B'={\tilde B}{\tilde B}_0$. Alice obtains a purification of $({\tilde\Psi}^p)^{{\bar R}_A}$ with a small error.
\item {\bf Alice's isometry}: Alice performs an isometry and obtains $|\Phi_d^{\otimes n}\rangle^{\bar{A}\bar{R}_A}$ within a small error. 
\end{enumerate}

The forward classical communication cost $nC_f$ is simply equal to $nR$ bits.  As for Step 4), we consider a state merging in which no entanglement is obtained afterward. Thus the total entanglement cost is equal to the amount of entanglement that Alice and Bob have initially shared, i.e., $nE=n(R+r)$ ebits of (\ref{eq:initialent}).  Applying Theorem \ref{thm:achieveqsm} and the rank inequality in (\ref{eq:rankineq}) for ${\tilde\Psi}^p$, the backward classical communication cost is bounded above by
\begin{eqnarray}
nC_b&\leq&\log{{\rm rank}[({\check\Psi}^p)^{B'}]}\:=\:\log{{\rm rank}[({\tilde\Psi}^p)^{\tilde B}]}+nr\nn\\
&\leq&\log{{\rm dim}[{\mathcal H}^{B_0}]}+nr\:=\:n(R+r).\nn
\end{eqnarray}
In total, we have $(E,C_f,C_b)=(R,R+r,R+r)$.

The total error is evaluated by counting errors of (\ref{eq:nene}), (\ref{eq:jean}) and one induced by state merging (see Appendix \ref{app:evaerror} for the detail). Due to Lemma \ref{lmm:decoupleisometryyy}, the first two errors are bounded above by $5\sqrt[4]{2\epsilon}$ and $3\sqrt[4]{2\epsilon}$, respectively. Theorem \ref{thm:achieveqsm} implies that the merging error $\epsilon_{\rm merg}$ is bounded as
\begin{eqnarray}
\epsilon_{\rm merg}\leq4\!\cdot\!2^{-nr/4}.\nn
\end{eqnarray}    
A simple calculation then yields an upper bound on the total error $\epsilon_{\rm tot}$:
\begin{align}
\epsilon_{\rm tot}\leq2\sqrt{3}\sqrt[8]{2\epsilon}+5\sqrt[4]{2\epsilon}+4\!\cdot\!2^{-nr/8}.\label{eq:toterror}
\end{align}
Since $\epsilon,r>0$ can be arbitrarily small, we conclude that a rate triplet $(E,C_f,C_b)=(R,R,R)$ is achievable if $R>M(U^\dagger)$.

\hfill$\blacksquare$

\subsection{Converse Part (Outline)}\label{sec:optimaltworound}

We prove the converse part of Theorem \ref{thm:misoptach} by combining (\ref{eq:markovmeasure3}) and (\ref{eq:qsmentccopt}).  Let ${\mathcal M}_n$ be a two-round LOCC protocol that satisfies Condition (\ref{eq:fidelityn}). We then have
\begin{eqnarray}
F({\rho}({\mathcal M}_n,U^{\dagger}),|\Phi_d^{\otimes n}\rangle^{{\bar A}{\bar R}_A}|\Phi_d^{\otimes n}\rangle^{{\bar B}{\bar R}_B}|\Phi_{L_n}\rangle^{A_1B_1})\geq1-\epsilon\!\!\!\!\!\!\!\!\nn\\\label{eq:convfid}
\end{eqnarray}
for
\beq
\rho({\mathcal M}_n,U^{\dagger}):={\mathcal M}_n(|\Psi^{\otimes n}\rangle^{{\bar A}{\bar R}_A{\bar B}{\bar R}_B}|\Phi_{K_n}\rangle^{A_0B_0}),\nn
\eeq 
corresponding to (\ref{eq:1shoterror2}). From Lemma \ref{lmm:markovifepsilon}, the map induced by Alice's measurement in ${\mathcal M}_n$ is $4\sqrt[4]{\epsilon}$-oblivious and $12\sqrt[4]{\epsilon}$-Markovianizing from $A'{\bar R}_A$. Then the first two conditions in Definition \ref{dfn:rofpsi} are satisfied by the following correspondence:
\begin{eqnarray}
&&\!\!\!\!\!\!{\bar A},{\bar B},{\bar C},A_0,G\rightarrow {\bar A},{\bar R}_A,({\bar B}{\bar R}_B),A_0,B_0,\nn\\
&&\!\!\!\!\!\!\!\!\!\!|\Psi\rangle^{ABC}\rightarrow|\Psi\rangle^{AR_A(BR_B)},\;\;\;\phi_{\rm res}\rightarrow\Phi_{K_n},\nn\\
&&\;\;\;\;\;\;\;\;\;\;\;\;\;\;\;\;\;\;\;\;\epsilon\rightarrow 12\sqrt[4]{\epsilon}.\label{correspondence}
\end{eqnarray}
Therefore,  from (\ref{eq:markovmeasure3}), we have 
\begin{eqnarray}
H(\{p_k\}_k)&\geq&\Delta S(A')_{av}\nn\\
&\geq&\Delta S(A')_{av}\!-\!\Delta{S}(B_0)_{av}\nn\\
&\geq&nM(U^\dagger)-n{\tilde\xi}( 12\sqrt[4]{\epsilon}\cdot n).\label{eq:markovmeasure33}
\end{eqnarray}

Suppose a rate triplet $(E,C_f,C_b)$ is achievable by  a two-round protocol. Due to Definition \ref{dfn:achievablerate} and Assumption (\ref{eq:condconv}), there exists a sequence of $(2,n,\epsilon_n)$-protocols for implementing $U$, with the entanglement cost $nE$ and classical communication costs $nC_f$ and $nC_b$ for each $n$, such that
\begin{eqnarray}
\lim_{n\rightarrow\infty}12\sqrt[4]{\epsilon_n}\cdot n=0.\label{eq:condconvv}
\end{eqnarray}
The optimality of the forward classical communication cost immediately follows from (\ref{eq:markovmeasure33}) and $nC_f\geq H(\{p_k\}_{k\in{\mathbb K}})$.

As for the backward classical communication cost, recall that Bob's measurement is decomposed into an isometry operation for obtaining $(\Phi_d^{\otimes n})^{{\bar B}{\bar R}_B}$ and a projective measurement on an ancillary system $B_E$. The latter forms state merging of ${\Psi_k^p}$, together with the backward classical communication and  the subsequent local operation by Alice. Thus the backward classical communication cost is equal to the one required for performing state merging of $\Psi_k^p$. Due to (\ref{eq:qsmentccopt}) with the correspondence $A\rightarrow A'$, $B\rightarrow B_1B_E$ and $R\rightarrow{\bar R}_A$, the cost is given by $I(B_1B_E:{\bar R}_A)_{\Psi_k^p}$. Because of $\Delta S(A')_{av}\gtrsim nM(U^\dagger)$ in (\ref{eq:markovmeasure33}), this cost turns out not to be smaller than $nM(U^\dagger)$. 

 The amount of entanglement obtained after state merging is bounded by (\ref{eq:qsmentccopt}) as $\log{L_n}\lesssim -S(B_1B_E|A')_{\Psi_k^p}$, which implies the optimality of the total cost of entanglement when combined with $\Delta S(A')_{av}-\Delta S(B_0)_{av}\gtrsim nM(U^\dagger)$. See Appendix \ref{app:prfconv} for a detailed proof.

\section{Properties of the Cost}\label{sec:algorithm}

In this section, we investigate properties of the Markovianizing cost of unitaries. The results obtained here will be used in the next section for analyzing examples.

Consider a tripartite pure state $|\Psi_U\rangle^{AR_A(\!BR_B\!)}$ defined by (\ref{eq:tripartiteu}). The {\it Petz recovery map} ${\mathcal R}_U:A\rightarrow A(BR_B)$ corresponding to $|\Psi_U\rangle^{AR_A(\!BR_B\!)}$ is defined by
\begin{eqnarray}
&&\!\!\!\!\!\!\!\!\!\!\!\!\!\!\!\!{\mathcal R}_U(\tau)=(\Psi_U^{A(BR_B)})^{\frac{1}{2}}(\Psi_U^{A})^{-\frac{1}{2}}\tau(\Psi_U^{A})^{-\frac{1}{2}}(\Psi_U^{A(BR_B)})^{\frac{1}{2}}\nn\\
&&=U^{AB}({\rm Tr}_B[U^{\dagger AB}(\tau^A\!\otimes I^B)U^{AB}]\otimes\Phi_d^{BR_B})U^{\dagger AB}\nn
\end{eqnarray}
for $\tau\in{\mathcal S}({\mathcal H}^A)$\cite{hayden04}. Define CPTP maps ${\mathcal E}_U$ and ${\mathcal E}_{U,\infty}$ on $A$ by 
\begin{align}
{\mathcal E}_U:={\rm Tr}_{BR_B}\circ{\mathcal R}_U,\;\;\;\;{\mathcal E}_{U,\infty}:=\lim_{N\rightarrow\infty}\frac{1}{N}\sum_{n=1}^N{\mathcal E}_U^n,
\label{def:einfty}
\end{align}
and consider the states
\begin{eqnarray}
\Psi_{U,\infty}^{AR_ABR_B}&:=&{\mathcal E}_{U,\infty}^A(|\Psi_U\rangle\!\langle\Psi_U|^{AR_ABR_B}),\nn\label{def:einfty2}\\
\Phi_{U,\infty}^{AR_A}&:=&{\mathcal E}_{U,\infty}^A(|\Phi_d\rangle\!\langle\Phi_d|^{AR_A}).\label{def:einfty3}
\end{eqnarray}
As we prove below, the map ${\mathcal E}_U$ is self-adjoint in the sense that ${\mathcal E}_U={\mathcal E}_U^*$. Therefore, due to Theorem 9 in \cite{waka15_markov}, the Markovianizing cost of $U$ is given by
\begin{align}
M(U)=M_{A|R_A}(\Psi_U^{AR_A(BR_B)})=S(\Psi_{U,\infty}^{AR_ABR_B}),\nn
\end{align} 
which can be computed by a finite-step algorithm proposed in \cite{waka15_markov} (see Section III therein). Due to the unitary invariance of the von Neumann entropy, it immediately follows that
\begin{align}
M(U)=S(\Phi_{U,\infty}^{AR_A}).\label{eq:miss2}
\end{align} 
As a consequence, the Markovianizing costs of unitaries $U_1$ and $U_2$ are equal if they are local unitarily equivalent, that is, if there exist unitaries $v,v'$ on ${\mathcal H}^A$ and $w,w'$ on ${\mathcal H}^B$ such that $U_1=(v\otimes w)U_2(v'\otimes w')$.

 Let us also analyze the Schmidt strength of unitaries. Consider Decomposition (\ref{eq:opsch}) of a bipartite unitary $U$. A CPTP map ${\mathcal E}_U$ defined by (\ref{def:einfty}) takes the form of
\begin{eqnarray}
{\mathcal E}_U(\tau)=\sum_{ss'}c_s^2c_{s'}^2E_s^\dagger E_{s'}\tau E_{s'}^\dagger E_s=\sum_{ss'}{\tilde c}_{ss'}^2{\tilde E}_{ss'}\tau {\tilde E}_{ss'}^\dagger,\label{eq:defcalE}
\end{eqnarray}
where we introduced notations ${\tilde c}_{ss'}:=c_sc_{s'}$ and ${\tilde E}_{ss'}:=E_s^\dagger E_{s'}$. It is straightforward to verify that ${\mathcal E}_U$ is self-adjoint, that is, it satisfies
\begin{eqnarray}
{\mathcal E}_U(\tau)={\mathcal E}_U^*(\tau):=\sum_{ss'}{\tilde c}_{ss'}^2{\tilde E}_{ss'}^\dagger\tau {\tilde E}_{ss'}.\nn
\end{eqnarray}
 Due to the orthonormality of $\{E_s|\Phi_d\rangle^{AR_A}\}_s$ and $\{F_s|\Phi_d\rangle^{BR_B}\}_s$, which follows from (\ref{eq:orthoef}), the eigen decomposition of $\Psi_U^{AR_A}$ is given by
\begin{eqnarray}
\Psi_U^{AR_A}=\sum_{s=0}^{d^2-1}c_s^2E_s^A|\Phi_d\rangle\!\langle\Phi_d|^{AR_A}E_s^{\dagger A}.\label{eq:psiuspect}
\end{eqnarray} 
Thus we have
\begin{eqnarray}
K(U)&=&S(AR_A)_{\Psi_U}.\label{eq:opsch2}
\end{eqnarray}

The following lemma provides a lower bound on the Markovianizing cost of unitaries.
\begin{lmm}\label{lmm:mgeqk}
$M(U)\geq K(U)$ holds for any bipartite unitary $U$.
\end{lmm}

\begin{prf}
Define quantum operations $e$ and $e^*$ on ${\mathcal S}({\mathcal H}^A)$ by
\begin{eqnarray}
e(\tau)=\sum_{s}c_s^2E_s\tau E_s^\dagger,\;e^*(\tau)=\sum_{s}c_s^2E_s^\dagger\tau E_s.\nn
\end{eqnarray} 
From (\ref{eq:psiuspect}) and (\ref{eq:opsch2}), the Schmidt strength of the unitary is given by
\begin{eqnarray}
K(U)=S(e^A(|\Phi_d\rangle\!\langle\Phi_d|^{AR_A})).\label{eq:kuisse}
\end{eqnarray} 
It immediately follows from (\ref{eq:defcalE}) that ${\mathcal E}_U=e^*\circ e$. We have
\begin{eqnarray}
e(\tau)&=&{\rm Tr}_B[U^{\dagger AB}(\tau^A\!\otimes I^B)U^{AB}]\nn\\
e^*(\tau)&=&{\rm Tr}_B[U^{AB}(\tau^A\!\otimes I^B)U^{\dagger AB}]\nn
\end{eqnarray}
due to (\ref{eq:opsch}) and (\ref{eq:orthoef}), which implies that $e$ and $e^*$ are unital, i.e., $e(I)=e^*(I)=I$. Therefore, owing to the monotonicity of the von Neumann entropy under unital maps, we have
\begin{eqnarray}
S(({\mathcal E}_U^n)^A(|\Phi_d\rangle\!\langle\Phi_d|^{AR_A}))\geq S(e^A(|\Phi_d\rangle\!\langle\Phi_d|^{AR_A}))\label{eq:engeqe}
\end{eqnarray}
for any $n\geq1$. Due to Definitions (\ref{def:einfty}), (\ref{def:einfty3}) and the concavity of the von Neumann entropy, we obtain
\begin{eqnarray}
S(\Phi_{U,\infty}^{AR_A})\geq\lim_{N\rightarrow\infty}\frac{1}{N}\sum_{n=1}^NS(({\mathcal E}_U^n)^A(|\Phi_d\rangle\!\langle\Phi_d|^{AR_A})).\label{eq:concvn}
\end{eqnarray}
Expressions (\ref{eq:miss2}), (\ref{eq:kuisse}), (\ref{eq:engeqe}) and (\ref{eq:concvn}) yields $M(U)\geq K(U)$.\:$\blacksquare$
\end{prf}

\section{Examples}\label{sec:exa}

In this section, we consider two classes of bipartite unitaries and  compute their Markovianizing costs.

\subsection{Two-Qubit Unitaries}\label{sec:twoqubit}

It is  proven in \cite{nielsen03} that all two-qubit unitaries are classified into the following categories:
\begin{enumerate}
\item Unitaries that can be written as a tensor product of local unitaries as $U=u^A\otimes u^B$. We do not consider this type of unitaries because of its triviality.
\item Unitaries that can be written in the form of
\begin{eqnarray}
U=\cos{\left(\frac{\theta}{2}\right)}I^A\otimes I^B+i\sin{\left(\frac{\theta}{2}\right)}\sigma_z^A\otimes \sigma_z^B\label{eq:twoqcu}
\end{eqnarray}
up to local unitaries, where $\theta\in(0,\pi/2]$. Controlled-unitary gates are examples of such unitaries.
\item Unitaries that can be written in the form of
\begin{eqnarray}
U&=&\!\!\!c_0e^{i\theta_0}I^A\!\otimes I^B+c_1e^{i\theta_1}\sigma_z^A\!\otimes \sigma_z^B\nonumber\\
&&\!\!\!\!\!\!+c_2e^{i\theta_2}\sigma_x^A\!\otimes \sigma_x^B+c_3e^{i\theta_3}\sigma_y^A\!\otimes \sigma_y^B\nn
\end{eqnarray}
up to local unitaries, where $c_s,\theta_s\in{\mathbb R}\;(s=0,1,2,3)$ are nonnegative real parameters satisfying $\sum_{s=0}^3c_s^2=1$. All two-qubit unitaries that  are not classified to the first two categories are of this category.
\end{enumerate}

Let us consider unitaries of the form (\ref{eq:twoqcu}), which is local unitarily equivalent to the following controlled phase gate:
\begin{eqnarray}
U_\theta=\proj{0}^A\otimes I^B+\proj{1}^A\otimes (e^{i\theta\sigma_z})^B.\nn
\end{eqnarray}
We have
\begin{eqnarray}
&&\!\!\!\!c_0=\cos{(\theta/2)},\;c_1=\sin{(\theta/2)},\nn\\
&&\!\!\!\!{\tilde c}_{00}^2+{\tilde c}_{11}^2=\frac{1}{2}(1+\cos^2{\theta}),\;{\tilde c}_{01}^2+{\tilde c}_{10}^2=\frac{1}{2}\sin^2{\theta},\nn\\
&&\!\!\!\!E_0=F_0=I,\;E_1=\sigma_z,\;F_1=i\sigma_z,\nn\\
&&\!\!\!\!{\tilde E}_{00}={\tilde E}_{11}=I,\;{\tilde E}_{01}={\tilde E}_{10}=\sigma_z.\nn
\end{eqnarray}
Thus a map corresponding to (\ref{eq:defcalE}) is given by
\begin{eqnarray}
{\mathcal E}(\tau)=\frac{1+\cos^2{\theta}}{2}\cdot\tau+\frac{1}{2}\sin^2{\theta}\cdot \sigma_z\tau \sigma_z,\nn
\end{eqnarray}
which leads to
\begin{eqnarray}
{\mathcal E}_{U,\infty}(\tau)=\frac{1}{2}(\tau+\sigma_z\tau\sigma_z)=\proj{0}\tau\proj{0}+\proj{1}\tau\proj{1}\nn
\end{eqnarray}
from (\ref{def:einfty}). Hence we have
\begin{align}
\Phi_{U,\infty}^{AR_A}=\frac{1}{2}(|0\rangle\!\langle0|\otimes|0\rangle\!\langle0|+|1\rangle\!\langle1|\otimes|1\rangle\!\langle1|)\nn
\end{align}
corresponding to (\ref{def:einfty3}), which implies $M(U)=1$ due to (\ref{eq:miss2}). Consequently, we obtain the following theorem:

\begin{thm}\label{thm:contunit}
A rate triplet $(E,C_f,C_b)$ is achievable by two-round protocols for implementing a two qubit controlled-unitary gate only if $E,C_f,C_b\geq1$, if we additionally require in Definition \ref{dfn:achievablerate} that Condition (\ref{eq:condconv}) holds.
\end{thm}
The above theorem implies that, counterintuitively, at least $1$ ebit of entanglement consumption per copy is necessary for implementing two-qubit controlled-unitary gate by two round protocols, regardless of how close the unitary is to the identity operation (i.e., regardless of how small $\theta$ is). In \cite{waka16_tradeoff}, we prove that a certain class of two-qubit controlled-unitary gates can be implemented by a {\it four}-round protocol with the entanglement cost strictly smaller than $1$ ebit per copy. Thereby we reveal a trade-off relation between the entanglement cost and the number of rounds for a LOCC task.

\subsection{Generalized Clifford Operators}\label{sec:gcliff}

The generalized Pauli operators $\sigma_{pq}\:(p,q\in\{1,\cdots,d\})$ on a $d$-dimensional Hilbert space is defined as
\begin{eqnarray}
&&\sigma_{p0}:=\sum_{t=1}^{d}\outpro{t-p}{t},\;\sigma_{0q}:=\sum_{t=1}^{d}e^{2\pi iqt/d}\outpro{t}{t},\nn\\&&\sigma_{pq}:=\sigma_{p0}\sigma_{0q},\label{eq:dfncliff}
\end{eqnarray}
 with a fixed basis $\{\ket{t}\}_{t=1}^d$. Here, subtraction is taken with mod $d$. A bipartite unitary $U$ is called a generalized Clifford operator if, for any $p$, $q$, $r$ and $s$, there exist $p'$, $q'$, $r'$, $s'$ and a phase $\theta_{pqrs}\in{\mathbb R}$ such that
 \begin{eqnarray}
 U(\sigma_{pq}\otimes\sigma_{rs})U^{\dagger}=e^{i\theta_{pqrs}}\sigma_{p'q'}\otimes\sigma_{r's'}.\nn
 \end{eqnarray}
The Markovianizing cost of generalized Clifford operators can be simply computed by the following theorem, a proof of which  is given in Appendix \ref{app:prfcliff}.
\begin{thm}\label{thm:cliff}
$M(U)=K(U)$ holds for any generalized Clifford operator $U$.
\end{thm}

As a corollary of Theorem \ref{thm:misoptach} and \ref{thm:cliff}, the Schmidt strength $K(U)$ is equal to the minimum  cost of entanglement and classical communication for implementing generalized Clifford operator by two-round protocols under additional assumption (\ref{eq:condconv}).  A stronger statement, represented by the following theorem, immediately follows from Theorem \ref{thm:lowerbound} and \ref{thm:cliff}.

\begin{thm}\label{thm:cliffku}
The following statements hold for any generalized Clifford operator $U$ and $r\geq2$.
\begin{itemize}
\item {\it Direct:} A rate triplet $(E,C_f,C_b)$ is achievable by $r$-round protocols for implementing $U$ if $E,C_f,C_b>K(U)$.
\item {\it Converse:} A rate triplet $(E,C_f,C_b)$ is achievable by $r$-round protocols for implementing $U$ only if $E,C_f,C_b\geq K(U)$.
\end{itemize}
\end{thm}

\hfill

\section{Open Problems}\label{sec:open}

We have derived a converse bound on the cost of entanglement and classical communication for implementing a bipartite unitary by two-round protocols. However, we do not know whether the converse bound remains  to hold when we  remove the additional requirement on the convergence speed of error, represented by Inequality (\ref{eq:condconv}). In this section, we investigate a relation between this open problem and  another regarding a property of approximate recoverability. 

In the proof of the converse part, Condition (\ref{eq:condconv}) is exploited in the form of Inequality (\ref{eq:condconvv}). This inequality is required to  prove the convergence of an error term in (\ref{eq:markovmeasure33}), which depends not only on $\epsilon$ but also on $n$. The $n$-dependence of the error term originates from that in Inequality (\ref{eq:markovmeasure3}), and the latter arises due to the fact that Condition (\ref{eq:condconvrec}) is required to prove (\ref{eq:mreqmrm}). In summary, we require Condition (\ref{eq:condconv}) to prove the converse part of Theorem \ref{thm:misoptach} because we require Condition (\ref{eq:condconvrec}) to prove (\ref{eq:mreqmrm}).

In \cite{waka15_rec}, we proved that Condition (\ref{eq:condconvrec}) in Theorem \ref{thm:mreqmrm} can be eliminated if a conjecture about approximate recoverability is true. The conjecture states that a tripartite quantum state $\rho^{ABC}$ is approximately recoverable from $\rho^{AB}$ by an operation from ${\mathcal R}:B\rightarrow BC$ if it is approximately recoverable from $\rho^{BC}$ by an operation ${\mathcal R}':B\rightarrow AB$, up to a {\it dimension-independent} rescaling of error of recovery. A rigorous statement is as follows:

\begin{cjt}\label{cjt:symrec}(Proposition 13 (unproven) in \cite{waka15_rec})
There exists a nonnegative function $g(\epsilon)$,  independent of the dimension of quantum systems and satisfies $\lim_{\epsilon\rightarrow0}g(\epsilon)=0$, such that the following statement holds for an arbitrary tripartite state $\rho^{ABC}$ and $\epsilon>0$: The state $\rho^{ABC}$ is $g(\epsilon)$-recoverable from $BC$ if it is $\epsilon$-recoverable from $AB$.
\end{cjt}
Condition (\ref{eq:condconv}) in Theorem  \ref{thm:misoptach} can be eliminated if the above conjecture is true. See also Appendix \ref{eq:appconsp}.

\hfill

\section{Communication Power of a LOCC protocol}\label{sec:ccpowerlocc}

In this section, we  analyze classical communication power of a LOCC protocol with an arbitrary preshared resource state. The results obtained here will be used in the next section to prove Theorem \ref{thm:lowerbound}. 

Consider the following scenario in which Alice aims to transmit $nR$ bits of classical message to Bob by a bidirectional LOCC protocol that transforms a preshared quantum state $\rho^{AB}$. 
\begin{itemize}
\item Alice and Bob initially share a bipartite quantum state $\rho^{AB}$.
\item Alice is given an array of uniformly random classical bits ${\vec X}=X_1\cdots X_{nR}$.
\item Alice and Bob transforms $\rho^{AB}$ by an LOCC protocol.
\item Alice's operations during the protocol, as well as the message from Alice to Bob, may depend on ${\vec X}$. 
\item After the completion of the protocol, Bob performs a measurement on $B$ to decode ${\vec X}$.
\end{itemize}
Let ${\vec X}'$ be the result of Bob's decoding measurement. The decoding error is defined by
\begin{align}
P_e:={\rm Pr}\{{\vec X}\neq{\vec X}'\}.\nn
\end{align}
In the following, we prove that the length $nR$ of classical message ${\vec X}$ does not exceed the total number of classical bits transmitted from Alice to Bob during the protocol, if the decoding error is vanishingly small.

\begin{figure}[t]
\begin{center}
\includegraphics[bb={0 0 697 1226}, scale=0.3]{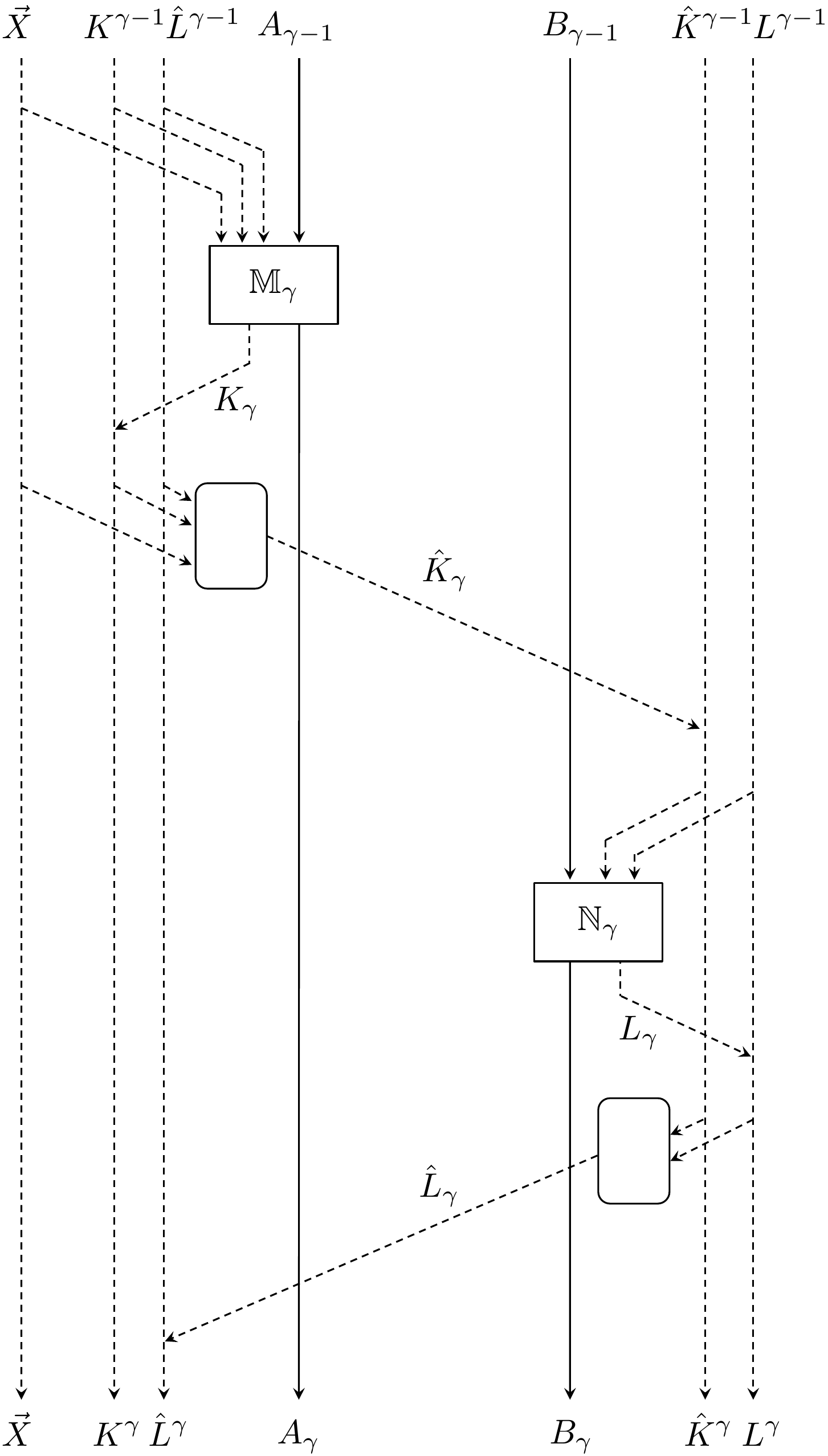}
\end{center}\caption{A graphical representation of the $\gamma$-th step in an LOCC protocol is depicted. We denote system $A$ and $B$ after the $\gamma$-th step by $A_\gamma$ and $B_\gamma$ for $\gamma=1,\cdots,\Gamma$, respectively.}
\label{fig:LOCC}
\end{figure}

Without loss of generality, we assume that the protocol proceeds as follows. Here, $\Gamma$ is a natural number,  and $K_\gamma$, $L_\gamma$, ${\hat K}_\gamma$, ${\hat L}_\gamma$ are random variables which take values in finite sets ${\mathcal K}_\gamma$, ${\mathcal L}_\gamma$, ${\hat{\mathcal K}}_\gamma$, ${\hat{\mathcal L}}_\gamma$, respectively. 
\begin{enumerate}
\item Alice and Bob recursively apply the following operation from $\gamma=1$ to $\gamma=\Gamma$: 
\begin{enumerate}
\item Alice performs a measurement ${\mathbb M}_\gamma$ on her system and obtains an outcome $K_\gamma$.
\item Alice transmits a classical message ${\hat K}_\gamma$ to Bob.
\item Bob performs a measurement ${\mathbb N}_\gamma$ on his system and obtains an outcome $L_\gamma$.
\item Bob transmits a classical message ${\hat L}_\gamma$ to Alice.
\end{enumerate}
\item Alice performs a quantum operation on her system.
\end{enumerate}
The total number of classical bits, transmitted from Alice to Bob during the protocol, is given by
\begin{align}
C_{\rm tot}:=\sum_{\gamma=1}^\Gamma\log{|{\hat{\mathcal K}}_\gamma|}.\nn
\end{align}

Let us introduce the following notations:
\begin{eqnarray}
&K^{\gamma}:=(K_1,\cdots,K_{\gamma}),&L^{\gamma}:=(L_1,\cdots,L_{\gamma})\nonumber\\
&\hat{K}^{\gamma}:=(\hat{K}_1,\cdots,\hat{K}_{\gamma}),&{\hat L}^{\gamma}:=({\hat L}_1,\cdots,{\hat L}_{\gamma})\nonumber
\end{eqnarray}
In general, Alice and Bob's measurement in the protocol, as well as classical messages, may dependent on the previous measurement outcomes and messages in the following way (Figure \ref{fig:LOCC}).
\begin{itemize}
\item ${\mathbb M}_\gamma$ depends on $({\vec X},K^{\gamma-1},{\hat L}^{\gamma-1})$.
\item ${\mathbb N}_\gamma$ depends on $(L^{\gamma-1},\hat{K}^{\gamma})$.
\item $\hat{K}_{\gamma}$ depends on $({\vec X},K^{\gamma},{\hat L}^{\gamma-1})$.
\item ${\hat L}_{\gamma}$ depends on $(L^{\gamma},\hat{K}^{\gamma-1})$.
\end{itemize}

The following lemma states that the mutual information between ${\vec X}$ and all that Bob has after the protocol is bounded above by the total amount of classical communication transmitted from Alice to Bob during the protocol. See Appendix \ref{sec:prfhoriemon} for a proof.

\begin{lmm}\label{lmm:horiemon}
The following inequalities hold:
\begin{align}
&I({\vec X}:B_\Gamma,L^\Gamma,\hat{K}^\Gamma)\leq C_{\rm tot},\label{eq:regain}\\
&nR\leq C_{\rm tot}+h(P_e)+nRP_e.\label{eq:regain2}
\end{align}
Here, $h(x)$ is the binary entropy defined by
\begin{align}
h(x):=-x\log{x}-(1-x)\log{(1-x)},\nn
\end{align} 
and $B_\Gamma$ denotes system $B$ after the $\Gamma$-th step of the protocol.
\end{lmm}

\begin{rmk}
An upper bound on the classical communication power of a two-way LOQC (local operations and {\it quantum} communication) protocol, which is similar to (\ref{eq:regain2}), has been  proven in \cite{nayak02}.
\end{rmk}

\section{Proof of Theorem \ref{thm:lowerbound}}
\label{sec:lowerbound}

We prove Theorem \ref{thm:lowerbound} in this section, based on the idea that the cost of entanglement and classical communication for implementing a unitary is not smaller than powers of the unitary for generating entanglement and transmitting classical information (\!\!\cite{stahlke11,soeda11,montanaro07}). 

Let us analyze power of a bipartite unitary for transmitting classical information. The following lemma states that the Schmidt strength is a lower bound on the classical communication power of a bipartite unitary. See Appendix \ref{app:drinkbar} for a proof.

\begin{lmm}\label{lmm:ccpowerofu}
For any $\epsilon\in(0,1]$ and sufficiently large $n$, let ${\mathcal U}_n$ be a quantum operation on ${\bar A}{\bar B}$ that satisfy
\begin{align}
F\left(\rho({\mathcal U}_n),|\Psi_U\rangle^{\otimes n}\right)\geq1-\epsilon\label{eq:mathcalu}
\end{align}
for
\begin{align}
\rho({\mathcal U}_n):={\mathcal U}_n(|\Phi_{d}^{AR_A}\rangle^{\otimes n}|\Phi_{d}^{BR_B}\rangle^{\otimes n}).\nn
\end{align}
Then ${\mathcal U}_n$ has a capacity to transmit $n(K(U)-\epsilon)$ bits of classical information from Alice to Bob up to an error $5\sqrt{\epsilon}$, when assisted by shared entanglement.
\end{lmm}
Theorem \ref{thm:lowerbound} is then proved as follows.\\

\noindent\;{\it Proof of Theorem \ref{thm:lowerbound}:}
Suppose a rate triplet $(E,C_f,C_b)$ is achievable. By definition, for any $\epsilon>0$ and sufficiently large $n$, there exist $K_n$ and $L_n$ that satisfy $\log{K_n}-\log{L_n}=nE$, and a LOCC protocol ${\mathcal M}_n$ that satisfies (\ref{eq:fidelityn}), with the forward and backward classical communication cost $nC_f$ and $nC_b$, respectively. 

Define a quantum operation ${\hat{\mathcal M}}_n$ on ${\bar A}{\bar B}$ by
\begin{align}
{\hat{\mathcal M}}_n:\tau\rightarrow{\rm Tr}_{A_1B_1}[{\mathcal M}_n(\tau^{{\bar A}{\bar B}}\otimes\Phi_{K_n}^{A_0B_0})].\nn
\end{align}
Due to Lemma \ref{lmm:ccpowerofu}, ${\hat{\mathcal M}}_n$ has a capacity to transmit $n(K(U)-\epsilon)$ bits of classical information from Alice to Bob up to an error $5\sqrt{\epsilon}$, when assisted by shared entanglement. By definition, ${\mathcal M}_n$ has the same capacity. Applying Lemma \ref{lmm:horiemon} yields
\begin{align}
n(K(U)-\epsilon)\leq nC_f+h(5\sqrt{\epsilon})+n\epsilon(K(U)-\epsilon),\nn
\end{align}
which leads to
\begin{align}
(1-\epsilon)(K(U)-\epsilon)\leq C_f+\frac{1}{n}h(5\sqrt{\epsilon}).\nn
\end{align}
Since $\epsilon>0$ can be arbitrarily small, we obtain $C_f\geq K(U)$. Exchanging roles of Alice and Bob, we also have $C_b\geq K(U)$.

To prove $E\geq K(U)$, we assume for simplicity that $K_n$ and $L_n$ is bounded above as
\begin{align}
\log{K_n},\;\log{L_n}\leq n\log{\kappa}\label{eq:arc035}
\end{align}
with a constant $\kappa>0$. We quantify entanglement of states between systems ${\bar A}{\bar R}_AA_0$ and ${\bar B}{\bar R}_BB_0$ (or between ${\bar A}{\bar R}_AA_1$ and ${\bar B}{\bar R}_BB_1$) by an entanglement measure that satisfy {\it asymptotic continuity} (\cite{plenio07}, see Appendix \ref{eq:entmeas}). We denote it by $\ca E$. Since $\ca E$ is equal to the entanglement entropy for pure states, we have
\begin{eqnarray}
&&{\ca E}(|\Psi_U\rangle^{\otimes n}|\Phi_{L_n}\rangle^{A_1B_1})=nK(U)+\log{L_n},\nn\\
&&{\ca E}(|\Phi_{d}^{AR_A}\rangle^{\otimes n}|\Phi_{d}^{BR_B}\rangle^{\otimes n}|\Phi_{K_n}\rangle^{A_0B_0})=\log{K_n}\nn
\end{eqnarray}
from (\ref{eq:opsch2}). Due to asymptotic continuity, Condition (\ref{eq:fidelityn}) and (\ref{eq:arc035}) implies
\begin{eqnarray}
{\ca E}(\rho({\mathcal M_n}))\geq nK(U)+\log{L_n}-n\delta(2\sqrt{\epsilon})\log{(d^4\kappa^2)},\label{eq:profJ}
\end{eqnarray}
where $\delta(\epsilon)$ is an $n$-independent nonnegative function that satisfies $\lim_{\epsilon\rightarrow0}\delta(\epsilon)=0$. Equality (\ref{eq:deffinst}) and the monotonicity of $\ca E$ under LOCC operations yield
\begin{eqnarray}
{\ca E}(\rho({\mathcal M_n}))\leq \log{K_n}.\label{eq:ninomae}
\end{eqnarray}
Combining (\ref{eq:profJ}) and (\ref{eq:ninomae}), we obtain
\begin{eqnarray}
E=\frac{1}{n}\left(\log{K_n}-\log{L_n}\right)\geq K(U)-\delta(2\sqrt{\epsilon})\log{(d^4\kappa^2)},\nn
\end{eqnarray}
which implies $E\geq K(U)$ by taking the limit of $\epsilon\rightarrow0$.\hfill$\blacksquare$

\section{Conclusion}
\label{sec:conclusion}
We have analyzed distributed quantum computation in terms of quantum Shannon theory  for the first time. We have considered an asymptotic scenario for entanglement-assisted LOCC implementations of bipartite unitaries. For protocols consisting of two-round LOCC, we have derived the achievable rate region for the costs of entanglement and classical communication under an additional requirement on the convergence speed of error.  We have also derived a general lower bound on the minimum cost of resources. The results can be straightforwardly generalized for cases where ${\rm dim}{{\mathcal H}^A}\neq{\rm dim}{{\mathcal H}^B}$. The problem formulated in this paper can be regarded as a quantum analog of `interactive coding for lossless computing' in classical information theory \cite{gamal11}.

\section*{Acknowledgments}

 Some parts of the contents of this paper (Theorem \ref{thm:lowerbound}, Theorem \ref{thm:cliffku}, Section \ref{sec:lowerbound}, Appendix \ref{app:drinkbar} and a part of Appendix \ref{app:okaimonohakochira}) were contained in our paper \cite{waka13}, which has been submitted to IEEE Transactions on Information Theory and withdrawn afterward. The authors thank the reviewers of that paper for valuable comments, which has been useful in preparing this manuscript.

In \cite{waka_dqc_isit} and the previous version of this manuscript, we failed to prove the converse part. The main weakness in the previous approach was that we exploit Markovianization in the version of \cite{waka15_markov}, rather than the one formulated in terms of approximate recoverability \cite{waka15_rec}. The authors thank  the referees of ISIT 2015 for pointing out the relevance of approximate recoverability to the problem addressed in this paper.

\appendices

\section{Mathematical Preliminaries}

In this appendix, we summarize technical tools that will be used in the following appendices. For the references, see e.g. \cite{nielsentext,hayashitext,wildetext}. See also Appendix A in \cite{waka15_markov} for basic properties of quantum entropies which are not presented here.

\subsection{Fidelity, Trace Distance and Uhlmann's Theorem}\label{app:fidelityetc}

The trace distance between two quantum states $\rho,\sigma\in{\mathcal S}({\mathcal H})$ is defined by
\begin{eqnarray}
\|\rho-\sigma\|_1={\rm Tr}\left[\sqrt{(\rho-\sigma)^2}\right].\nonumber
\end{eqnarray}
 It satisfies
\begin{align}
0\leq\|\rho-\sigma\|_1\leq2\nn
\end{align}
and
\begin{eqnarray}
\|\rho-\sigma\|_1=2\max_{\Lambda}{\rm Tr}[\Lambda(\rho-\sigma)],\label{eq:defTD2}
\end{eqnarray}
where the maximization is taken over all linear operators $\Lambda$ on $\mathcal H$ that satisfy $0\leq\Lambda\leq I$.

For $\rho,\sigma,\tau\in{\mathcal S}({\mathcal H})$, we have
\begin{eqnarray}
\left\|\rho-\tau\right\|_1\leq\left\|\rho-\sigma\right\|_1+\left\|\sigma-\tau\right\|_1,\label{eq:triangle}
\end{eqnarray}
which is called the {\it triangle inequality}.  For two ensembles $\{p_i,\rho_i\}$ and $\{p_i,\sigma_i\}$, we have
\begin{align}
\sum_ip_i\left\|\rho_i-\sigma_i\right\|_1\geq\left\|\sum_ip_i\rho_i-\sum_ip_i\sigma_i\right\|_1.\label{eq:convTD}
\end{align}
The trace distance takes a simple form under tensor product, i.e., for any $\rho,\sigma\in{\mathcal S}({\mathcal H}^A)$ and $\tau\in{\mathcal S}({\mathcal H}^B)$, we have
 \begin{eqnarray}
\|\rho^A\otimes\tau^B-\sigma^A\otimes\tau^B\|_1=\|\rho^A-\sigma^A\|_1.\label{eq:tdtensor}
\end{eqnarray}

The fidelity between two quantum states $\rho,\sigma\in{\mathcal S}({\mathcal H})$ is defined by
\begin{eqnarray}
F(\rho,\sigma):=\left({\rm Tr}\left[\sqrt{\sqrt{\rho}\sigma\sqrt{\rho}}\right]\right)^2,\nonumber
\end{eqnarray}
and satisfies
\begin{align}
0\leq F(\rho,\sigma)\leq1.\nn
\end{align}
The fidelity takes a simple form for pure states as
\begin{eqnarray}
F(|\psi\rangle,|\phi\rangle)=|\langle\psi|\phi\rangle|^2\label{eq:cooppure}
\end{eqnarray}
and
\begin{eqnarray}
F(\rho,|\phi\rangle)=\langle\phi|\rho|\phi\rangle,\label{eq:fidpm}
\end{eqnarray}
the latter of which yields
\begin{align}
\sum_kp_kF(\rho_k,|\phi\rangle)=F\left(\sum_kp_k\rho_k,\:|\phi\rangle\right)\label{eq:fidpens}
\end{align}
for any ensemble $\{p_k,\rho_k\}_k$.

Let $|\psi_\rho\rangle,|\psi_\sigma\rangle\in{\mathcal H}^A\otimes{\mathcal H}^B$ be arbitrary purifications of $\rho,\sigma\in{\mathcal S}({\mathcal H}^A)$, respectively. Due to Uhlmann's theorem \cite{uhlmann76}, we have
\begin{eqnarray}
F(\rho,\sigma)&=&\max_{W}|\langle\psi_\rho|(I^A\otimes W^B)|\psi_\sigma\rangle|^2\label{eq:cooppuree}\\
&=&\max_{\psi_\sigma'}|\langle\psi_\rho|\psi_\sigma'\rangle|^2.\label{eq:cooppureeee}
\end{eqnarray}
Here, the maximization in the first line is taken over all unitaries $W$ acting on ${\mathcal H}^B$, and that in the second line over all purifications $|\psi_\sigma'\rangle\in{\mathcal H}^A\otimes{\mathcal H}^B$ of $\sigma$. It immediately follows that, for an arbitrary pure states $|\Psi\rangle\in{\mathcal H}^A\otimes{\mathcal H}^B$ and $|\phi\rangle\in{\mathcal H}^A$, we have
\begin{eqnarray}
F\left(\Psi^A,|\phi\rangle\right)=\max_{|\varphi\rangle\in{\mathcal H}^B}F\left(|\Psi\rangle^{AB},|\phi\rangle^A|\varphi\rangle^B\right),\label{eq:fidpp}
\end{eqnarray}
where the maximization is taken over all pure states on system $B$.

The trace distance and the fidelity are monotonic under quantum operations, i.e., it satisfies
\begin{eqnarray}
&&\left\|\rho-\sigma\right\|_1\geq\left\|{\mathcal E}(\rho)-{\mathcal E}(\sigma)\right\|_1,\label{eq:monottd}\\
&&F(\rho,\sigma)\leq F({\mathcal E}(\rho),{\mathcal E}(\sigma))\nonumber
\end{eqnarray}
for any $\rho,\sigma\in{\mathcal S}({\mathcal H})$ and any linear CPTP map ${\mathcal E}:{\mathcal S}({\mathcal H})\rightarrow{\mathcal S}({\mathcal H}')$.  In particular, the two functions are monotonic under under taking the partial trace, that is, for any $\rho,\sigma\in{\mathcal S}({\mathcal H}^A\otimes{\mathcal H}^B)$ we have
\begin{eqnarray}
&&\left\|\rho^{AB}-\sigma^{AB}\right\|_1\geq\left\|\rho^A-\sigma^A\right\|_1,\label{eq:monottdd}\\
&&F(\rho^{AB},\sigma^{AB})\leq F(\rho^A,\sigma^A).\label{eq:fidmonPT}
\end{eqnarray}
The two functions are invariant under unitary operations, namely, for any unitary $U$ acting on ${\mathcal H}$ we have
\begin{eqnarray}
&&\left\|\rho-\sigma\right\|_1=\left\|U\rho U^\dagger-U\sigma U^\dagger\right\|_1,\label{eq:fuwatororo}\\
&&F(\rho,\sigma)=F(U\rho U^\dagger,U\sigma U^\dagger).\label{eq:fuwatoro}
\end{eqnarray}
The trace distance and the fidelity satisfy the following relation in general:
\begin{eqnarray}
1-\sqrt{F(\rho,\sigma)}\leq\frac{1}{2}\left\|\rho-\sigma\right\|_1\leq\sqrt{1-F(\rho,\sigma)}.\label{eq:tracefid}
\end{eqnarray}
Therefore, if $F(\rho,\sigma)\geq1-\epsilon$ then $\left\|\rho-\sigma\right\|_1\leq2\sqrt{\epsilon}$. Conversely, if $\left\|\rho-\sigma\right\|_1\leq\epsilon$ then $F(\rho,\sigma)\geq1-\epsilon$. 

 Let $\mathcal E$ be a quantum operation on a system described by a $d$-dimensional Hilbert space ${\mathcal H}$, and $U$ be a unitary acting on ${\mathcal H}$. How precisely ${\mathcal E}$ approximates a unitary operation ${\mathcal U}(\cdot)=U(\cdot)U^\dagger$ is evaluated by the {\it average fidelity} and the {\it entanglement fidelity}, which are defined as
\begin{align}
F_{\rm av}({\mathcal E};U):=\int_{\rm Haar} p(d\psi)\:F({\mathcal E}(|\psi\rangle),U|\psi\rangle)\label{eq:dfnavefid}
\end{align}
and
\begin{align}
F_{\rm en}({\mathcal E};U):=F(({\mathcal E}\circ{\rm id})(|\Phi\rangle),(U\otimes I)|\Phi\rangle),\label{eq:dfnentfid}
\end{align}
respectively. Here, the integral in (\ref{eq:dfnavefid}) is taken with respect to the Haar measure on ${\mathcal H}$, and $|\Phi\rangle\in{\mathcal H}\otimes{\mathcal H}$ is a maximally entangled state with the Schmidt rank $d$. As proved in \cite{nielsen2002simple}, it holds that
\begin{align}
F_{\rm av}({\mathcal E};U)=\frac{dF_{\rm en}({\mathcal E};U)+1}{d+1}.\label{eq:saruganseki}
\end{align}

Let us introduce two lemmas that will be used in the following Appendices.

\begin{lmm}\label{lmm:uhlmann}
For any two bipartite pure states $|\psi\rangle\in{\mathcal H}^A\otimes{\mathcal H}^B$ and $|\phi\rangle\in{\mathcal H}^A\otimes{\mathcal H}^{B'}$ that satisfy
\begin{align}
\left\|\psi^A-\phi^A\right\|_1\leq\epsilon,\label{eq:mitusmi}
\end{align}
the following statements hold:
\begin{enumerate}
\item There exists a linear CPTP map ${\mathcal T}:B\rightarrow B'$ that satisfies
\begin{align}
\left\|{\mathcal T}(|\psi\rangle\!\langle\psi|)-|\phi\rangle\!\langle\phi|\right\|_1\leq2\sqrt{\epsilon}.\label{eq:kuzo}
\end{align}
\item If ${\rm dim}{\mathcal H}^B\leq{\rm dim}{\mathcal H}^{B'}$, then there exists an isometry ${\tilde W}:{\mathcal H}^B\rightarrow {\mathcal H}^{B'}$ that satisfies
\begin{align}
\left\|{\tilde W}|\psi\rangle\!\langle\psi|{\tilde W}^\dagger-|\phi\rangle\!\langle\phi|\right\|_1\leq2\sqrt{\epsilon}.\label{eq:zokuzoku}
\end{align}
\end{enumerate}
\end{lmm}

\begin{prf}
To prove 2), let $|{\tilde\psi}\rangle\in{\mathcal H}^A\otimes{\mathcal H}^{B'}$ be a purification of $\psi^A$. Since all purifications are equivalent up to a local isometry, there exists an isometry $W_1:{\mathcal H}^B\rightarrow {\mathcal H}^{B'}$ that satisfies $W_1|\psi\rangle=|{\tilde\psi}\rangle$. From (\ref{eq:mitusmi}) and (\ref{eq:tracefid}), the states satisfy
\begin{align}
F({\tilde\psi}^A,\phi^A)\geq1-\epsilon.\nn
\end{align}
Due to (\ref{eq:cooppure}) and (\ref{eq:cooppuree}), there exists a unitary $W_2$ acting on ${\mathcal H}^{B'}$ such that
\begin{align}
F(W_2|{\tilde\psi}\rangle,|\phi\rangle)\geq1-\epsilon.\nn
\end{align}
Using (\ref{eq:tracefid}) once again, we obtain
\begin{eqnarray}
&&\left\|W_2W_1|\psi\rangle\!\langle\psi|W_1^\dagger W_2^\dagger-|\phi\rangle\!\langle\phi|\right\|_1\nn\\
&&=\left\|W_2|{\tilde\psi}\rangle\!\langle{\tilde\psi}|W_2^\dagger-|\phi\rangle\!\langle\phi|\right\|_1\leq2\sqrt{\epsilon},\nn
\end{eqnarray}
which implies (\ref{eq:zokuzoku}) by ${\tilde W}:=W_2W_1$.

To prove 1), let $B''$ be an ancillary system such that
\begin{align}
{\rm dim}{\mathcal H}^B\leq{\rm dim}{\mathcal H}^{B'}\times{\rm dim}{\mathcal H}^{B''}.\nn
\end{align}
Due to 2), for any $|\varphi\rangle\in{\mathcal H}^{B''}$ there exists an isometry ${\tilde W}:{\mathcal H}^B\rightarrow {\mathcal H}^{B'}\otimes{\mathcal H}^{B''}$ that satisfies
\begin{align}
\left\|{\tilde W}|\psi\rangle\!\langle\psi|{\tilde W}^\dagger-|\phi\rangle\!\langle\phi|^{AB'}\otimes|\varphi\rangle\!\langle\varphi|^{B''}\right\|_1\leq2\sqrt{\epsilon}.\label{eq:ayasan}
\end{align}
Define a linear CPTP map ${\mathcal T}:B\rightarrow B'$ by
\begin{align}
{\mathcal T}:\tau\rightarrow{\rm Tr}_{B''}[{\tilde W}\tau{\tilde W}^\dagger].\nn
\end{align}
 From (\ref{eq:ayasan}) and (\ref{eq:monottdd}), we obtain (\ref{eq:kuzo}).\hfill$\blacksquare$
\end{prf}

\begin{lmm}\label{lmm:prodfid}
For any pure states $|\psi\rangle\in{\mathcal H}^A$, $|\phi\rangle\in{\mathcal H}^B$ and any state $\rho\in{\mathcal S}({\mathcal H}^A\otimes{\mathcal H}^B)$ satisfying
\begin{align}
&1-F(\rho^A,|\psi\rangle)\leq\epsilon,\label{eq:fidprod1}\\
&1-F(\rho^B,|\phi\rangle)\leq\epsilon,\label{eq:fidprod2}
\end{align}
we have
\begin{align}
1-F(\rho^{AB},|\psi\rangle^A|\phi\rangle^B)\leq6\sqrt{\epsilon}.\label{eq:fidprod3}
\end{align}
\end{lmm}

\begin{prf}
Let $|\chi_\rho\rangle\in{\mathcal H}^A\otimes{\mathcal H}^B\otimes{\mathcal H}^E$ be a purification of $\rho^{AB}$, and $|\tau\rangle\in{\mathcal H}^E$ be an arbitrary pure state. Due to Equality (\ref{eq:cooppuree}), Condition (\ref{eq:fidprod1}) and the fact that $|\chi_\rho\rangle$ is a purification of $\rho^{A}$, there exists a unitary $W$ on ${\mathcal H}^B\otimes{\mathcal H}^E$ such that 
\begin{align}
F(|\chi_\rho\rangle,|\psi\rangle^A(W|\phi\rangle|\tau\rangle)^{BE})\geq1-\epsilon,\nn
\end{align}
which implies
\begin{align}
\left\|\proj{\chi_\rho}-\proj{\psi}^A\otimes W(\proj{\phi}^B\otimes\proj{\tau}^E)W^\dagger\right\|_1\leq2\sqrt{\epsilon}\nn
\end{align}
due to (\ref{eq:tracefid}). From the monotonicity of the trace distance under the partial trace (\ref{eq:monottd}), we have
\begin{align}
\left\|\rho^{AB}-\proj{\psi}^A\otimes {\tilde\phi}^B\right\|_1\leq2\sqrt{\epsilon},\quad\left\|\rho^{B}-{\tilde\phi}^B\right\|_1\leq2\sqrt{\epsilon},\nn
\end{align} 
where we defined ${\tilde\phi}:={\rm Tr}_E[W(\proj{\phi}^B\otimes\proj{\tau}^E)W^\dagger]$. The triangle inequality (\ref{eq:triangle}) and (\ref{eq:tdtensor}) yield
\begin{align}
&\left\|\rho^{AB}-\proj{\psi}^A\otimes\proj{\phi}^B\right\|_1\nn\\
&\leq\left\|\rho^{AB}-\proj{\psi}^A\otimes {\tilde\phi}^B\right\|_1\nn\\
&\quad+\left\|\proj{\psi}^A\otimes {\tilde\phi}^B-\proj{\psi}^A\otimes\proj{\phi}^B\right\|_1\nn\\
&\leq2\sqrt{\epsilon}+\left\|{\tilde\phi}^B-\proj{\phi}\right\|_1\nn\\
&\leq2\sqrt{\epsilon}+\left\|{\tilde\phi}^B-\rho^{B}\right\|_1+\left\|\rho^{B}-\proj{\phi}\right\|_1\nn\\
&\leq2\sqrt{\epsilon}+2\sqrt{\epsilon}+2\sqrt{\epsilon}\nn\\
&=6\sqrt{\epsilon},\nn
\end{align}
where we used Condition (\ref{eq:fidprod2}) and (\ref{eq:tracefid}) in the last inequality. Using (\ref{eq:tracefid}) once again, we obtain (\ref{eq:fidprod3}). \hfill$\blacksquare$
\end{prf}

\subsection{Gentle measurement lemma}

The gentle measurement lemma (Lemma 9.4.1 in \cite{wildetext}) states that for any $\rho\in{\mathcal S}({\mathcal H})$, $X\in{\mathcal L}({\mathcal H})$ and $\epsilon\geq0$ such that $0\leq X\leq I$ and ${\rm Tr}[\rho X]\geq1-\epsilon$, we have
\begin{eqnarray}
\left\|\rho-\frac{\sqrt{X}\rho\sqrt{X}}{{\rm Tr}[\rho X]}\right\|_1\leq2\sqrt{\epsilon}.\nn
\end{eqnarray}
Let us introduce extensions of the gentle measurement lemma. Although similar lemmas have been used in  the literature, we provide rigorous proofs for completeness.

\begin{lmm}\label{lmm:gentlemeasurement} For any $\rho\in{\mathcal S}({\mathcal H})$, $X,Y\in{\mathcal L}({\mathcal H})$ and $\epsilon\in[0,1]$ such that 
\begin{align}
0\leq X\leq I,\;0\leq Y\leq I\nn
\end{align}
and
\begin{align}
{\rm Tr}[\rho X]\geq1-\epsilon,\;{\rm Tr}[\rho Y]\geq1-\epsilon,\label{eq:xyrho}
\end{align}
define
\begin{eqnarray}
D_{XY}&:=&{\rm Tr}[\sqrt{Y}\sqrt{X}\rho\sqrt{X}\sqrt{Y}],\nn\\
\rho_{XY}&:=&\frac{\sqrt{Y}\sqrt{X}\rho\sqrt{X}\sqrt{Y}}{D_{XY}}.\nn
\end{eqnarray}
Then we have
\begin{eqnarray}
D_{XY}\geq1-2\sqrt{\epsilon},\;\left\|\rho-\rho_{XY}\right\|_1\leq5\sqrt[4]{\epsilon}.\nn
\end{eqnarray}
\end{lmm}
\begin{prf}
Define
\begin{eqnarray}
\rho_{X}:=\frac{\sqrt{X}\rho\sqrt{X}}{{\rm Tr}[\sqrt{X}\rho\sqrt{X}]}.\nn
\end{eqnarray}
Due to the gentle measurement lemma, Condition (\ref{eq:xyrho}) implies
\begin{eqnarray}
\left\|\rho-\rho_{X}\right\|_1\leq2\sqrt{\epsilon}.\nn
\end{eqnarray}
Consequently, we have
\begin{eqnarray}
D_{XY}&=&{\rm Tr}[\rho_{X}Y]\nn\\
&\geq&{\rm Tr}[\rho Y]-\frac{1}{2}\left\|\rho-\rho_{X}\right\|_1\nn\\
&\geq&1-\epsilon-\sqrt{\epsilon}\geq1-2\sqrt{\epsilon},\nn
\end{eqnarray}
 where the second line follows from (\ref{eq:defTD2}), which leads to
\begin{eqnarray}
\left\|\rho_{X}-\rho_{XY}\right\|_1\leq2\sqrt{2\sqrt{\epsilon}}\nn
\end{eqnarray}
by (\ref{eq:xyrho}). Thus we obtain
\begin{eqnarray}
\left\|\rho-\rho_{XY}\right\|_1&\leq&\left\|\rho-\rho_{X}\right\|_1+\left\|\rho_{X}-\rho_{XY}\right\|_1\nn\\
&\leq&2\sqrt{\epsilon}+2\sqrt{2\sqrt{\epsilon}}\leq5\sqrt[4]{\epsilon}.\nn
\end{eqnarray}
\hfill$\blacksquare$
\end{prf}

\begin{lmm}\label{lmm:3f}
Suppose $\rho,\sigma\in{\mathcal S}({\mathcal H})$ satisfy $\|\rho-\sigma\|_1\leq\epsilon$. Let $\{\lambda_i^{\downarrow}\}_{i=1}^d$ be the eigenvalues of $\sigma$ sorted in decreasing order, where $d:={\rm dim}{\mathcal H}$, and let
\begin{align}
\sigma=\sum_{i=1}^d\lambda_i^{\downarrow}\proj{i}\nn
\end{align}
be the eigen decomposition of $\sigma$. Define a projection operator
\begin{align}
\tilde\Pi:=\sum_{i=1}^{{\rm rank}[\rho]}\proj{i}.\nn
\end{align}
Then we have
\begin{eqnarray}
{\rm Tr}[\tilde\Pi\sigma]\geq1-\epsilon.\nn
\end{eqnarray}
\end{lmm}

\begin{prf}
Let $\Pi_\rho$ be the projection onto ${\rm supp}[\rho]$, and $\Pi_\rho^\perp$ be that onto its orthogonal complement (i.e., $\Pi_\rho^\perp=I-\Pi_\rho$). Then we have
\begin{eqnarray}
\epsilon&\geq&\left\|\rho-\sigma\right\|_1\geq\left\|\rho-\Pi_\rho\sigma\Pi_\rho-\Pi_\rho^\perp\sigma\Pi_\rho^\perp\right\|_1\nonumber\\
&=&\left\|\rho-\Pi_\rho\sigma\Pi_\rho\right\|_1+{\rm Tr}[\Pi_\rho^\perp\sigma]\geq1-{\rm Tr}[\Pi_\rho\sigma],\label{eq:ayano}
\end{eqnarray}
 where the second inequality follows from the monotonicity of the trace distance under a linear CPTP map defined by
\begin{align}
\tau\rightarrow\Pi_\rho\tau\Pi_\rho+\Pi_\rho^\perp\tau\Pi_\rho^\perp.\nonumber
\end{align}
We also have
\begin{eqnarray}
&&{\rm Tr}[{\tilde\Pi}\sigma]=\sum_{i=1}^{{\rm rank}[\rho]}\lambda_i^{\downarrow}\nonumber\\
&&=\sum_{i=1}^{{\rm rank}[\rho]}\lambda_i^{\downarrow}\langle i|\Pi_\rho|i\rangle+\sum_{i=1}^{{\rm rank}[\rho]}\lambda_i^{\downarrow}(1-\langle i|\Pi_\rho|i\rangle)\nonumber\\
&&\geq\sum_{i=1}^{{\rm rank}[\rho]}\lambda_i^{\downarrow}\langle i|\Pi_\rho|i\rangle+\lambda_{{\rm rank}[\rho]}^{\downarrow}\sum_{i=1}^{{\rm rank}[\rho]}(1-\langle i|\Pi_\rho|i\rangle)\nonumber\\
&&=\sum_{i=1}^{{\rm rank}[\rho]}\lambda_i^{\downarrow}\langle i|\Pi_\rho|i\rangle+\lambda_{{\rm rank}[\rho]}^{\downarrow}\sum_{i={\rm rank}[\rho]+1}^d\langle i|\Pi_\rho|i\rangle\nonumber\\
&&\geq\sum_{i=1}^d\lambda_i^{\downarrow}\langle i|\Pi_\rho|i\rangle={\rm Tr}[\Pi_\rho\sigma],\label{eq:iori}
\end{eqnarray}
where the fourth line follows due to
\begin{eqnarray}
\sum_{i=1}^d\langle i|\Pi_\rho|i\rangle={\rm Tr}[\Pi_\rho]={\rm rank}[\rho].\nonumber
\end{eqnarray}
From (\ref{eq:ayano}) and (\ref{eq:iori}), we obtain
\begin{align}
\quad\qquad\qquad\qquad{\rm Tr}[{\tilde\Pi}\sigma]\geq{\rm Tr}[\Pi_\rho\sigma]\geq1-\epsilon.\qquad\qquad\qquad\blacksquare\nonumber
\end{align}
\end{prf}

\subsection{Continuity of Quantum Entropies}\label{app:content}

Define
\begin{eqnarray}
\eta_0(x):=\begin{cases}
-x\log{x}&(x\leq 1/e)\\
\frac{1}{e}&(x\geq 1/e)
\end{cases}\nn
\end{eqnarray}
 and $\eta(x)=x+\eta_0(x)$, where $e$ is the base of the natural logarithm.  Define also
 \begin{align}
 h(x):=-x\log{x}-(1-x)\log{(1-x)}.\nn
 \end{align}
 For two states $\rho$ and $\sigma$ in a $d$-dimensional quantum system ($d<\infty$) such that $\|\rho-\sigma\|_1\leq\epsilon$, we have
\begin{eqnarray}
|S(\rho)-S(\sigma)|\leq\epsilon\log{d}+\eta_0(\epsilon),\label{eq:contvNent}
\end{eqnarray}
which is called the {\it Fannes inequality}\cite{fannes73}.  A simple calculation then yields
\begin{eqnarray}
|S(\rho)-S(\sigma)|\leq\eta(\epsilon)\log{d}.\label{eq:fannes}
\end{eqnarray}
For two bipartite states $\rho,\sigma\in{\mathcal S}({\mathcal H}^A\otimes{\mathcal H}^B)$ such that  $\|\rho-\sigma\|_1\leq\epsilon<1$, we have
\begin{eqnarray}
|S(A|B)_\rho-S(A|B)_\sigma|\leq4\epsilon\log{d_A}+2h(\epsilon).\nn
\end{eqnarray}
 This inequality is called the {\it Alicki-Fannes inequality}\cite{alicki04}, and leads to 
\begin{eqnarray}
|S(A|B)_\rho-S(A|B)_\sigma|\leq4\eta(\epsilon)\log{d_A}.\label{eq:alicki}
\end{eqnarray}
Note that the upper bound in (\ref{eq:alicki}) does not depend on $d_B$. As a consequence, we have
\begin{eqnarray}
&&|I(A:B)_\rho-I(A:B)_\sigma|\nn\\
&\leq&|S(A)_\rho-S(A)_\sigma|+|S(A|B)_\rho-S(A|B)_\sigma|\nn\\
&\leq&5\eta(\epsilon)\log{d_A}.\label{eq:contmi}
\end{eqnarray}

The following lemma will be used for evaluating average errors.
\begin{lmm}\label{lmm:yoshiki}
Let $c\in(0,\infty)$ be a constant, $f:[0,c]\rightarrow{\mathbb R}$ be a monotonically nondecreasing function that  satisfies $f(c)<\infty$, and $\{p_k\}_{k\in{\mathbb K}}$ be a probability distribution on a countable set ${\mathbb K}$.  Suppose $\epsilon_k\:(k\in{\mathbb K})$ satisfies $\epsilon_k\in[0,c]$, and $\sum_{k\in{\mathbb K}}p_k\epsilon_k\leq\epsilon$ for a given $\epsilon\in(0,c^2]$. Then we have
\begin{eqnarray}
\sum_{k\in{\mathbb K}}p_kf(\epsilon_k)\leq f(\sqrt{\epsilon})+f( c)\cdot\sqrt{\epsilon}.\label{eq:detectivewall}
\end{eqnarray}
\end{lmm}
\begin{prf}
Define ${\mathbb K}(\lambda):=\{k\in{\mathbb K}\:|\:\epsilon_k\leq\lambda\}$. For any $t>0$, we have
\begin{eqnarray}
\sum_{k\in{\mathbb K}\setminus{\mathbb K}(t\epsilon)}p_k&=&\frac{1}{t\epsilon}\sum_{k\in{\mathbb K}\setminus{\mathbb K}(t\epsilon)}p_kt\epsilon\;\;\leq\;\;\frac{1}{t\epsilon}\sum_{k\in{\mathbb K}\setminus{\mathbb K}(t\epsilon)}p_k\epsilon_k\nonumber\\
&\leq&\frac{1}{t\epsilon}\sum_{k\in {\mathbb K}}p_k\epsilon_k\leq\frac{1}{t},\nonumber
\end{eqnarray}
and consequently,
\begin{eqnarray}
\sum_{k\in{\mathbb K}}p_kf(\epsilon_k)&=&\sum_{k\in{\mathbb K}(t\epsilon)}p_kf(\epsilon_k)+\sum_{k\in{\mathbb K}\setminus{\mathbb K}(t\epsilon)}p_kf(\epsilon_k)\nn\\
&\leq&f(t\epsilon)+\frac{f( c)}{t}.\nn
\end{eqnarray}
Choosing $t=1/\sqrt{\epsilon}$, we obtain (\ref{eq:detectivewall}).\hfill$\blacksquare$
\end{prf}

\subsection{Entanglement Measures}\label{eq:entmeas}

A function ${\ca E}:{\mathcal S}({\mathcal H}^A\otimes{\mathcal H}^B)\rightarrow[0,\infty)$ is called an {\it entanglement measure} if it satisfies the following three properties\cite{plenio07}:
\begin{enumerate}
\item If $\rho$ is a pure state on $AB$, then ${\ca E}(\rho)=S(\rho^A)$.
\item If $\rho$ is a separable state on $AB$, then ${\ca E}(\rho)=0$.
\item ${\ca E}(\rho)$ does not increase on average under LOCC, i.e., if an ensemble $\{p_i,\rho_i\}\:(\rho_i\in{\mathcal S}({\mathcal H}^{A}\otimes{\mathcal H}^{B}))$ is obtained from $\rho\in{\mathcal S}({\mathcal H}^A\otimes{\mathcal H}^B)$ by an LOCC transformation between $A$ and $B$, then $\sum_ip_i{\ca E}(\rho_i)\leq {\ca E}(\rho)$.
\end{enumerate}
An entanglement measure $\ca E$ is said to be {\it asymptotically continuous}, if there exists an $n$-independent nonnegative function $\delta(\epsilon)$ that satisfies $\lim_{\epsilon\rightarrow0}\delta(\epsilon)=0$, and it holds that
\begin{align}
\left|{\ca E}(\rho)-{\ca E}(\sigma)\right|\leq \delta(\epsilon)\log{d_Ad_B}\nn
\end{align}
for all $\rho,\sigma\in{\mathcal S}({\mathcal H}^A\otimes{\mathcal H}^B)$ satisfying $\|\rho-\sigma\|_1\leq\epsilon$. Examples of asymptotically continuous entanglement measures are entanglement of formation \cite{wooters98,nielsen00}, the relative entropy of entanglement \cite{vedral97,vedral98,matthew99} and squashed entanglement \cite{christandl04,alicki04}.

\section{Equivalence of Conditions (\ref{eq:bFapprox}) and (\ref{eq:Feapprox})}\label{app:equivbe}

We prove the equivalence of Conditions (\ref{eq:bFapprox}) and (\ref{eq:Feapprox}) by showing that
\begin{align}
&\frac{d^{2n}}{36(d^{2n}+1)}(1-F_{\rm en}({\mathcal M}_n,K_n,L_n;U^{\otimes n}))^2\nn\\
&\leq1-F_{\rm av}({\mathcal M}_n,K_n,L_n;U^{\otimes n})\label{eq:massugu1}\\
&\leq8\sqrt[3]{1-F_{\rm en}({\mathcal M}_n,K_n,L_n;U^{\otimes n})}.\label{eq:massugu2}
\end{align}
In the following, we prove these inequalities for the case of $n=1$. It is straightforward to generalize the proof for an arbitrary $n$.

Define states $\rho({\mathcal M}_1)$ and $\rho({\mathcal M}_1,\psi)$ by
\begin{align}
&\rho({\mathcal M}_1):={\mathcal M}_1(|\Phi_d\rangle^{AR_A}|\Phi_d\rangle^{BR_B}|\Phi_{K_1}\rangle^{A_0B_0}),\nn\\
&\rho({\mathcal M}_1,\psi):={\mathcal M}_1(|\psi\rangle^{AB}|\Phi_{K_1}\rangle^{A_0B_0}),
\end{align}
and define quantum operations ${\hat{\mathcal M}}_1:AB\rightarrow AB$ and ${\check{\mathcal M}}_1:AB\rightarrow A_1B_1$ by
\begin{align}
&{\hat{\mathcal M}}_1(\tau)={\rm Tr}_{A_1B_1}[{\mathcal M}_1(\tau^{AB}\otimes \Phi_{K_1}^{A_0B_0})],\nn\\
&{\check{\mathcal M}}_1(\tau)={\rm Tr}_{AB}[{\mathcal M}_1(\tau^{AB}\otimes \Phi_{K_1}^{A_0B_0})],\nn
\end{align}
respectively. Using (\ref{eq:fidpens}) and
\begin{align}
\int_{\rm Haar} p(d\psi)\proj{\psi}=\pi_d^A\otimes\pi_d^B,\nn
\end{align}
where the integral is taken with respect to the Haar measure on ${\mathcal H}^A\otimes{\mathcal H}^B$, it is straightforward to verify that
\begin{align}
&\int_{\rm Haar} p(d\psi)\:F(\rho({\mathcal M_1},\psi)^{A_1B_1},|\Phi_{L_1}\rangle)\nn\\
=&\int_{\rm Haar} p(d\psi)\:F({\check{\mathcal M}}_1(|\psi\rangle),|\Phi_{L_1}\rangle)\nn\\
=&F\left({\check{\mathcal M}}_1(\pi_d^A\otimes\pi_d^B),|\Phi_{L_1}\rangle\right)\nn\\
=&F\left(\rho({\mathcal M_1})^{A_1B_1}, |\Phi_{L_1}\rangle\right).\label{eq:urusakute}
\end{align}
From (\ref{eq:dfnavefid}), (\ref{eq:fidmonPT}), and (\ref{eq:ensfide}), we have
\begin{align}
&F_{\rm av}({\hat{\mathcal M_1}};U)\nn\\
&=\int_{\rm Haar} p(d\psi)\:F({\hat{\mathcal M}}_1(|\psi\rangle), U|\psi\rangle)\nn\\
&=\int_{\rm Haar} p(d\psi)\:F(\rho({\mathcal M_1},\psi)^{AB}, U|\psi\rangle)\label{eq:haybadboy}\\
&\geq \int_{\rm Haar} p(d\psi)\:F(\rho({\mathcal M_1},\psi), (U|\psi\rangle)^{AB}|\Phi_{L_1}\rangle^{A_1B_1})\nn\\
&=F_{\rm av}({\mathcal M}_1,K_1,L_1;U)\label{eq:plusplus}
\end{align}
and
\begin{align}
&\int_{\rm Haar} p(d\psi)\:F(\rho({\mathcal M_1},\psi)^{A_1B_1}, |\Phi_{L_1}\rangle)\nn\\
&\geq \int_{\rm Haar} p(d\psi)\:F(\rho({\mathcal M_1},\psi), (U|\psi\rangle)^{AB}|\Phi_{L_1}\rangle^{A_1B_1})\nn\\
&= F_{\rm av}({\mathcal M}_1,K_1,L_1;U).\label{eq:plusplusplus}
\end{align}
Similarly, from (\ref{eq:dfnentfid}), (\ref{eq:fidmonPT}) and (\ref{eq:entfide}), we have
\begin{align}
&F_{\rm en}({\hat{\mathcal M_1}};U)\nn\\
&=F({\hat{\mathcal M}}_1(|\Phi_d\rangle^{AR_A}|\Phi_d\rangle^{BR_B}), |\Psi_U\rangle)\nn\\
&=F(\rho({\mathcal M_1})^{AB}, |\Psi_U\rangle)\label{eq:evevev}\\
&\geq F(\rho({\mathcal M_1}), |\Psi_U\rangle^{AB}|\Phi_{L_1}\rangle^{A_1B_1})\nn\\
&=F_{\rm en}({\mathcal M}_1,K_1,L_1;U)\label{eq:nininini}
\end{align}
and
\begin{align}
&F(\rho({\mathcal M_1})^{A_1B_1}, |\Phi_{L_1}\rangle)\nn\\
&\geq F(\rho({\mathcal M_1}), |\Psi_U\rangle^{AB}|\Phi_{L_1}\rangle^{A_1B_1})\nn\\
&=F_{\rm en}({\mathcal M}_1,K_1,L_1;U).\label{eq:akirame}
\end{align}
Due to (\ref{eq:saruganseki}), we also have
\begin{align}
1-F_{\rm en}({\hat{\mathcal M_1}};U)&=\left(1+\frac{1}{d^2}\right)(1-F_{\rm av}({\hat{\mathcal M_1}};U))\label{eq:teruyuki}\\
&\geq1-F_{\rm av}({\hat{\mathcal M_1}};U).\label{eq:tommytommy}
\end{align}

Suppose we have
\begin{align}
\epsilon=1-F_{\rm av}({\mathcal M}_1,K_1,L_1;U).\nn
\end{align}
From (\ref{eq:plusplus}), (\ref{eq:teruyuki}) and (\ref{eq:evevev}), we have
\begin{align}
\left(1+\frac{1}{d^2}\right)\epsilon&\geq\left(1+\frac{1}{d^2}\right)(1-F_{\rm av}({\hat{\mathcal M_1}};U))\nn\\
&=1-F_{\rm en}({\hat{\mathcal M_1}};U)\nn\\
&=1-F(\rho({\mathcal M_1})^{AB}, |\Psi_U\rangle).\nn
\end{align}
Due to (\ref{eq:plusplusplus}) and (\ref{eq:urusakute}), we also have
\begin{align}
\epsilon\geq1-F\left(\rho({\mathcal M_1})^{A_1B_1}, |\Phi_{L_1}\rangle\right).\nn
\end{align}
Therefore, from (\ref{eq:entfide}) and Lemma \ref{lmm:prodfid}, we obtain that
\begin{align}
&1-F_{\rm en}({\mathcal M}_1,K_1,L_1;U)\nn\\
&=1-F\left(\rho({\mathcal M_1}), |\Psi_U\rangle^{AB}|\Phi_{L_1}\rangle^{A_1B_1}\right)\nn\\
&\leq6\sqrt{\epsilon}\sqrt{1+\frac{1}{d^2}},\nn
\end{align}
which implies (\ref{eq:massugu1}) for $n=1$.

To prove Inequality (\ref{eq:massugu2}), let us define indicator functions $\theta_1$ and $\theta_2$ by
\begin{align}
\theta_1(\psi|x):=
\begin{cases}
1 &\text{if }1-F(\rho({\mathcal M_1},\psi)^{AB}, U|\psi\rangle)\leq x\\
0 &\text{othewise}
\end{cases}\nn
\end{align}
and
\begin{align}
\theta_2(\psi|x):=
\begin{cases}
1 & \text{if }1-F(\rho({\mathcal M_1},\psi)^{A_1B_1}, |\Phi_{L_1}\rangle)\leq x\\
0 & \text{otherwise}
\end{cases},\nn
\end{align}
respectively, for any positive number $x$. It follows from Lemma \ref{lmm:prodfid} that for any $\psi$ satisfying $\theta_1(\psi|x)=\theta_2(\psi|x)=1$, we have
\begin{align}
1-F(\rho({\mathcal M_1},\psi), (U|\psi\rangle)^{AB}|\Phi_{L_1}\rangle^{A_1B_1})\leq6\sqrt{x}.\nn
\end{align}
Thus we obtain
\begin{align}
&1-F_{\rm av}({\mathcal M}_1,K_1,L_1;U)\nn\\
&=\int_{\rm Haar} p(d\psi)\:\left(1-F(\rho({\mathcal M_1},\psi), (U|\psi\rangle)^{AB}|\Phi_{L_1}\rangle^{A_1B_1})\right)\nn\\
&=\int_{\rm Haar} p(d\psi)\:\theta_1(\psi|x)\theta_2(\psi|x)\nn\\
&\quad\quad\quad\quad\quad\times\left(1-F(\rho({\mathcal M_1},\psi), (U|\psi\rangle)^{AB}|\Phi_{L_1}\rangle^{A_1B_1})\right)\nn\\
&\quad+\int_{\rm Haar} p(d\psi)\:(1-\theta_1(\psi|x)\theta_2(\psi|x))\nn\\
&\quad\quad\quad\quad\quad\times\left(1-F(\rho({\mathcal M_1},\psi), (U|\psi\rangle)^{AB}|\Phi_{L_1}\rangle^{A_1B_1})\right)\nn\\
&\leq6\sqrt{x}\int_{\rm Haar} p(d\psi)\:\theta_1(\psi|x)\theta_2(\psi|x)\nn\\
&\quad+\int_{\rm Haar} p(d\psi)\:(1-\theta_1(\psi|x)\theta_2(\psi|x))\nn\\
&\leq6\sqrt{x}+\int_{\rm Haar} p(d\psi)\:(1-\theta_1(\psi|x)\theta_2(\psi|x))\label{eq:antamona}
\end{align}
for any $x>0$.

Suppose that we have
\begin{align}
\epsilon=1-F_{\rm en}({\mathcal M}_1,K_1,L_1;U).\nn
\end{align}
From (\ref{eq:nininini}), (\ref{eq:tommytommy}) and (\ref{eq:haybadboy}), we have
\begin{align}
\epsilon&\geq1-F_{\rm en}({\hat{\mathcal M_1}};U)\nn\\
&\geq1-F_{\rm av}({\hat{\mathcal M_1}};U)\nn\\
&=\int_{\rm Haar} p(d\psi)\:\left(1-F(\rho({\mathcal M_1},\psi)^{AB}, U|\psi\rangle)\right).\label{eq:ikedachara}
\end{align}
Due to (\ref{eq:akirame}) and (\ref{eq:urusakute}), we also have
\begin{align}
\epsilon&\geq1-F\left(\rho({\mathcal M_1})^{A_1B_1}, |\Phi_{L_1}\rangle\right)\nn\\
&=\int_{\rm Haar} p(d\psi)\:\left(1-F(\rho({\mathcal M_1},\psi)^{A_1B_1}, |\Phi_{L_1}\rangle)\right).\label{eq:ikeikeike}
\end{align}
It follows from Inequality (\ref{eq:ikedachara}) that
\begin{align}
\epsilon&\geq\int_{\rm Haar} p(d\psi)\:(1-\theta_1(\psi|x))\left(1-F(\rho({\mathcal M_1},\psi)^{AB}, U|\psi\rangle)\right)\nn\\
&\geq x\int_{\rm Haar} p(d\psi)\:(1-\theta_1(\psi|x)),\nn
\end{align}
and from (\ref{eq:ikeikeike}) that
\begin{align}
\epsilon&\geq\int_{\rm Haar} \!p(d\psi)\:(1-\theta_2(\psi|x))\!\left(1\!-\!F(\rho({\mathcal M_1},\psi)^{A_1B_1},U|\psi\rangle)\right)\nn\\
&\geq x\int_{\rm Haar} p(d\psi)\:(1-\theta_2(\psi|x)).\nn
\end{align}
Thus we obtain
\begin{align}
\frac{2\epsilon}{x}&\geq\int_{\rm Haar} p(d\psi)(1-\theta_1(\psi|x))\nn\\
&\quad\quad+\int_{\rm Haar} p(d\psi)(1-\theta_2(\psi|x))\nn\\
&\geq\int_{\rm Haar} p(d\psi)\:(1-\theta_1(\psi|x)+\theta_1(\psi|x)(1-\theta_2(\psi|x)))\nn\\
&=\int_{\rm Haar} p(d\psi)\:(1-\theta_1(\psi|x)\theta_2(\psi|x)).\nn
\end{align}
Combining this with (\ref{eq:antamona}) and substituting $x=\epsilon^{2/3}$, we arrive at 
\begin{align}
1-F_{\rm av}({\mathcal M}_1,K_1,L_1;U)\leq8\sqrt[3]{\epsilon},\nn
\end{align}
which completes the proof of Inequality (\ref{eq:massugu2}).\hfill$\blacksquare$

\section{Proofs of Lemma \ref{thm:recov} and Inequalities (\ref{eq:qsmentccopt})}

\subsection{Proof of  Lemma \ref{thm:recov}}\label{app:prfrecov}

Let $V:{\bar A}A_0\rightarrow A'E_0$ be an isometry such that the Naimark extension of $\{M_{k}\}_{k\in{\mathbb K}}$ is given by $M_{k}=\bra{k}^{E_0}V$, and let \begin{eqnarray}
\Psi'^{A'E_0}:=\sum_{k\in{\mathbb K}}\proj{k}^{E_0}V((\Psi^{\otimes n})^{\bar A}\otimes{\varrho}^{A_0})V^\dagger\proj{k}^{E_0}.\nn
\end{eqnarray}
We have
\begin{eqnarray}
&&H(\{p_k\}_{k\in{\mathbb K}})=S(E_0)_{\Psi'}=S(A'E_0)_{\Psi'}-S(A'|E_0)_{\Psi'}\nn\\
&&\geq S({\bar A}A_0)_{\Psi^{\otimes n}\otimes{\varrho}}-\sum_{k\in{\mathbb K}}p_kS(A')_{\Psi_{k}}=\Delta S(A')_{ave},\nn
\end{eqnarray}
where the second line follows due to the von Neumann entropy nondecreasing under dephasing operations. Hence we obtain the first inequality in (\ref{eq:markovmeasure3}). The second inequality is due to $\Delta{S}(G)_{av}\geq0$, which follows from the  concavity of the von Neumann entropy.

As for the third inequality, we first prove that there exists a nondecreasing function ${\tilde\eta}(\epsilon)$, satisfying $\lim_{\epsilon \rightarrow0}{\tilde\eta}(\epsilon)=0$, such that
\begin{eqnarray}
&&\Delta S(A')_{ave}-\Delta S(G)_{ave}\nn\\
&\geq&I({\bar B}{\bar C}:G)_{ave}-n{\tilde\eta}(\epsilon)\log{(d_Bd_C)}.\nn
\end{eqnarray}
Define
\begin{eqnarray}
\epsilon_k:=\left\|(\Psi^{\otimes n})^{{\bar B}{\bar C}}-\Psi_{k}^{{\bar B}{\bar C}}\right\|_1.\nn
\end{eqnarray}
 Using (\ref{eq:fannes}), we have 
\begin{eqnarray}
&&\!\!\!\!\!\!\!\!\!\!\!\!\!\!\!\!\!\!\!\!\!\!\!\!\Delta S(A')_k  := nS(A)_{\Psi}+S(A_0)_{\varrho}-S(A')_{\Psi_{k}}\nn\\
&=&S({\bar B}{\bar C})_{\Psi^{\otimes n}}+S(A_0)_{\varrho}-S({\bar B}{\bar C}G)_{\Psi_{k}}\nn\\
		     &\geq& S({\bar B}{\bar C})_{\Psi_k}+S(A_0)_{\varrho}-S({\bar B}{\bar C}G)_{\Psi_{k}}\nn\\
		     &&\;\;\;\;\;\;-n\eta(\epsilon_k)\log{(d_Bd_C)}\nn\\
		     &=&S(A_0)_{\varrho}-S(G|{\bar B}{\bar C})_{\Psi_{k}}-n\eta(\epsilon_k)\log{(d_Bd_C)}\!\!\!\!\nn\\
		     &=&S(G)_{\varrho}-S(G)_{\Psi_{k}}+I({\bar B}{\bar C}:G)_{\Psi_{k}}\nn\\
		     &&\;\;\;\;\;\;-n\eta(\epsilon_k)\log{(d_Bd_C)}\nn\\
		     &=&\Delta S(G)_{k}+I({\bar B}{\bar C}:G)_{\Psi_{k}}-n\eta(\epsilon_k)\log{(d_Bd_C)},\nn
\end{eqnarray}
 where  we defined $\Delta S(G)_{k}:=S(G)_{\varrho}-S(G)_{\Psi_{k}}$. In the fifth line, we used the fact that $\varrho$ is a pure state on $A_0G$. Averaging over $k$, we obtain
\begin{eqnarray}
&&\Delta S(A')_{ave}-\Delta S(G)_{ave}\nn\\
&\geq&\sum_{k\in{\mathbb K}}p_{k}\left(I({\bar B}{\bar C}:G)_{\Psi_{k}}-n\eta(\epsilon_k)\log{(d_Bd_C)}\right).\nn
\end{eqnarray}
Applying Lemma \ref{lmm:yoshiki}  for $c=2$ together with $\epsilon_k\leq2$ and $\sum_{k\in{\mathbb K}}p_{k}\epsilon_k\leq\epsilon<1/8<2^2$ yields
\begin{eqnarray}
&&\Delta S(A')_{ave}-\Delta S(G)_{ave}\nn\\
&\geq&I({\bar B}{\bar C}:G)_{ave}-n{\tilde\eta}(\epsilon)\log{(d_Bd_C)},\nn
\end{eqnarray}
where we defined
\begin{eqnarray}
{\tilde\eta}(\epsilon):=\eta(\sqrt{\epsilon})+\eta(2)\cdot\sqrt{\epsilon}.\nn
\end{eqnarray}
 Second we prove that there exists a function ${\xi}( x)$, satisfying $\lim_{ x\rightarrow0}{\xi}(x)=0$, such that we have
\begin{align}
&I({\bar B}{\bar C}:G)_{ave}\geq nM^{R,m}_{A|AB}(\Psi^{ABC})-n{\xi}(n\epsilon)\log{(d_Ad_Bd_C)}.\nn\\&\label{eq:iavelow}
\end{align}
This simply follows from the results in \cite{waka15_rec} (see Theorem 15 and  Inequalities (69) and (71) therein). Defining
\begin{eqnarray}
{\tilde\xi}( x):={\xi}(x)+{\tilde\eta}(x),\label{eq:wanabee}
\end{eqnarray}
and noting ${\tilde\eta}(\epsilon)\leq{\tilde\eta}(n\epsilon)$, we obtain the last inequality in (\ref{eq:markovmeasure3}).

\hfill$\blacksquare$

\subsection{Proof of Inequalities (\ref{eq:qsmentccopt})}\label{app:statemerging}

The following theorem is essentially the same, but technically different from what is proved in \cite{horo05}. We give a rigorous proof for completeness.
\begin{thm}\label{thm:qsmoptimal}
Let ${\mathcal N}$ be state merging of $\Psi$ with error $\epsilon\in(0,1/4]$. Entanglement cost and classical communication cost of ${\mathcal N}$ are bounded below as
\begin{eqnarray}
\:\log{K}-\log{L}&\geq&S(B|A)_{\Psi}-\eta'(2\sqrt{\epsilon})\log{(d_RL)},\nn\\
C&\geq&I(B:R)_{\Psi}-5\eta(2\sqrt{\epsilon})\log{d_R},\nn
\end{eqnarray}
where
\begin{eqnarray}
\eta'(x):=\frac{5}{2}\eta(x)+\eta(2\sqrt{x})+\eta(2)\sqrt{x}.\label{eq:dfnetapr}
\end{eqnarray}
\end{thm}

\begin{prf}
Without loss of generality, we assume that the protocol $\mathcal N$ consists of (i) Bob's measurement described by $\{N_l^{BB_0\rightarrow B_1}\}_l$, (ii) communication of $l$ from Bob to Alice, and (iii) Alice's operation described by a CPTP map ${\mathcal O}_l:AA_0\rightarrow AA_{\!B}A_1$. The final state is given by
\beq
{\rho}({\mathcal N})^{AA_{\!B}RA_1B_1}=\sum_lp_l{\hat\Psi}_l,\label{eq:nihonrikujooo}
\eeq
where
\begin{align}
p_l:=\|N_l|\Psi\rangle|\Phi_{K}\rangle\|_1^2,\;|\Psi_l\rangle:=p_l^{-1/2}N_l|\Psi\rangle|\Phi_{K}\rangle\nn
\end{align}
and ${\hat\Psi}_l:={\mathcal O}_l(\Psi_l)$.  From (\ref{eq:fidelityqsm}) and (\ref{eq:tracefid}), we have
\begin{align}
\left\|{\rho}({\mathcal N})-{\Psi}^{AA_{\!B}R}\otimes{\Phi_L}^{A_1B_1}\right\|_1\leq2\sqrt{\epsilon}.\label{eq:jirit}
\end{align}

Define
\begin{eqnarray}
\epsilon_l:=\left\|{\hat\Psi}_l-{\Psi}^{AA_{\!B}R}\otimes{\Phi_L}^{A_1B_1}\right\|_1\label{eq:324}
\end{eqnarray}
and
\begin{eqnarray}
f_l:=F({\hat\Psi}_l,{\Psi}^{AA_{\!B}R}\otimes{\Phi_L}^{A_1B_1})\nn
\end{eqnarray}
for each $l$.  Due to the convexity of the square function, Inequality (\ref{eq:tracefid}), Equalities (\ref{eq:fidpens}), (\ref{eq:nihonrikujooo}) and Inequality (\ref{eq:fidelityqsm}), we have
\begin{eqnarray}
&&\!\!\!\!\!\!\!\!\!\left(\sum_{l}p_{l}\epsilon_{l}\right)^2\:\leq\:\sum_{l}p_{l}\epsilon_{l}^2\:\leq\:4\sum_{l}p_{l}(1-f_{l})\nn\\
&&\!\!\!\!\!\!\!\!\!=4-4F(\rho({\mathcal N}), |\Psi\rangle^{AA_{\!B}R}|\Phi_{L}\rangle^{A_1B_1})\leq4\epsilon,\nn
\end{eqnarray}
which yields
\begin{eqnarray}
\sum_{l}p_{l}\epsilon_{l}\leq2\sqrt{\epsilon}.\nn
\end{eqnarray}

Consider the following protocol, which is as a whole equivalent to the protocol described above.
\begin{enumerate}
\item Bob performs a CPTP map ${\mathcal E}_1:BB_0\rightarrow B_1C$ defined by ${\mathcal E}_1(\tau)=\sum_l|l\rangle\!\langle l|^{C}\otimes N_l\tau N_l^{\dagger}$. The state after this operation is $\Psi'=\sum_lp_l|l\rangle\!\langle l|^{C}\otimes|\Psi_l\rangle\!\langle\Psi_l|^{AA_0B_1R}$.
\item Bob transmits system $C$ to Alice.
\item  Alice performs a CPTP map ${\mathcal E}_2:CAA_0\rightarrow AA_{\!B}A_1$ defined as ${\mathcal E}_2(\tau)=\sum_l{\mathcal O}_l(\langle l|^C\tau^{CAA_0}|l\rangle^C)$. The state after the operation is ${\mathcal E}_2(\Psi')={\rho}({\mathcal N})$.
\end{enumerate}
By the chain rule and the data processing inequality, we have 
\begin{eqnarray}
&&2S(A)_\Psi+2\log{K}\nn\\
&=&I(AA_0:BB_0R)_{\Psi\otimes\Phi_K}\nn\\
&\geq&I(AA_0:B_1CR)_{\Psi'}\nn\\
&=&I(AA_0:C)_{\Psi'}+I(AA_0:B_1R|C)_{\Psi'}\nn\\
&\geq&I(AA_0C:B_1R)_{\Psi'}-I(C:B_1R)_{\Psi'}\nn\\
&\geq& I(AA_{\!B}A_1:B_1R)_{{\rho}({\mathcal N})}-I(C:B_1R)_{\Psi'}.\label{eq:kyuukabo}
\end{eqnarray}
Due to Inequality (\ref{eq:jirit}) and (\ref{eq:contmi}), we have
\begin{eqnarray}
&&I(AA_{\!B}A_1:B_1R)_{{\rho}({\mathcal N})}\nn\\
&\geq& I(AA_{\!B}A_1:B_1R)_{\Psi\otimes\Phi_L}-5\eta(2\sqrt{\epsilon})\log{(d_RL)}\nn\\
&=&I(AA_{\!B}:R)_{\Psi}+I(A_1:B_1)_{\Phi_L}-5\eta(2\sqrt{\epsilon})\log{(d_RL)}\nn\\
&=&2S(R )_{\Psi}+2\log{L}-5\eta(2\sqrt{\epsilon})\log{(d_RL)}.\label{eq:taromadogiwa}
\end{eqnarray}
From (\ref{eq:jirit}), (\ref{eq:324}), (\ref{eq:contmi}) and Lemma \ref{lmm:yoshiki}, we also have
\begin{eqnarray}
&&I(C:B_1R)_{\Psi'}\nn\\
&=&S(B_1R)_{\Psi'}-S(B_1R|C)_{\Psi'}\nn\\
&=&S(B_1R)_{{\rho}({\mathcal N})}-\sum_lp_lS(B_1R)_{{\hat\Psi}_l}\nn\\
&=&S(B_1R)_{{\rho}({\mathcal N})}-S(B_1R)_{\Psi\otimes\Phi_L}\nn\\
&&\;\;+\sum_lp_l\left(S(B_1R)_{\Psi\otimes\Phi_L}-S(B_1R)_{{\hat\Psi}_l}\right)\nn\\
&\leq&\sum_lp_l\left(\eta(\epsilon_l)+\eta(2\sqrt{\epsilon_l})\right)\log{(d_RL)}\nn\\
&\leq&2\sum_lp_l\eta(2\sqrt{\epsilon_l})\log{(d_RL)}\nn\\
&\leq&2\left(\eta\left(2\sqrt{\epsilon}\right)+\eta(2)\cdot\sqrt{2\sqrt{\epsilon}}\right)\log{(d_RL)}.\label{eq:ayataka}
\end{eqnarray}
From (\ref{eq:kyuukabo}), (\ref{eq:taromadogiwa}) and (\ref{eq:ayataka}), we obtain
 \begin{eqnarray}
&&\log{K}-\log{L}\nn\\
&\geq& S(R )_{\Psi}-S(A)_\Psi-\eta'(2\sqrt{\epsilon})\log{(d_RL)}\nn\\
&=&S(AB)_{\Psi}-S(A)_\Psi-\eta'(2\sqrt{\epsilon})\log{(d_RL)}\nn\\
&=&S(B|A)_{\Psi}-\eta'(2\sqrt{\epsilon})\log{(d_RL)}\nn
\end{eqnarray}
for the entanglement cost. As for the classical communication cost,  from (\ref{eq:jirit}) and (\ref{eq:contmi}), we have
\begin{eqnarray}
&&2S(R)_\Psi\nn\\
&=&I(AA_{\!B}:R)_\Psi\nn\\
&\leq& I(AA_{\!B}:R)_{{\rho}({\mathcal N})}+5\eta(2\sqrt{\epsilon})\log{d_R}\nn\\
&\leq&I(AA_0C:R)_{\Psi'}+5\eta(2\sqrt{\epsilon})\log{d_R}\nn\\
&=&I(AA_0:R)_{\Psi'}+I(C:R|AA_0)_{\Psi'}+5\eta(2\sqrt{\epsilon})\log{d_R}\nn\\
&=&I(AA_0:R)_{\Psi\otimes\Phi_K}+I(C:AA_0R)_{\Psi'}\nn\\
&&\;\;\;-I(C:AA_0)_{\Psi'}+5\eta(2\sqrt{\epsilon})\log{d_R}\nn\\
&\leq& I(A:R)_{\Psi}+S( C)_{\Psi'}+5\eta(2\sqrt{\epsilon})\!\log{d_R}\nn\\
&=&I(A:R)_{\Psi}+H(\{p_l\}_l)+5\eta(2\sqrt{\epsilon})\log{d_R}.\nn
\end{eqnarray}
 Here, the fifth line follows from the fact that Bob's measurement does not change the average reduced state of $AA_0R$. Thus we obtain
\begin{eqnarray}
C&\geq& H(\{p_l\}_l)\nn\\
&\geq& 2S(R )_\Psi-I(A:R)_{\Psi}-5\eta(2\sqrt{\epsilon})\log{d_R}\nn\\
&=&S(R )_\Psi+S(AR)_\Psi-S(A)_\Psi-5\eta(2\sqrt{\epsilon})\log{d_R}\nn\\
&=&S(R )_\Psi+S(B)_\Psi-S(BR)_\Psi-5\eta(2\sqrt{\epsilon})\log{d_R}\nn\\
&=&I(B:R)_\Psi-5\eta(2\sqrt{\epsilon})\log{d_R},\nn
\end{eqnarray}
which concludes the proof. \hfill$\blacksquare$\end{prf}

\section{Proof of Lemma \ref{lmm:decoupleifmarkov}, \ref{lmm:markovifdecouple} and \ref{lmm:decoupleisometry}}

\subsection{Proof of Lemma \ref{lmm:decoupleifmarkov}}\label{app:prfdecmar}

We prove that an $M$-induced map is $(3\varsigma+2\nu)$-decoupling between $A'R_A$ and $R_B$ if it is $\varsigma$-oblivious and $\nu$-Markovianizing from $R_ABR_B$, which implies Lemma \ref{lmm:decoupleifmarkov}. 

 Due to Equalities (\ref{eq:tdtensor}), (\ref{eq:fuwatororo}), (\ref{eq:notfound}) and (\ref{eq:sommisama}), we have
\begin{eqnarray}
&&\left\|\Phi_{M}^{R_A}-\pi_d^{R_A}\right\|_1\nn\\
&=&\left\|\Phi_{M}^{R_A}\otimes\Phi_d^{BR_B}-\pi_d^{R_A}\otimes\Phi_d^{BR_B}\right\|_1\nn\\
&=&\left\|{\hat U}^{R_AR_B}(\Phi_{M}^{R_A}\otimes\Phi_d^{BR_B}){\hat U}^{\dagger R_AR_B}\right.\nn\\
&&\;\;\left.-{\hat U}^{R_AR_B}(\pi_d^{R_A}\otimes\Phi_d^{BR_B}){\hat U}^{\dagger R_AR_B}\right\|_1\nn\\
&=&\left\|\Psi_{M}^{R_ABR_B}-\Psi^{R_ABR_B}\right\|_1.\nn
\end{eqnarray}
Thus the condition of $\varsigma$-obliviousness is equivalent to
\begin{eqnarray}
\left\|\Psi_{M}^{R_ABR_B}-\Psi^{R_ABR_B}\right\|_1\leq\varsigma,\nn
\end{eqnarray}
which implies
\begin{eqnarray}
\left\|\Psi_{M}^{R_AR_B}-\pi_d^{R_A}\otimes\pi_d^{R_B}\right\|_1\leq\varsigma\nn
\end{eqnarray}
and
\begin{eqnarray}
\left\|\Psi_{M}^{R_A}-\pi_d^{R_A}\right\|_1\leq\varsigma,\;\left\|\Psi_{M}^{R_B}-\pi_d^{R_B}\right\|_1\leq\varsigma,
\label{eq:mentalist}
\end{eqnarray}
 due to (\ref{eq:monottdd}). From (\ref{eq:triangle}) and (\ref{eq:tdtensor}), it follows that
\begin{eqnarray}
&&\left\|\Psi_{M}^{R_A}\otimes\Psi_{M}^{R_B}-\pi_d^{R_A}\otimes\pi_d^{R_B}\right\|_1\nn\\
&\leq&\left\|\Psi_{M}^{R_A}\otimes\Psi_{M}^{R_B}-\pi_d^{R_A}\otimes\Psi_{M}^{R_B}\right\|_1\nn\\
&&+\left\|\pi_d^{R_A}\otimes\Psi_{M}^{R_B}-\pi_d^{R_A}\otimes\pi_d^{R_B}\right\|_1\nn\\
&=&\left\|\Psi_{M}^{R_A}-\pi_d^{R_A}\right\|_1+\left\|\Psi_{M}^{R_B}-\pi_d^{R_B}\right\|_1\nn\\
&\leq&2\varsigma,\nn
\end{eqnarray}
and that
\begin{eqnarray}
&&\left\|\Psi_{M}^{R_AR_B}-\Psi_{M}^{R_A}\otimes\Psi_{M}^{R_B}\right\|_1\nn\\
&\leq&\left\|\Psi_{M}^{R_AR_B}-\pi_d^{R_A}\otimes\pi_d^{R_B}\right\|_1\nn\\
&&+\left\|\Psi_{M}^{R_A}\otimes\Psi_{M}^{R_B}-\pi_d^{R_A}\otimes\pi_d^{R_B}\right\|_1\nn\\
&\leq&3\varsigma.
\label{eq:atohappen}
\end{eqnarray}

Suppose $\Psi_{M}^{A'R_A(BR_B)}$ is $\nu$-recoverable from $R_ABR_B$. By definition, there exists a linear CPTP map ${\mathcal R}:R_A\rightarrow A'R_A$ such that
\begin{eqnarray}
\left\|\Psi_{M}^{A'R_A(BR_B)}-{\mathcal R}(\Psi_{M}^{R_ABR_B})\right\|_1\leq\nu,\nn
\end{eqnarray}
which implies
\begin{eqnarray}
\left\|\Psi_{M}^{A'R_A}-{\mathcal R}(\Psi_{M}^{R_A})\right\|_1\leq\left\|\Psi_{M}^{A'R_AR_B}-{\mathcal R}(\Psi_{M}^{R_AR_B})\right\|_1\leq\nu.\!\!\!\!\!\!\!\!\!\!\!\!\nn\\\label{eq:ayanolove}
\end{eqnarray}
 Due to (\ref{eq:monottd}), Inequality (\ref{eq:atohappen}) implies
\begin{eqnarray}
\left\|{\mathcal R}(\Psi_{M}^{R_AR_B})-{\mathcal R}(\Psi_{M}^{R_A})\otimes\Psi_{M}^{R_B}\right\|_1\leq3\varsigma.
\label{eq:orenidoushiroto}
\end{eqnarray}
From  (\ref{eq:triangle}), (\ref{eq:tdtensor}), (\ref{eq:ayanolove}) and (\ref{eq:orenidoushiroto}), we obtain
\beq
&&\left\|\Psi_{M}^{A'R_AR_B}-\Psi_{M}^{A'R_A}\otimes\Psi_{M}^{R_B}\right\|_1\nn\\
&\leq&\left\|\Psi_{M}^{A'R_AR_B}-{\mathcal R}(\Psi_{M}^{R_AR_B})\right\|_1\nn\\
&&+\left\|{\mathcal R}(\Psi_{M}^{R_AR_B})-{\mathcal R}(\Psi_{M}^{R_A})\otimes\Psi_{M}^{R_B}\right\|_1\nn\\
&&+\left\|{\mathcal R}(\Psi_{M}^{R_A})\otimes\Psi_{M}^{R_B}-\Psi_{M}^{A'R_A}\otimes\Psi_{M}^{R_B}\right\|_1\nn\\
&\leq&3\varsigma+2\nu,\nn
\eeq
which completes the proof.
\hfill$\blacksquare$

\subsection{Proof of Lemma \ref{lmm:markovifdecouple}}\label{app:prfmardec}
We prove that an $M$-induced map is $(\varsigma+\mu)$-Markovianizing  from $A'R_A$ if it is $\varsigma$-oblivious and $\mu$-decoupling between $A'R_A$ and $R_B$, which implies Lemma \ref{lmm:markovifdecouple}. Let $\Xi:R_AR_B\rightarrow R_ABR_B$ be a linear CPTP map defined by
\begin{eqnarray}
\Xi(\tau)=d^{2}\cdot(\Psi^{R_ABR_B})^{\frac{1}{2}}(\tau^{R_AR_B}\otimes I^{B})(\Psi^{R_ABR_B})^{\frac{1}{2}}.\nn
\end{eqnarray}
This is indeed CPTP since we have
\begin{eqnarray}
{\rm Tr}[\Xi(\tau)]&=&d^{2}\cdot{\rm Tr}[\tau^{R_AR_B}(\Psi^{R_ABR_B})]\nn\\
&=&d^{2}\cdot{\rm Tr}[\tau^{R_AR_B}(\pi_d^{R_A}\otimes\pi_d^{R_B})]\nn\\
&=&{\rm Tr}[\tau].\nn
\end{eqnarray}
Using the relation
\begin{eqnarray}
&&(\Psi^{R_ABR_B})^{\frac{1}{2}}\nn\\
&=&\left(U^{*R_AR_B}(\pi_d^{R_A}\otimes|\Phi_d\rangle\!\langle\Phi_d|^{BR_B})U^{tR_AR_B}\right)^{\frac{1}{2}}\nn\\
&=&\frac{1}{\sqrt{d}}\:U^{*R_AR_B}(I^{R_A}\otimes|\Phi_d\rangle\!\langle\Phi_d|^{BR_B})U^{tR_AR_B},\nn
\end{eqnarray}
we have
\begin{eqnarray}
\Xi^{R_AR_B}(\Psi^{AR_AR_B})={\Psi}^{AR_ABR_B},\nn
\end{eqnarray}
which leads to
\begin{eqnarray}
{\Psi}_{M}^{A'R_ABR_B}=\Xi^{R_AR_B}(\Psi_{M}^{A'R_AR_B}).\nn
\end{eqnarray}
It is straightforward to verify that a map ${\mathcal R}':R_A\rightarrow R_ABR_B$ defined by
\begin{eqnarray}
{\mathcal R}'(\tau)=\Xi(\tau^{R_A}\otimes\pi_d^{R_B})\;\;(\forall\tau\in{\mathcal S}({\mathcal H}^{R_A}))\nn
\end{eqnarray}
is CPTP as well. Therefore,  from (\ref{eq:monottd}), (\ref{eq:triangle}) and (\ref{eq:tdtensor}), we have
\begin{eqnarray}
&&\left\|\Psi_{M}^{A'R_ABR_B}-{\mathcal R}'({\Psi}_{M}^{A'R_A})\right\|_1\nn\\
&=&\left\|\Xi^{R_AR_B}(\Psi_M^{A'R_AR_B})-\Xi^{R_AR_B}(\Psi_{M}^{A'R_A}\otimes\pi_d^{R_B})\right\|_1\nn\\
&\leq&\left\|\Psi_{M}^{A'R_AR_B}-\Psi_{M}^{A'R_A}\otimes\pi_d^{R_B}\right\|_1\nn\\
&\leq&\left\|\Psi_{M}^{A'R_AR_B}-\Psi_{M}^{A'R_A}\otimes\Psi_{M}^{R_B}\right\|_1\nn\\
&&+\left\|\Psi_{M}^{A'R_A}\otimes\Psi_{M}^{R_B}-\Psi_{M}^{A'R_A}\otimes\pi_d^{R_B}\right\|_1\nn\\
&=&\left\|\Psi_{M}^{A'R_AR_B}-\Psi_{M}^{A'R_A}\otimes\Psi_{M}^{R_B}\right\|_1+\left\|\Psi_{M}^{R_B}-\pi_d^{R_B}\right\|_1\nn\\
&\leq&\varsigma+\mu,\nn
\end{eqnarray}
where the last line follows from the assumption and (\ref{eq:mentalist}). 
\hfill$\blacksquare$

\subsection{Proof of Lemma \ref{lmm:decoupleisometry}}\label{app:decoupleisometry}
Suppose that an $M_k$-induced map is $\mu_k$-decoupling between $A'R_A$ and $R_B$ for each $k\in{\mathbb K}$, and that $\sum_{k\in{\mathbb K}}p_k\mu_k\leq\mu$. Due to (\ref{eq:dfnedec}) and  Equality (\ref{eq:cooppuree}), there exist pure states $|\Psi_k^p\rangle^{A'R_A{\tilde B}}$, $|\Psi_k^q\rangle^{BR_B}$ and isometries $W_k^{BB_0\rightarrow B{\tilde B}}\:(k\in{\mathbb K})$ such that
\begin{eqnarray}
&&\left\|\Psi_{k}'-(\Psi^p_k)^{A'R_A{\tilde B}}\otimes(\Psi^q_k)^{BR_B}\right\|_1\leq2\sqrt{\mu_k},\nn\\
&&(\Psi^p_k)^{A'R_A}=\Psi_{M_k}^{A'R_A},\;(\Psi^q_k)^{R_B}=\Psi_{M_k}^{R_B},\label{eq:csd}
\end{eqnarray}
where $|\Psi_{k}'\rangle:=W_k|\Psi_{M_k}\rangle$. Suppose in addition that $\mathbb M$ is $0$-oblivious. Then we have
\begin{align}
(\Psi^p_k)^{R_A}=\pi_d^{R_A},\;(\Psi^q_k)^{R_B}=\pi_d^{R_B}\label{eq:pqkindd}
\end{align}
for each $k$, due to (\ref{eq:mentalist})  and (\ref{eq:csd}). The latter of (\ref{eq:pqkindd}) implies we can choose $|\Psi_k^q\rangle^{BR_B}=|\Phi_d\rangle^{BR_B}$ by an appropriate choice of $W_k$, since all purifications are  local-isometry equivalent. If $\Psi_{M_k}^{A'R_AR_B}$ does not depend on $k$, neither does $\Psi_{M_k}^{A'R_A}$. Thus the $k$-dependence of $|\Psi_{k}'\rangle$ and $|\Psi_k^p\rangle$ can be dropped by an appropriate choice of $W_k$ for the same reason. Hence we obtain
\begin{eqnarray}
\left\||\Psi'\rangle\!\langle\Psi'|-(\Psi^p)^{A'R_A{\tilde B}}\otimes\Phi_d^{BR_B}\right\|_1\leq2\sqrt{\mu_k},\nn
\end{eqnarray}
for any $k\in{\mathbb K}$, which leads to
\begin{eqnarray}
\left\||\Psi'\rangle\!\langle\Psi'|-(\Psi^p)^{A'R_A{\tilde B}}\otimes\Phi_d^{BR_B}\right\|_1\leq2\sqrt{\mu}\label{eq:invisible}
\end{eqnarray}
and
\begin{eqnarray}
\left\|(\Psi')^{{\tilde B}B}-(\Psi^p)^{\tilde B}\otimes\pi_d^{B}\right\|_1\leq2\sqrt{\mu}\label{eq:ssss}
\end{eqnarray}
 due to (\ref{eq:monottdd}).

Let $\{\lambda_i^{\downarrow}\}_{i=1}^{{\rm dim}{\tilde B}}$ be the eigenvalues of $(\Psi^p)^{\tilde B}$ sorted in decreasing order, let
\begin{align}
(\Psi^p)^{\tilde B}=\sum_{i=1}^{{\rm dim}{\tilde B}}\lambda_i^{\downarrow}\proj{i}\nn
\end{align}
be the eigen decomposition of $(\Psi^p)^{\tilde B}$, and define  a linear operator $\tilde\Pi$ on ${\mathcal H}^{\tilde B}$ by
\begin{align}
\Pi:=\sum_{i=1}^{{\rm dim}B_0}\proj{i}.\nn
\end{align}
Due to Lemma \ref{lmm:3f} and (\ref{eq:ssss}), we have
\begin{align}
{\rm Tr}[{\tilde\Pi}(\Psi^p)^{\tilde B}]={\rm Tr}[({\tilde\Pi}^{\tilde B}\otimes I^B)((\Psi^p)^{\tilde B}\otimes\pi_d^{B})]\geq1-2\sqrt{\mu}.\label{eq:strawberry}
\end{align}
Note that, by definition, we have
\begin{align}
{\rm rank}[(\Psi')^{{\tilde B}B}]\leq{\rm dim}B_0\times{\rm dim}B.\nn
\end{align}
Define
\begin{align}
|{\tilde\Psi}^p\rangle^{A'R_A{\tilde B}}:=\frac{{\tilde\Pi}^{\tilde B}|\Psi^p\rangle}{\|{\tilde\Pi}^{\tilde B}|\Psi^p\rangle\|}.\nn
\end{align}
From (\ref{eq:strawberry}) and the gentle measurement lemma, we obtain
\begin{eqnarray}
\left\||{\tilde\Psi}^p\rangle\!\langle{\tilde\Psi}^p|-\proj{\Psi^p}\right\|_1\leq2\sqrt{2\sqrt{\mu}}.\label{eq:F}
\end{eqnarray}
 From (\ref{eq:triangle}), (\ref{eq:tdtensor}), (\ref{eq:invisible}), (\ref{eq:F}) and $\mu\in(0,1]$, we see that
\begin{eqnarray}
&&\left\||\Psi'\rangle\!\langle\Psi'|-({\tilde\Psi}^p)^{A'R_A{\tilde B}}\otimes\Phi_d^{BR_B}\right\|_1\nn\\
&\leq&\left\||\Psi'\rangle\!\langle\Psi'|-(\Psi^p)^{A'R_A{\tilde B}}\otimes\Phi_d^{BR_B}\right\|_1\nn\\
&&+\left\|(\Psi^p)^{A'R_A{\tilde B}}\otimes\Phi_d^{BR_B}-({\tilde\Psi}^p)^{A'R_A{\tilde B}}\otimes\Phi_d^{BR_B}\right\|_1\nn\\
&\leq&2\sqrt{\mu}+2\sqrt{2\sqrt{\mu}}\:\leq\:2(1+\sqrt{2})\sqrt[4]{\mu}\nn\\
&\leq&5\sqrt[4]{\mu},\nn
\end{eqnarray}
which implies (\ref{eq:ohlonesomeme}). From (\ref{eq:pqkindd}), $\Psi_k^p=\Psi^p$, (\ref{eq:F}) and (\ref{eq:monottdd}), we also have
\begin{align}
\left\|({\tilde\Psi}^p)^{R_A}-\pi_d^{R_A}\right\|_1\leq2\sqrt{2\sqrt{\mu}}\leq3\sqrt[4]{\mu}.\nn
\end{align}
\hfill$\blacksquare$

\section{Proofs of Lemma \ref{lmm:markovifepsilon} and \ref{thm:abcaabcc}}

\subsection{Settings}\label{app:setting}
As we described in Section \ref{sec:singletworound}, we assume without loss of generality that the protocol $\mathcal M$ proceeds as follows. (See also Remark  at the end of this appendix.)
\begin{itemize}\setlength{\leftskip}{0.1cm}
\item[I-1.] Alice performs a measurement $\{{ M}_{k}^{AA_0\rightarrow A'}\}_{k\in{\mathbb K}}$. The probability of obtaining measurement outcome $k$ is given by $p_{k}=\|{ M}_{k}|\Psi\rangle|\phi_{\rm res}\rangle\|_1^2$, and the state after the measurement is $|\Psi_{k}\rangle=p_{k}^{-1/2}{ M}_{k}|\Psi\rangle|\phi_{\rm res}\rangle$.
\item[I-2.] Alice communicates the measurement outcome $k$ to Bob.
\item[I-3.] Bob performs a measurement $\{{ N}_{l|k}^{BB_0\rightarrow BB_1}\}_{l}$. The probability of obtaining measurement outcome $l$, conditioned by $k$, is given by $p_{l|k}=\|{ N}_{l|k}|\Psi_{k}\rangle\|_1^2$ and the state after Bob's measurement is $|\Psi_{kl}\rangle=p_{l|k}^{-1/2}{ N}_{l|k}|\Psi_{k}\rangle$.
\item[I-4.] Bob communicates the measurement outcome $l$ to Alice.
\item[I-5.] Alice performs an operation which is described by a CPTP map ${\mathcal O}_{kl}:A'\rightarrow AA_1$. The final state is given by ${\hat\Psi}_{kl}={\mathcal O}_{kl}(\Psi_{kl})$. 
\end{itemize}

\begin{figure}[t]
\begin{center}
\includegraphics[bb={0 0 541 963}, scale=0.35]{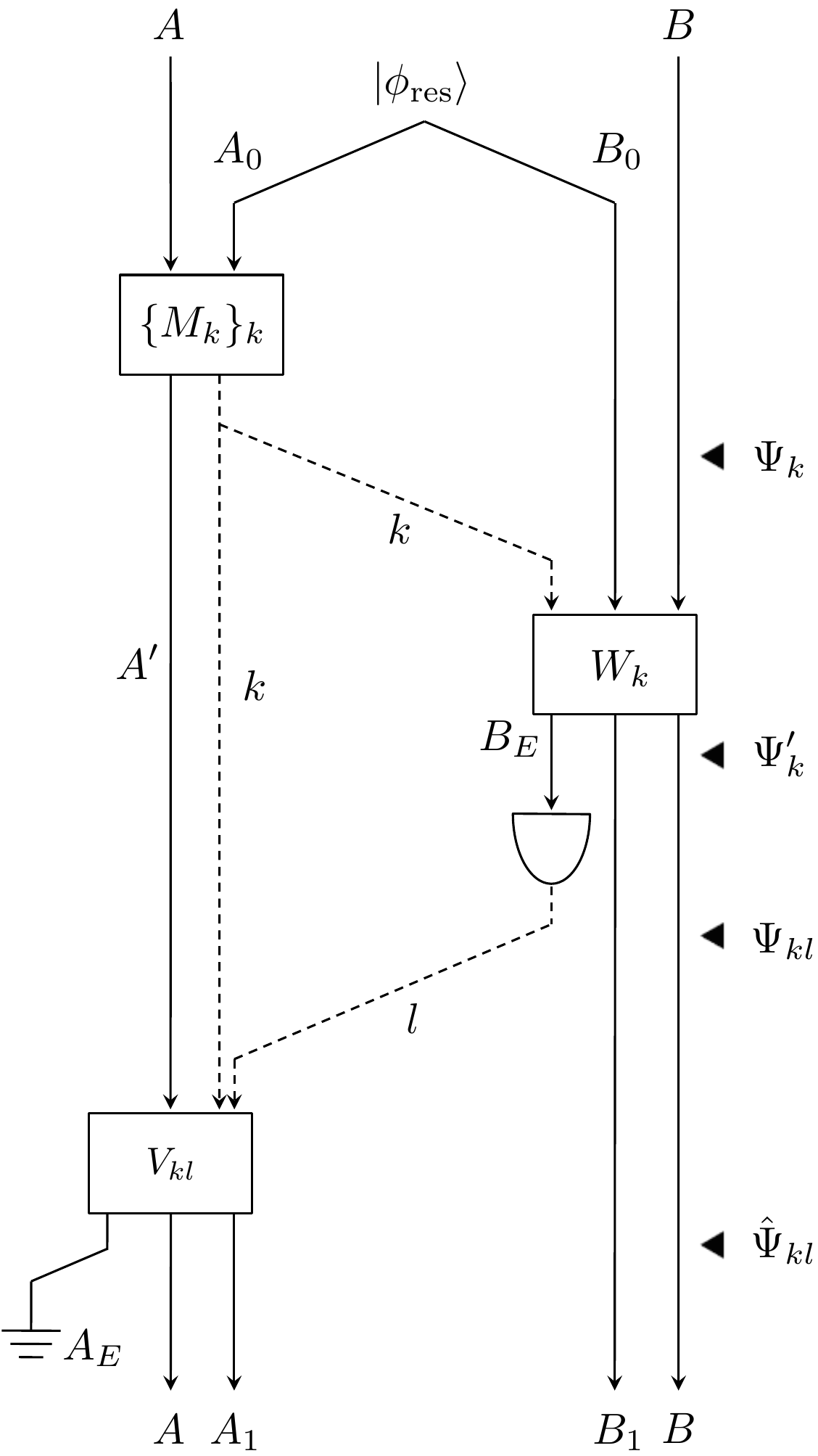}
\end{center}\caption{A graphical representation of Procedure I\!I-1$\sim$7 in Appendix \ref{app:setting}. The parabolic symbol in the middle represents the projective measurement in the basis $\{|l\rangle\}_l$. }
\label{fig:singleshotformm}
\end{figure}

Let $A_E$ and $B_E$ be  an ancillary system of Alice and Bob, respectively. Let $W_{k}:BB_0\rightarrow BB_1B_E\;({k\in{\mathbb K}})$ be isometries such that the Naimark extension of Bob's measurement is given by ${N}_{l|k}=\bra{l}^{B_E}W_{k}$, and let ${V}_{kl}:A'\rightarrow AA_1A_E$ be an isometry such that the Stinespring dilation of ${\mathcal O}_{kl}$ is given by ${\mathcal O}_{kl}(\tau)={\rm Tr}_{A_E}[{V}_{kl}\tau^{A'}{V}_{kl}^{\dagger}]$. Consider the following protocol, which is equivalent to the protocol given by I-1$\sim$5 (Figure \ref{fig:singleshotformm}).
\begin{itemize}\setlength{\leftskip}{0.1cm}
\item[I\!I-1.] Alice performs a measurement $\mathbb M$ and obtains measurement outcome $k$. The state after the measurement is $|\Psi_{k}\rangle^{A'R_ABR_BB_0}$.
\item[I\!I-2.] Alice communicates $k$ to Bob.
\item[I\!I-3.] Bob performs $W_{k}$. The state becomes
\beq
\ket{\Psi_{k}'}^{A'R_ABR_BB_1B_E}:=W_{k}\ket{\Psi_{k}}.\label{eq:prpr}
\eeq
\item[I\!I-4.] Bob performs a projective measurement on $B_E$ in the basis $\{|l\rangle\}_{l}$, and obtains outcome $l$ with probability $p_{l|k}$. The state after the measurement is 
\begin{eqnarray}
\ket{\Psi_{kl}}^{A'R_ABR_BB_1}:=p_{l|k}^{-1/2}\bra{l}^{B_E}\ket{\Psi_{k}'}.\label{eq:psikl}
\end{eqnarray}
\item[I\!I-5.] Bob communicates $l$ to Alice.
\item[I\!I-6.] Alice performs ${V}_{kl}$. The state becomes
\beq
|{\hat\Psi}_{kl}\rangle^{AA_1A_ER_ABR_BB_1}:={V}_{kl}\ket{\Psi_{kl}}.\label{eq:psikl2}
\eeq
\item[I\!I-7.] Alice discards $A_E$.
\end{itemize}

\begin{rmk}
In the description of $\mathcal M$ by I-1$\sim$5, we assume that all information about the outcome of Alice's measurement, represented by $k\in{\mathbb K}$, is communicated to Bob. In  a general protocol, however, not all information about the measurement outcome need to be communicated. In such cases, the measurement outcomes are represented as $(k_1,k_2)\in{\mathbb K}_1\times{\mathbb K}_2$ by two countable sets ${\mathbb K}_1$ and ${\mathbb K}_2$. The ${\mathbb K}_1$ part of the outcome is communicated to Bob, whereas the ${\mathbb K}_2$ part is kept on Alice's register until she performs the last operation. We show that such protocols can also be described by II-1$\sim$7 as follows. Let ${\tilde V}:A\rightarrow A'A_{E_1}A_{E_2}$ be an isometry such that the Naimark extension of Alice's measurement is given by $M_{k_1k_2}=\langle k_1|^{A_{E_1}}\langle k_2|^{A_{E_2}}{\tilde V}$,  and let $V_{ k_1k_2l}:A'\rightarrow AA_1A_E$ be an isometry such that the Stinespring delation of Alice's last operation is given by ${\mathcal O}_{k_1k_2l}(\tau)={\rm Tr}_{A_E}[{V}_{k_1k_2l}\tau^{A'}{V}_{k_1k_2l}^{\dagger}]$. The procedure II-1$\sim$7 then gives a description of the general protocol by the following correspondence:
\begin{eqnarray}
k&\rightarrow&k_1\nn\\
A'&\rightarrow&A'A_{E_2}\nn\\
M_k^{AA_0\rightarrow A'}&\rightarrow&M_{k_1}^{AA_0\rightarrow A'A_{E_2}}=\langle k_1|^{A_{E_1}}{\tilde V}\nn\\
V_{kl}&\rightarrow&{V_{k_1l}=\sum_{k_2\in{\mathbb K}_2}|k_2\rangle\!\langle k_2|^{A_{E_2}}\otimes V_{k_1k_2l}}\nn\\
A_E&\rightarrow&A_{E_2}A_E.\nn
\end{eqnarray}
\end{rmk}

\subsection{Proof of Lemma \ref{lmm:markovifepsilon}}\label{app:prfmarkovifepsilon}
We prove that the measurement $\mathbb M$ is $4\sqrt[4]{\epsilon}$-oblivious and $8\sqrt[4]{\epsilon}$-decoupling between $A'R_A$ and $R_B$, which implies Lemma \ref{lmm:markovifepsilon} combined with Lemma \ref{lmm:markovifdecouple}. From (\ref{eq:1shoterror2}) and Equality (\ref{eq:fidpp}), we have
\begin{eqnarray}
&&\!\!\!\!\!\!1-\epsilon\leq\sum_{kl}p_{kl}F\left({\hat\Psi}_{kl}^{ABR_AR_B},\ket{\Phi_d}^{AR_A}\ket{\Phi_d}^{BR_B}\right)\nn\\
&&\!\!\!\!\!\!=\sum_{kl}p_{kl}F\left(|{\hat\Psi}_{kl}\rangle,\ket{\Phi_d}^{AR_A}\ket{\Phi_d}^{BR_B}\ket{\phi_{kl}}^{A_1B_1A_E}\right)\nn
\end{eqnarray}
for some states $\phi_{kl}$, which leads to
\begin{eqnarray}
\sum_{kl}p_{kl}\epsilon_{kl}\leq\epsilon\nn
\end{eqnarray}
 for
\begin{eqnarray}
\epsilon_{kl}:=1-F\left(|{\hat\Psi}_{kl}\rangle,\ket{\Phi_d}^{AR_A}\ket{\Phi_d}^{BR_B}\ket{\phi_{kl}}^{A_1B_1A_E}\right).\label{eq:defepskl}
\end{eqnarray}
Due to Lemma \ref{lmm:yoshiki}  and $\epsilon_{kl}\in[0,1]$, we have
\begin{eqnarray}
\sum_{kl}p_{kl}\sqrt{\epsilon_{kl}}\leq\sqrt[4]{\epsilon}+\sqrt{\epsilon}\leq2\sqrt[4]{\epsilon}.\label{eq:yakushoku}
\end{eqnarray}
Therefore,  by using (\ref{eq:tracefid}), (\ref{eq:monottdd}) and (\ref{eq:convTD}), we obtain
\begin{eqnarray}
	&&\!\!\!\!\!\!\!\!\!4\sqrt[4]{\epsilon}\geq2\sum_{kl}p_{kl}\sqrt{\epsilon_{kl}}\nn\\
	&&\!\!\!\!\!\!\geq\sum_{kl}p_{kl}\left\||{\hat\Psi}_{kl}\rangle\!\langle{\hat\Psi}_{kl}|-{\Phi_d}^{AR_A}\otimes\Phi_d^{BR_B}\otimes{\phi_{kl}}^{A_1B_1A_E}\right\|_1\nn\\
	&&\!\!\!\!\!\!\geq\sum_{kl}p_{kl}\left\|{\hat\Psi}_{kl}^{AR_AR_BA_1A_E}-{\Phi_d}^{AR_A}\otimes\pi_d^{R_B}\otimes{\phi_{kl}}^{A_1A_E}\right\|_1\nn\\
	&&\!\!\!\!\!\!=\sum_{kl}p_{kl}\left\|\Psi_{kl}^{A'R_AR_B}-{V}_{kl}^{\dagger}(\Phi_d^{AR_A}\otimes\phi_{kl}^{A_1A_E}){V}_{kl}\otimes\pi_d^{R_B}\right\|_1\nn\\
	&&\!\!\!\!\!\!\geq\sum_{k}p_{k}\left\|\sum_{l}p_{l|k}\Psi_{kl}^{A'R_AR_B}-\psi_{k}^{A'R_A}\otimes\pi_d^{R_B}\right\|_1\nn\\
	&&\!\!\!\!\!\!=\sum_{k}p_{k}\left\|\Psi_{k}'^{A'R_AR_B}-\psi_{k}^{A'R_A}\otimes\pi_d^{R_B}\right\|_1\nn\\
	&&\!\!\!\!\!\!=\sum_{k}p_{k}\left\|\Psi_{k}^{A'R_AR_B}-\psi_{k}^{A'R_A}\otimes\pi_d^{R_B}\right\|_1,
\label{eq:bbyyy}
\end{eqnarray}
where we defined
\begin{align}
\psi_{k}^{A'R_A}:=\sum_{l}p_{l|k}{V}_{kl}^{\dagger}(\Phi_d^{AR_A}\otimes\phi_{kl}^{A_1A_E}){V}_{kl}.\label{eq:IPOD}
\end{align}
Hence,  from (\ref{eq:triangle}), (\ref{eq:tdtensor}) and (\ref{eq:monottd}), we obtain
\begin{eqnarray}
&&\sum_{k}p_{k}\left\|\Psi_{k}^{A'R_AR_B}-\Psi_{k}^{A'R_A}\otimes\Psi_k^{R_B}\right\|_1\nn\\
&&\!\!\!\!\leq\sum_{k}p_{k}\left\|\Psi_{k}^{A'R_AR_B}-\psi_{k}^{A'R_A}\otimes\pi_d^{R_B}\right\|_1\nn\\
&&+\sum_{k}p_{k}\left\|\psi_{k}^{A'R_A}\otimes\pi_d^{R_B}-\Psi_{k}^{A'R_A}\otimes\pi_d^{R_B}\right\|_1\nn\\
&&+\sum_{k}p_{k}\left\|\Psi_{k}^{A'R_A}\otimes\pi_d^{R_B}-\Psi_{k}^{A'R_A}\otimes\Psi_k^{R_B}\right\|_1\nn\\
&&\!\!\!\!=\sum_{k}p_{k}\left\|\Psi_{k}^{A'R_AR_B}-\psi_{k}^{A'R_A}\otimes\pi_d^{R_B}\right\|_1\nn\\
&&+\sum_{k}p_{k}\left\|\psi_{k}^{A'R_A}-\Psi_{k}^{A'R_A}\right\|_1+\sum_{k}p_{k}\left\|\pi_d^{R_B}-\Psi_k^{R_B}\right\|_1\nn\\
&&\!\!\!\!\leq3\sum_{k}p_{k}\left\|\Psi_{k}^{A'R_AR_B}-\psi_{k}^{A'R_A}\otimes\pi_d^{R_B}\right\|_1\leq12\sqrt[4]{\epsilon},\nn
\end{eqnarray}
Thus Alice's measurement is $12\sqrt[4]{\epsilon}$-decoupling between $A'R_A$ and $R_B$. From (\ref{eq:bbyyy}), (\ref{eq:monottdd}), (\ref{eq:IPOD}), (\ref{eq:fuwatororo}), (\ref{eq:notfound}), (\ref{eq:hatupipi}) and (\ref{eq:tdtensor}), we also have
\begin{eqnarray}
4\sqrt[4]{\epsilon}&\geq&\sum_{k}p_{k}\left\|\Psi_{k}^{R_AR_B}-\psi_{k}^{R_A}\otimes\pi_d^{R_B}\right\|_1\nn\\
		&=&\sum_{k}p_{k}\left\|\Psi_{k}^{R_AR_B}-\pi_d^{R_A}\otimes\pi_d^{R_B}\right\|_1\nn\\
		&=&\sum_{k}p_{k}\left\|{\hat U}^{R_AR_B}(\Phi_{M_k}^{R_A}\otimes\pi_d^{R_B}){\hat U}^{\dagger R_AR_B}\right.\nn\\
		&&\;\;\;\;\;\;\;\;\;\;\;\;\left.-{\hat U}^{R_AR_B}(\pi_d^{R_A}\otimes\pi_d^{R_B}){\hat U}^{\dagger R_AR_B}\right\|_1\nn\\
		&=&\sum_{k}p_{k}\left\|\Phi_{M_k}^{R_A}\otimes\pi_d^{R_B}-\pi_d^{R_A}\otimes\pi_d^{R_B}\right\|_1\nn\\
		&=&\sum_{k}p_{k}\left\|\Phi_{M_k}^{R_A}-\pi_d^{R_A}\right\|_1,\nn
\end{eqnarray}
which implies that Alice's measurement is $4\sqrt[4]{\epsilon}$-oblivious.
\hfill$\blacksquare$

\subsection{Proof of Lemma \ref{thm:abcaabcc}}\label{app:prfoblivi}
 From (\ref{eq:defepskl}), (\ref{eq:fidmonPT}) and (\ref{eq:psikl2}), we have
\begin{eqnarray}
	1-\epsilon_{kl}\leq F({\hat\Psi}_{kl}^{BR_B},\Phi_d^{BR_B})=F({\Psi}_{kl}^{BR_B},\Phi_d^{BR_B}).\nn
\end{eqnarray}
By using (\ref{eq:fidpens}), we have
\begin{eqnarray}
1-\sum_{l}p_{l|k}\epsilon_{kl}&\leq&F\left({\scriptstyle \sum_{l}}p_{l|k}\Psi_{kl}^{BR_B},\Phi_d^{BR_B}\right)\nn\\
	&=&F(\Psi_{k}'^{BR_B},\Phi_d^{BR_B})\nn
\end{eqnarray}
because of (\ref{eq:psikl}).  Due to Equality (\ref{eq:fidpp}), there exist pure states $|\Psi_{k}^p\rangle^{A'R_AB_1B_E}\;(k\in{\mathbb K})$ such that
\begin{eqnarray}
1-\sum_{l}p_{l|k}\epsilon_{kl}\leq F(|\Psi_{k}'\rangle,(\Psi_{k}^p)^{A'R_AB_1B_E}\otimes\Phi_d^{BR_B})\nn
\end{eqnarray}
for each $k$, which leads to
\begin{eqnarray}
2\sqrt{\sum_{l}p_{l|k}\epsilon_{kl}}\geq\left\|\Psi_{k}'-(\Psi_{k}^p)^{A'R_AB_1B_E}\otimes\Phi_d^{BR_B}\right\|_1\nn
\end{eqnarray}
 due to (\ref{eq:tracefid}). Thus we have
\begin{eqnarray}
&&\sum_kp_k\left\|\Psi_{k}'-(\Psi_{k}^p)^{A'R_AB_1B_E}\otimes\Phi_d^{BR_B}\right\|_1\nn\\
&&\leq2\sum_kp_k\sqrt{\sum_{l}p_{l|k}\epsilon_{kl}}\leq2\sum_{kl}p_{kl}\sqrt{\epsilon_{kl}}\leq4\sqrt[4]{\epsilon},\nn
\end{eqnarray}
where the last line follows from the concavity of the square root function and Inequality (\ref{eq:yakushoku}). This completes the proof of Lemma \ref{thm:abcaabcc}. \hfill$\blacksquare$

\section{Proof of Inequality (\ref{eq:toterror})}\label{app:evaerror}

In this Appendix, we describe an evaluation of the total error, which have appeared in Section \ref{sec:direct} in the proof of the direct part of Theorem  \ref{thm:misoptach}. 

From Lemma \ref{lmm:decoupleisometryyy}, we have 
\begin{align}
\left\||\Psi'\rangle\!\langle\Psi'|-({\tilde\Psi}^p)^{{\bar A}{\bar R}_A{\tilde B}}\otimes(\Phi_d^{\otimes n})^{{\bar B}{\bar R}_B}\right\|_1\leq5\sqrt[4]{2\epsilon}\label{eq:together}
\end{align}
and
\begin{align}
\!\!\left\|({\tilde\Psi}^p)^{{\bar R}_A}-(\pi_d^{\otimes n})^{{\bar R}_A}\right\|_1\leq3\sqrt[4]{2\epsilon},\label{eq:myworld}
\end{align}
corresponding to (\ref{eq:nene}) and (\ref{eq:jean}), respectively. Let $A_{\!B'}$ be Alice's register which is identical to $B'$, and ${\mathcal N}:A'B'\rightarrow A'A_{\!B'}$ be state merging of $|{\check\Psi}^p\rangle^{A'\bar{R}_AB'}$. Define the merging error $\epsilon_{\rm merg}$ by
\begin{align}
\epsilon_{\rm merg}:=\left\|{\mathcal N}(({\check\Psi}^p)^{A'\bar{R}_AB'})-({\check\Psi}^p)^{A'\bar{R}_AA_{\!B'}}\right\|_1.\label{eq:defmergerr}
\end{align}

From (\ref{eq:together}) and (\ref{eq:defpsicheck}), we have
\begin{align}
\left\||\Psi'\rangle\!\langle\Psi'|\otimes\Phi_{2^{nr}}^{{\tilde A}_0{\tilde B}_0}-({\check\Psi}^p)^{A'{\bar R}_AB'}\otimes(\Phi_d^{\otimes n})^{{\bar B}{\bar R}_B}\right\|_1\leq5\sqrt[4]{2\epsilon},\nn
\end{align}
which leads to
\begin{eqnarray}
&&\!\!\!\!\!\!\!\!\left\|{\mathcal N}\left(|\Psi'\rangle\!\langle\Psi'|\otimes\Phi_{2^{nr}}^{{\tilde A}_0{\tilde B}_0}\right)-\right.\nn\\
&&\left.{\mathcal N}(({\check\Psi}^p)^{A'{\bar R}_AB'})\otimes(\Phi_d^{\otimes n})^{{\bar B}{\bar R}_B}\right\|_1\leq5\sqrt[4]{2\epsilon},\label{eq:forgotten}
\end{eqnarray}
By definition, we have $({\tilde\Psi}^p)^{{\bar R}_A}=({\check\Psi}^p)^{{\bar R}_A}$. Therefore, due to Inequality (\ref{eq:myworld}) and Lemma \ref{lmm:uhlmann}, there exists a quantum operation ${\mathcal O}:A'A_{\!B'}\rightarrow{\bar A}$ such that
\begin{align}
\left\|{\mathcal O}^{A'A_{\!B'}}(({\check\Psi}^p)^{A'\bar{R}_AA_{\!B'}})-(\Phi_d^{\otimes n})^{{\bar A}{\bar R}_A}\right\|_1\leq2\sqrt{3}\sqrt[8]{2\epsilon}.\label{eq:takemeaway}
\end{align}
Define  a quantum operation ${\mathcal N}':A'B'\rightarrow{\bar A}$ by ${\mathcal N}':={\mathcal O}\circ{\mathcal N}$. From (\ref{eq:defmergerr}), (\ref{eq:forgotten})  and (\ref{eq:monottd}), we obtain
\begin{eqnarray}
&&\!\!\!\!\!\!\!\!\!\!\!\!\!\!\!\!\left\|{\mathcal N}'(({\check\Psi}^p)^{A'\bar{R}_AB'})-{\mathcal O}^{A'A_{\!B'}}(({\check\Psi}^p)^{A'\bar{R}_AA_{\!B'}})\right\|_1\leq\epsilon_{\rm merg},\nn\\
&&\label{eq:defmergerr2}\\
&&\!\!\!\!\!\!\!\!\!\!\!\!\!\!\!\!\left\|{\mathcal N}'\left(|\Psi'\rangle\!\langle\Psi'|\otimes\Phi_{2^{nr}}^{{\tilde A}_0{\tilde B}_0}\right)-\right.\nn\\
&&\left.{\mathcal N}'(({\check\Psi}^p)^{A'{\bar R}_AB'})\otimes(\Phi_d^{\otimes n})^{{\bar B}{\bar R}_B}\right\|_1\leq5\sqrt[4]{2\epsilon}.\label{eq:forgotten2}
\end{eqnarray}
From  (\ref{eq:triangle}), (\ref{eq:tdtensor}), (\ref{eq:takemeaway}), (\ref{eq:defmergerr2}) and (\ref{eq:forgotten2}), we see that
\begin{eqnarray}
\!\!\!\!\!\!\!\!\!\!\!\!\!\!\!&&\left\|{\mathcal N}'\left(|\Psi'\rangle\!\langle\Psi'|\otimes\Phi_{2^{nr}}^{{\tilde A}_0{\tilde B}_0}\right)-(\Phi_d^{\otimes n})^{{\bar A}{\bar R}_A}\otimes(\Phi_d^{\otimes n})^{{\bar B}{\bar R}_B}\right\|\nn\\
&\leq&\left\|{\mathcal N}'\left(|\Psi'\rangle\!\langle\Psi'|\otimes\Phi_{2^{nr}}^{{\tilde A}_0{\tilde B}_0}\right)\right.\nn\\
&&\qquad\qquad\left.-{\mathcal N}'(({\check\Psi}^p)^{A'{\bar R}_AB'})\otimes(\Phi_d^{\otimes n})^{{\bar B}{\bar R}_B}\right\|_1\nn\\
&&+\left\|{\mathcal N}'(({\check\Psi}^p)^{A'\bar{R}_AB'})\otimes(\Phi_d^{\otimes n})^{{\bar B}{\bar R}_B}\right.\nn\\
&&\qquad\qquad\left.-{\mathcal O}^{A'A_{\!B'}}(({\check\Psi}^p)^{A'\bar{R}_AA_{\!B'}})\otimes(\Phi_d^{\otimes n})^{{\bar B}{\bar R}_B}\right\|_1\nn\\
&&+\left\|{\mathcal O}^{A'A_{\!B'}}(({\check\Psi}^p)^{A'\bar{R}_AA_{\!B'}})\otimes(\Phi_d^{\otimes n})^{{\bar B}{\bar R}_B}\right.\nn\\
&&\qquad\qquad\left.-(\Phi_d^{\otimes n})^{{\bar A}{\bar R}_A}\otimes(\Phi_d^{\otimes n})^{{\bar B}{\bar R}_B}\right\|_1\nn\\
&\leq&\left\|{\mathcal N}'\left(|\Psi'\rangle\!\langle\Psi'|\otimes\Phi_{2^{nr}}^{{\tilde A}_0{\tilde B}_0}\right)-\right.\nn\\
&&\qquad\qquad\left.{\mathcal N}'(({\check\Psi}^p)^{A'{\bar R}_AB'})\otimes(\Phi_d^{\otimes n})^{{\bar B}{\bar R}_B}\right\|_1\nn\\
&&+\left\|{\mathcal N}'(({\check\Psi}^p)^{A'\bar{R}_AB'})-{\mathcal O}^{A'A_{\!B'}}(({\check\Psi}^p)^{A'\bar{R}_AA_{\!B'}})\right\|_1\nn\\
&&+\left\|{\mathcal O}^{A'A_{\!B'}}(({\check\Psi}^p)^{A'\bar{R}_AA_{\!B'}})-(\Phi_d^{\otimes n})^{{\bar A}{\bar R}_A}\right\|_1\nn\\
&\leq&2\sqrt{3}\sqrt[8]{2\epsilon}+5\sqrt[4]{2\epsilon}+\epsilon_{\rm merg}.\nn
\end{eqnarray}
Due to (\ref{eq:fidelityqsm}), (\ref{eq:mergerrbound}) and (\ref{eq:tracefid}), we have
\begin{eqnarray}
\epsilon_{\rm merg}\leq2\left(2\sqrt{2}\sqrt{2^{-\frac{nr}{2}}+2^{-n(R+r)}}\right)^{\frac{1}{2}}\leq4\!\cdot\!2^{-nr/8}.\nn
\end{eqnarray}   
Thus we obtain (\ref{eq:toterror}).\hfill$\blacksquare$

\section{Proof of the Converse Part}\label{app:prfconv}

 Fix arbitrary $n\in{\mathbb N}$ and $\epsilon>0$ such that
\begin{eqnarray}
16\cdot12\sqrt[4]{\epsilon}< n\leq\frac{1}{4\cdot12\sqrt[4]{\epsilon}},\label{eq:deldel}
\end{eqnarray}
and let ${\mathcal M}_n$ be a $(2,n,\epsilon)$-protocol for implementing $U$ with the entanglement cost $nE$, the classical communication cost $nC_f$ and the backward classical communication cost $nC_b$.  We assume here for simplicity that $K_n$ and $L_n$ is bounded above as
\begin{align}
\log{K_n},\;\log{L_n}\leq n\log{\kappa}\label{eq:elel}
\end{align}
with a constant $\kappa>0$. As we prove below, the following inequalities hold for any such ${\mathcal M}_n$:
\begin{eqnarray}
&&\!\!\!\!\!\!\!\!\!\!\!\!\!\!\!\!\!\!\!\!\!\!\!\!C_f\geq M(U^\dagger)-{\tilde\xi}( 12\sqrt[4]{\epsilon}\cdot n)\log{d},\label{eq:b}\\
&&\!\!\!\!\!\!\!\!\!\!\!\!\!\!\!\!\!\!\!\!\!\!\!\!n\log{d}+\log{K_n}-\sum_{{k\in{\mathbb K}}}p_{k}S(A')_{\Psi_{k}}\nn\\
&&\geq nM(U^{\dagger})-n{\tilde\xi}(12\sqrt[4]{\epsilon}\cdot n)\log{d},\label{eq:l}\\
&&\!\!\!\!\!\!\!\!\!\!\!\!\!\!\!\!\!\!\!\!\!\!\!\!\frac{1}{n}(\log{K_n}-\log{L_n})\nn\\
&&\geq M(U^{\dagger})-\xi_1(12\sqrt[4]{\epsilon}\cdot n)\log{(d\kappa)},\label{eq:a}\\
&&\!\!\!\!\!\!\!\!\!\!\!\!\!\!\!\!\!\!\!\!\!\!\!\!C_b\geq M(U^{\dagger})-\xi_2(12\sqrt[4]{\epsilon}\cdot n)\log{(d\kappa\!\cdot\!2^{C_b})}\label{eq:c}.
\end{eqnarray}
Here, ${\tilde\xi}$ is a function defined by (\ref{eq:wanabee}), and $\xi_1$, $\xi_2$ are  nonnegative functions that are independent of $n$ and $d$, and satisfy $\lim_{ x\rightarrow0}\xi_1(x)=\lim_{x\rightarrow0}\xi_2( x)=0$.  The converse part of Theorem \ref{thm:misoptach} immediately follows by substituting $\epsilon_n$ to $\epsilon$ in Inequalities (\ref{eq:b}), (\ref{eq:a}) and (\ref{eq:c}), and by taking the limit of $n\rightarrow\infty$. Note that Assumption (\ref{eq:condconv}) implies $\lim_{n\rightarrow\infty}12\sqrt[4]{\epsilon_n}\cdot n=0$.

Let us prove Inequalities (\ref{eq:b})$\sim$(\ref{eq:c}). From Lemma \ref{lmm:markovifepsilon}, Alice's measurement in ${\mathcal M}_n$ is $4\sqrt[4]{\epsilon}$-oblivious and $12\sqrt[4]{\epsilon}$-Markovianizing from $A'R_A$. Hence Conditions 1) and 2) in Definition \ref{dfn:rofpsi} are satisfied by the correspondence described by (\ref{correspondence}). Thus we can apply Lemma \ref{thm:recov} to obtain the above four inequalities.

Inequality (\ref{eq:b}) follows from $nC_f\geq H(\{p_k\}_{k\in{\mathbb K}})$ and Inequality (\ref{eq:markovmeasure3}). Inequality (\ref{eq:l}) follows from Inequality (\ref{eq:markovmeasure3}) on $\Delta S(A')_{av}$.  We prove Inequalities (\ref{eq:a}) and (\ref{eq:c}) in the following subsections.

\subsection{Proof of Inequality (\ref{eq:a})}

Let ${\tilde{\mathcal{M}}}_{n,k}:A'B_1B_E\rightarrow{\bar A}A_1B_1$ be a CPTP map that describes the procedure I\!I-4$\sim$7, presented in Appendix \ref{app:setting}, averaged over the measurement outcome $l$. The final state is given by
\begin{eqnarray}
\rho({\mathcal M}_n,U^\dagger)=\sum_{k\in{\mathbb K}}p_{k}{\tilde{\mathcal{M}}}_{n,k}(\Psi_{k}').\label{eq:nihonrikujo}
\end{eqnarray}
Define
\begin{align}
\epsilon_{k,1}:=\left\|{\tilde{\mathcal{M}}}_{n,k}(\Psi_{k}')-(\Phi_d^{\otimes n})^{{\bar A}{\bar R}_A}\otimes(\Phi_d^{\otimes n})^{{\bar B}{\bar R}_B}\otimes\Phi_{L_n}^{A_1B_1}\right\|.\label{eq:osaifu}
\end{align}
and
\begin{align}
f_{k,1}:=F({\tilde{\mathcal{M}}}_{n,k}(\Psi_{k}'),|\Phi_d^{\otimes n}\rangle^{{\bar A}{\bar R}_A}|\Phi_d^{\otimes n}\rangle^{{\bar B}{\bar R}_B}|\Phi_{L_n}\rangle^{A_1B_1})\nn
\end{align}
for ${k\in{\mathbb K}}$. Due to the convexity of the square function,  Inequality (\ref{eq:tracefid}), Equalities (\ref{eq:fidpens}), (\ref{eq:nihonrikujo}) and Inequality (\ref{eq:convfid}), we have
\begin{eqnarray}
&&\!\!\!\!\!\!\!\!\!\left(\sum_{k\in{\mathbb K}}p_{k}\epsilon_{k,1}\right)^2\:\leq\:\sum_{k\in{\mathbb K}}p_{k}\epsilon_{k,1}^2\:\leq\:4\sum_{k\in{\mathbb K}}p_{k}(1-f_{k,1})\nn\\
&&\!\!\!\!\!\!\!\!\!=4-4F(\rho({\mathcal M}_n, U^\dagger),|\Phi_d^{\otimes n}\rangle^{{\bar A}{\bar R}_A}|\Phi_d^{\otimes n}\rangle^{{\bar B}{\bar R}_B}|\Phi_{L_n}\rangle^{A_1B_1})\nn\\
&&\!\!\!\!\!\!\!\!\!\leq4\epsilon,\nn
\end{eqnarray}
which yields
\begin{eqnarray}
\sum_{k\in{\mathbb K}}p_{k}\epsilon_{k,1}\leq2\sqrt{\epsilon}.\label{eq:higeki}
\end{eqnarray}

From Lemma \ref{thm:abcaabcc}, there exist pure states $|\Psi_{k}^p\rangle^{A'{\bar R}_AB_1B_E}$ such that
\begin{align}
\sum_{k}p_{k}\left\|\Psi_{k}'-(\Psi_{k}^p)^{A'{\bar R}_AB_1B_E}\otimes(\Phi_d^{\otimes n})^{{\bar B}{\bar R}_B}\right\|_1\leq4\sqrt[4]{\epsilon}.\nn
\end{align}
Defining
\begin{eqnarray}
\epsilon_{k,2}:=\left\|\Psi_{k}'-(\Psi_{k}^p)^{A'{\bar R}_AB_1B_E}\otimes(\Phi_d^{\otimes n})^{{\bar B}{\bar R}_B}\right\|_1,\label{eq:defek2}
\end{eqnarray}
we obtain
\begin{eqnarray}
\sum_{k\in{\mathbb K}}p_{k}\epsilon_{k,2}\leq4\sqrt[4]{\epsilon}.\label{eq:hideki}
\end{eqnarray}
From (\ref{eq:higeki}) and (\ref{eq:hideki}), we have
\begin{align}
\sum_{k\in{\mathbb K}}p_{k}\epsilon_{k}\leq2\sqrt{\epsilon}+4\sqrt[4]{\epsilon}\leq6\sqrt[4]{\epsilon}\label{eq:qsmsim}
\end{align}
for
\begin{align}
\epsilon_k:=\epsilon_{k,1}+\epsilon_{k,2}.\label{eq:eqeqeq}
\end{align}
 Defining ${\mathbb K}(\lambda):=\{k\in{\mathbb K}\:|\:\epsilon_k\leq\lambda\}$, this leads to
\begin{eqnarray}
\sum_{k\in{\mathbb K}\setminus{\mathbb K}(\lambda)}p_k\leq\frac{1}{\lambda}\sum_{k\in{\mathbb K}\setminus{\mathbb K}(\lambda)}p_k\epsilon_k\leq\frac{1}{\lambda}\sum_{k\in{\mathbb K}}p_k\epsilon_k\leq\frac{6\sqrt[4]{\epsilon}}{\lambda}\label{eq:rokuyon}
\end{eqnarray}
for any $\lambda>0$.

From (\ref{eq:defek2}) and (\ref{eq:monottd}), we have
\begin{align}
\left\|{\tilde{\mathcal{M}}}_{n,k}(\Psi_{k}')-{\tilde{\mathcal{M}}}_{n,k}(\Psi_{k}^p)^{{\bar A}{\bar R}_AA_1B_1}\otimes(\Phi_d^{\otimes n})^{{\bar B}{\bar R}_B}\right\|_1\leq\epsilon_{k,2}\label{eq:saifu}
\end{align}
 for each $k$, which implies
\begin{align}
\left\|(\Psi_{k})^{A'}-(\Psi_{k}^p)^{A'}\right\|_1=\left\|(\Psi_{k}')^{A'}-(\Psi_{k}^p)^{A'}\right\|_1\leq\epsilon_{k,2}\leq\epsilon_{k}\label{eq:defekk2}
\end{align}
due to (\ref{eq:monottdd}) and (\ref{eq:prpr}). By (\ref{eq:tdtensor}), (\ref{eq:triangle}), (\ref{eq:saifu}), (\ref{eq:osaifu}) and (\ref{eq:eqeqeq}), we see that
\begin{eqnarray}
&&\left\|({\tilde{\mathcal{M}}}_{n,k}(\Psi_{k}^p))^{{\bar A}{\bar R}_AA_1B_1}-(\Phi_d^{\otimes n})^{{\bar A}{\bar R}_A}\otimes\Phi_{L_n}^{A_1B_1}\right\|_1\nn\\
&=&\left\|({\tilde{\mathcal{M}}}_{n,k}(\Psi_{k}^p))^{{\bar A}{\bar R}_AA_1B_1}\otimes(\Phi_d^{\otimes n})^{{\bar B}{\bar R}_B}\right.\nn\\
&&\quad\quad\left.-(\Phi_d^{\otimes n})^{{\bar A}{\bar R}_A}\otimes(\Phi_d^{\otimes n})^{{\bar B}{\bar R}_B}\otimes\Phi_{L_n}^{A_1B_1}\right\|_1\nn\\
&\leq&\left\|{\tilde{\mathcal{M}}}_{n,k}(\Psi_{k}')-{\tilde{\mathcal{M}}}_{n,k}(\Psi_{k}^p)^{{\bar A}{\bar R}_AA_1B_1}\otimes(\Phi_d^{\otimes n})^{{\bar B}{\bar R}_B}\right\|_1\nn\\
&&+\left\|{\tilde{\mathcal{M}}}_{n,k}(\Psi_{k}')-(\Phi_d^{\otimes n})^{{\bar A}{\bar R}_A}\otimes(\Phi_d^{\otimes n})^{{\bar B}{\bar R}_B}\otimes\Phi_{L_n}^{A_1B_1}\right\|\nn\\
&\leq&\epsilon_k,\label{eq:defepspri}
\end{eqnarray}
which leads to
\begin{eqnarray}
\left\|(\Psi_k^p)^{{\bar R}_A}-(\pi_d^{\otimes n})^{{\bar R}_A}\right\|_1\leq{\epsilon_k}.\label{eq:gokui}
\end{eqnarray}
by (\ref{eq:monottdd}). Therefore, due to  Lemma \ref{lmm:uhlmann}, there exists a quantum operation ${\mathcal T}_{n,k}:{\bar A}\rightarrow A'A_{B_1B_E}$, where $A_{B_1\!B_E}$ is a quantum system which is identical to $B_1B_E$, such that
\begin{align}
\left\|(\Psi_{k}^p)^{A'{\bar R}_AA_{B_1\!B_E}}-{\mathcal T}_{n,k}((\Phi_d^{\otimes n})^{{\bar A}{\bar R}_A})\right\|_1\leq2\sqrt{\epsilon_k}.\label{sekaishi}
\end{align}
Define
\begin{align}
{\tilde{\mathcal{M}}}_{n,k}':={\mathcal T}_{n,k}\circ{\tilde{\mathcal{M}}}_{n,k}.\label{eq:chunchun}
\end{align}
Owing to (\ref{eq:triangle}), (\ref{eq:chunchun}), (\ref{eq:monottd}), (\ref{eq:tdtensor}) and Inequalities (\ref{eq:defepspri}), (\ref{sekaishi}), we have 
\begin{eqnarray}
&&\left\|({\tilde{\mathcal{M}}}_{n,k}'(\Psi_{k}^p))^{A'{\bar R}_AA_{B_1\!B_E}A_1B_1}\right.\nn\\
&&\;\;\;\;\;\;\;\;\left.-(\Psi_{k}^p)^{A'{\bar R}_AA_{B_1\!B_E}}\otimes\Phi_{L_n}^{A_1B_1}\right\|_1\nn\\
&\leq&\left\|({\tilde{\mathcal{M}}}_{n,k}'(\Psi_{k}^p))^{A'{\bar R}_AA_{B_1\!B_E}A_1B_1}\right.\nn\\
&&\;\;\;\;\;\;\;\;\left.-{\mathcal T}_{n,k}((\Phi_d^{\otimes n})^{{\bar A}{\bar R}_A})\otimes\Phi_{L_n}^{A_1B_1}\right\|_1\nn\\
&&\!\!\!\!\!+\left\|{\mathcal T}_{n,k}((\Phi_d^{\otimes n})^{{\bar A}{\bar R}_A})\otimes\Phi_{L_n}^{A_1B_1}\right.\nn\\
&&\;\;\;\;\;\;\;\;\left.-(\Psi_{k}^p)^{A'{\bar R}_AA_{B_1\!B_E}}\otimes\Phi_{L_n}^{A_1B_1}\right\|_1\nn\\
&\leq&\left\|({\tilde{\mathcal{M}}}_{n,k}(\Psi_{k}^p))^{{\bar A}{\bar R}_AA_1B_1}-(\Phi_d^{\otimes n})^{{\bar A}{\bar R}_A}\otimes\Phi_{L_n}^{A_1B_1}\right\|_1\nn\\
&&+\left\|{\mathcal T}_{n,k}((\Phi_d^{\otimes n})^{{\bar A}{\bar R}_A})-(\Psi_{k}^p)^{A'{\bar R}_AA_{B_1\!B_E}}\right\|_1\nn\\
&\leq&\epsilon_k+2\sqrt{\epsilon_k},\nn
\end{eqnarray}
 which leads to
\begin{eqnarray}
&&F\left({\tilde{\mathcal{M}}}_{n,k}'(\Psi_{k}^p),\:|\Psi_{k}^p\rangle^{A'{\bar R}_AA_{B_1\!B_E}}|\Phi_{L_n}\rangle^{A_1B_1}\right)\nn\\
&&\geq1- \epsilon_k-2\sqrt{\epsilon_k}.\nn
\end{eqnarray}
Hence ${\tilde{\mathcal{M}}}_{n,k}'$ is a state merging of $|\Psi_{k}^p\rangle^{A'{\bar R}_A(B_1B_E)}$ with the error $ \epsilon_k+2\sqrt{\epsilon_k}$ and the entanglement cost $-\log{L_n}$ (see Definition \ref{dfn:sm} and Inequality (\ref{eq:tracefid})). 

Note that we have
\begin{align}
{\rm dim}{\mathcal H}^{A'}\leq {\rm dim}{\mathcal H}^{\bar A}\times{\rm dim}{\mathcal H}^{A_0}\leq d^n\times\kappa^n,\label{eq:shima}
\end{align}
which corresponds to (\ref{eq:dleqdd}), and Assumption (\ref{eq:elel}). Therefore, we apply Theorem \ref{thm:qsmoptimal}, Inequalities (\ref{eq:defekk2}), (\ref{eq:gokui}) and (\ref{eq:contvNent}) to obtain
\begin{eqnarray}
&&-\log{L_n}+n\eta'\left(2\sqrt{ \epsilon_k+2\sqrt{\epsilon_k}}\right)\log{(d\kappa)}\nn\\
&&\geq S(B_1B_E|A')_{\Psi_{k}^p}\nn\\
&&=S({\bar R}_A)_{\Psi_{k}^p}-S(A')_{\Psi_{k}^p}\label{sq:kikuta}\nn\\
&&\geq n\log{d}-S(A')_{\Psi_{k}}-2n\eta({\epsilon_k})\log{(d\kappa)}\label{eq:shikainosoto}
\end{eqnarray}
for each $k$  such that $\epsilon_k+2\sqrt{\epsilon_k}\leq1/4$, where $\eta'$ is a function defined by (\ref{eq:dfnetapr}).  Substituting $\lambda^*:=(9-4\sqrt{5})/4$ to $\lambda$ in (\ref{eq:rokuyon}), and noting that $\epsilon_k+2\sqrt{\epsilon_k}\leq1/4$ if and only if $\epsilon_k\in{\mathbb K}(\lambda^*)$, we also have
\begin{eqnarray}
\sum_{k:\:\epsilon_k+2\sqrt{\epsilon_k}>1/4}p_k=\sum_{k\in{\mathbb K}\setminus{\mathbb K}(\lambda^*)}p_k\leq\frac{4\cdot6\sqrt[4]{\epsilon}}{9-4\sqrt{5}}<432\sqrt[4]{\epsilon}.\label{eq:shikainonaka}
\end{eqnarray}
Defining
\begin{eqnarray}
\eta''(\epsilon):=\eta'\left(2\sqrt{\epsilon+2\sqrt{\epsilon}}\right)+2\eta(\epsilon),\label{eq:andou}
\end{eqnarray}
Inequalities (\ref{eq:shikainosoto}) and (\ref{eq:shikainonaka}) yield
\begin{align}
&-\log{L_n}\nn\\
&=-\sum_{k\in{\mathbb K}(\lambda^*)}p_k\log{L_n}-\sum_{k\in{\mathbb K}\setminus{\mathbb K}(\lambda^*)}p_k\log{L_n}\nn\\
&\geq \sum_{k\in{\mathbb K}(\lambda^*)}p_{k}\left(n\log{d}-S(A')_{\Psi_{k}}-\eta''(\epsilon_k)\log{(d\kappa)}\right)\nn\\
&\quad\quad-\sum_{k\in{\mathbb K}\setminus{\mathbb K}(\lambda^*)}p_{k} n\log{\kappa}\nn\\
&\geq n\log{d}-\sum_{k\in{\mathbb K}(\lambda^*)}p_{k}S(A')_{\Psi_{k}}\nn\\
&\quad\quad-n\sum_{k\in{\mathbb K}(\lambda^*)}p_{k}\eta''(\epsilon_k)\log{(d\kappa)}\nn\\
&\quad\quad-\sum_{k\in{\mathbb K}\setminus{\mathbb K}(\lambda^*)}p_{k} n(\log{d}+\log{\kappa})\nn\\
&\geq n\log{d}-\sum_{k\in{\mathbb K}}p_{k}S(A')_{\Psi_{k}}-n\sum_{k\in{\mathbb K}}p_{k}\eta''(\epsilon_k)\log{(d\kappa)}\nn\\
&\quad\quad-432\sqrt[4]{\epsilon}\cdot n\log{(d\kappa)}.\label{eq:NSN}
\end{align}

Let us define a function $\eta_1$ by
\begin{eqnarray}
\eta_1(x):=\eta''\left(\sqrt{x/2}\right)+\eta''( 4)\cdot\sqrt{x/2}+36x,\label{eq:www}
\end{eqnarray}
which satisfies $\lim_{x\rightarrow0}\eta_1(\epsilon)=0$. From Inequality (\ref{eq:NSN}), Lemma \ref{lmm:yoshiki} and (\ref{eq:qsmsim}), we have
\begin{eqnarray}
-\log{L_n}\geq n\log{d}-\sum_{k\in{\mathbb K}}p_{k}S(A')_{\Psi_{k}}-\eta_1( 12\sqrt[4]{\epsilon})\log{(d\kappa)},\nn
\end{eqnarray}
where we used the fact that we have $\epsilon_k\leq4$ from (\ref{eq:eqeqeq}) and $6\sqrt[4]{\epsilon}\leq16=4^2$ from (\ref{eq:deldel}). Combining this with Inequality (\ref{eq:l}), and defining $\xi_1(x):={\tilde\xi}(x)+\eta_1(x)$, we obtain Inequality (\ref{eq:a}).

\subsection{Proof of Inequality (\ref{eq:c})}

 To prove Inequality (\ref{eq:c}), note that ${\tilde{\mathcal{M}}}_{n,k}'$ is a state merging of $|\Psi_{k}^p\rangle^{A'{\bar R}_A(B_1B_E)}$ with the error $\epsilon_k+2\sqrt{\epsilon_k}$ and the classical communication cost $nC_b$. Therefore, from Theorem \ref{thm:qsmoptimal}, Inequalities (\ref{eq:gokui}) and (\ref{eq:fannes}), we have
\begin{eqnarray}
&&nC_b+5n\eta\left(2\sqrt{\epsilon_k+2\sqrt{\epsilon_k}}\right)\log{(d\kappa)}\nn\\
&\geq&I(B_1B_E:{\bar R}_A)_{\Psi_{k}^p}\nn\\
&=&S(B_1B_E)_{\Psi_{k}^p}+S({\bar R}_A)_{\Psi_{k}^p}-S(B_1B_E{\bar R}_A)_{\Psi_{k}^p}\nn\\
&\geq& S(B_1B_E)_{\Psi_{k}^p}+n\log{d}-S(A')_{\Psi_{k}^p}-n\eta(\epsilon_k)\log{d}\nn\\\label{eq:samaritan}
\end{eqnarray}
for each $k\in{\mathbb K}(\lambda^*)$. From (\ref{eq:defek2}), (\ref{eq:contvNent}), (\ref{eq:prpr}), (\ref{eq:eqeqeq}) and
\begin{eqnarray}
\log{\rm dim}{\mathcal H}^{B_E}=nC_b,\;\Psi_{k}^{{\bar B}B_0}=(\pi_d^{\otimes n})^{{\bar B}}\otimes\Psi_{k}^{B_0},\nn
\end{eqnarray} 
we have
\begin{eqnarray}
&&\!\!\!\!\!\!\!\!\!\!\!\!\!S(B_1B_E)_{\Psi_{k}^p}=S(B_1B_E{\bar B})_{\Psi_{k}^p\otimes\Phi_d^{\otimes n}}-S({\bar B})_{\Phi_d^{\otimes n}}\nn\\
					&&\!\!\!\!\!\!\!\!\!\!\!\!\!\geq S(B_1B_E{\bar B})_{\Psi_{k}'}-n\log{d}-n\eta(\epsilon_{k,2})\log{(d\kappa\!\cdot\!2^{C_b})}\nn\\
					&&\!\!\!\!\!\!\!\!\!\!\!\!\!= S({\bar B}B_0)_{\Psi_{k}}-n\log{d}-n\eta(\epsilon_{k,2})\log{(d\kappa\!\cdot\!2^{C_b})}\nn\\
					&&\!\!\!\!\!\!\!\!\!\!\!\!\!=S({\bar B})_{\pi_d^{\otimes n}}+S(B_0)_{\Psi_{k}}-n\log{d}-n\eta(\epsilon_{k,2})\log{(d\kappa\!\cdot\!2^{C_b})}\nn\\
					&&\!\!\!\!\!\!\!\!\!\!\!\!\!=S(B_0)_{\Psi_{k}}-n\eta(\epsilon_{k,2})\log{(d\kappa\!\cdot\!2^{C_b})}\nn\\
					&&\!\!\!\!\!\!\!\!\!\!\!\!\!\geq S(B_0)_{\Psi_{k}}-n\eta(\epsilon_{k})\log{(d\kappa\!\cdot\!2^{C_b})}.\label{eq:simon}
\end{eqnarray}
Using (\ref{eq:defek2}), (\ref{eq:shima}), (\ref{eq:prpr}) and (\ref{eq:eqeqeq}), we also have
\begin{eqnarray}
\!\!\!\!\!\!\!\!\!S(A')_{\Psi_{k}^p}&\leq&S(A')_{\Psi_{k}'}+n\eta(\epsilon_{k,2})\log{(d\kappa)}\nn\\
&\leq&S(A')_{\Psi_{k}}+n\eta(\epsilon_{k})\log{(d\kappa)}.\label{eq:lees}
\end{eqnarray}
From (\ref{eq:samaritan}), (\ref{eq:simon}), (\ref{eq:lees}) and (\ref{eq:andou}), we see that
\begin{eqnarray}
&&\!\!\!\!\!\!\!\!\!\!nC_b\nn\\
&&\!\!\!\!\!\!\!\!\geq n\log{d}-S(A')_{\Psi_{k}}+S(B_0)_{\Psi_{k}}\nn\\
&&\!\!\!-n\left(5\eta\left(2\sqrt{\epsilon+2\sqrt{\epsilon}}\right)+3\eta(\epsilon_k)\right)\log{(d\kappa\!\cdot\!2^{C_b})}\nn\\
&&\!\!\!\!\!\!\!\!\geq n\log{d}-S(A')_{\Psi_{k}}+S(B_0)_{\Psi_{k}}-5n\eta''(\epsilon_k)\log{(d\kappa\!\cdot\!2^{C_b})}\nn
\end{eqnarray}
for each $k\in{\mathbb K}(\lambda^*)$. Thus we have
\begin{align}
&\!\!\!nC_b\geq\sum_{k\in{\mathbb K}(\lambda^*)}p_knC_b\nn\\
&\geq\sum_{k\in{\mathbb K}(\lambda^*)}p_k\left(n\log{d}-S(A')_{\Psi_{k}}+S(B_0)_{\Psi_{k}}\right)\nn\\
&\quad-5n\sum_{k\in{\mathbb K}(\lambda^*)}p_k\eta''(\epsilon_k)\log{(d\kappa\!\cdot\!2^{C_b})}\nn\\
&\geq \sum_{k\in{\mathbb K}}p_k\left(n\log{d}-S(A')_{\Psi_{k}}+S(B_0)_{\Psi_{k}}\right)\nn\\
&\quad-5n\sum_{k\in{\mathbb K}}p_k\eta''(\epsilon_k)\log{(d\kappa\!\cdot\!2^{C_b})}\nn\\
&\quad-\sum_{k\in{\mathbb K}\setminus{\mathbb K}(\lambda^*)}p_k\left(n\log{d}-S(A')_{\Psi_{k}}+S(B_0)_{\Psi_{k}}\right).\label{eq:sonomukou}
\end{align}
Noting that $\dim{{\mathcal H}^{B_0}}=\kappa$ and
\begin{align}
&\sum_{k\in{\mathbb K}\setminus{\mathbb K}(\lambda^*)}p_k\left(n\log{d}-S(A')_{\Psi_{k}}+S(B_0)_{\Psi_{k}}\right)\nn\\
&\leq\sum_{k\in{\mathbb K}\setminus{\mathbb K}(\lambda^*)}p_kn(\log{d}+\log{\kappa})\leq432\sqrt[4]{\epsilon}\cdot n\log{(d\kappa)}\nn
\end{align}
from (\ref{eq:shikainonaka}), Inequality (\ref{eq:sonomukou}) leads to
\begin{align}
nC_b\geq&\sum_{k\in{\mathbb K}}p_k\left(n\log{d}-S(A')_{\Psi_{k}}+S(B_0)_{\Psi_{k}}\right)\nn\\
&-5\eta_1(12\sqrt[4]{\epsilon})\log{(d\kappa\!\cdot\!2^{C_b})}\nn
\end{align}
where $\eta_1$ is a function defined by (\ref{eq:www}). Thus, from Inequality (\ref{eq:markovmeasure3}) on $\Delta S(A')_{av}\!-\!\Delta S(G)_{av}$, we have
\begin{align}
nC_b\geq& \:\:nM(U^{\dagger})\nn\\
&-n\left({\tilde\xi}(12\sqrt[4]{\epsilon}\!\cdot\! n)+5\eta_1(12\sqrt[4]{\epsilon})\right)\log{(d\kappa\!\cdot\!2^{C_b})}.
\end{align}
Defining $\xi_2(x):={\tilde\xi}(x)+5\eta_1(x)$, we obtain Inequality (\ref{eq:c}).

\hfill$\blacksquare$

\subsection{On the Convergence Speed of the Error}\label{eq:appconsp}

We prove that the converse part of Theorem \ref{thm:misoptach} holds even when we drop Condition (\ref{eq:condconv}), if Conjecture \ref{cjt:symrec} is true. First, as we proved in Appendix E of \cite{waka15_rec} (see Remark therein), the function ${\xi}(n\epsilon)$ in (\ref{eq:iavelow}) can be replaced by another function $\xi'(\epsilon)$, which is independent of $n$ and satisfies $\lim_{\epsilon\rightarrow0}\xi'(\epsilon)=0$. Consequently, functions ${\tilde\xi}(12\sqrt[4]{\epsilon}\!\cdot\! n)$, $\xi_1(12\sqrt[4]{\epsilon}\!\cdot\! n)$ and $\xi_2(12\sqrt[4]{\epsilon}\!\cdot\!n)$ in Inequalities (\ref{eq:b}) $\sim$ (\ref{eq:c}) can be replaced by different functions ${\tilde\xi}'(\epsilon)$, $\xi_1'(\epsilon)$ and $\xi_2'(\epsilon)$, respectively, which do not depend on $n$ and vanishes in the limit of $\epsilon\rightarrow0$. Inequalities (\ref{eq:b}) $\sim$ (\ref{eq:c}) then hold for any $n$ and $\epsilon$, which implies that the converse part holds without additional assumption (\ref{eq:condconv}).

\section{Proof of Theorem \ref{thm:cliff}}\label{app:prfcliff}

We prove Theorem \ref{thm:cliff} after introducing a theorem regarding the cost of randomness for destroying correlations in a bipartite quantum state.

\subsection{Decoupling}\label{app:decoupling}

The following lemma is obtained as a corollary of Proposition 2 in \cite{berry08}, except an evaluation of the convergence speed of the error. 

\begin{lmm}\label{lmm:decoupling}
Let $\pi^A$ be the maximally mixed state on ${\mathcal H}^A$, and suppose a bipartite state $\rho^{AB}\in{\mathcal S}({\mathcal H}^A\otimes{\mathcal H}^B)$ satisfies $\rho^A=\pi^A$. There exists a constant $c>0$ that satisfies the following properties for any $R>I(A:B)_{\rho}$, sufficiently small $\delta>0$ and sufficiently large $n$. That is, for an arbitrary ensemble of unitaries on $({\mathcal H}^{A})^{\otimes n}$ satisfying
\begin{eqnarray}
\forall\ket{\phi}\in({\mathcal H}^{A})^{\otimes n};\;\; \int_{V}p(dV)V\proj{\phi}V^{\dagger}=(\pi^{A})^{\otimes n},\label{eq:pdvcond}
\end{eqnarray}
there exists a set of unitaries $\{V_k\}^{2^{nR}}_{k=1}$ on the support of $p(dV)$, such that a random unitary operation ${\mathcal V}_n$ on ${\mathcal S}({\mathcal H}^{\bar A})$ defined by
\begin{eqnarray}
{\mathcal V}_n(\cdot)=\frac{1}{2^{nR}}\sum_{k=1}^{2^{nR}}V_k(\cdot)V_k^{\dagger}\label{eq:defvn}
\end{eqnarray}
satisfies
\begin{align}
\left\|{\mathcal V}_n((\rho^{AB})^{\otimes n})-(\pi^{A})^{\otimes n}\otimes (\rho^B)^{\otimes n}\right\|_1\nn\\\leq2^{-n\delta}+14\exp{\left(-\frac{c\delta^2n}{4}\right)}.\label{eq:speedy}
\end{align}
\end{lmm}

\begin{prf}
The proof is basically the same as that of Proposition 2 in \cite{berry08}. Fix an arbitrary $\delta>0$. Let ${\mathcal H}_{n,\delta}^{{\bar A}{\bar B}}\subset{({\mathcal H}^{AB})^{\otimes n}}$ and ${\mathcal H}_{n,\delta}^{\bar B}\subset{({\mathcal H}^B)^{\otimes n}}$ be the $\delta$-weakly typical subspace with respect to $(\rho^{AB})^{\otimes n}$ and $(\rho^B)^{\otimes n}$, and let $\Pi_{n,\delta}^{{\bar A}{\bar B}}$ and $\Pi_{n,\delta}^{\bar B}$ be the projection onto those subspaces, respectively. There exists a $\delta$-independent constant $c>0$ such that we have
\begin{eqnarray}
{\rm Tr}[\Pi_{n,\delta}^{{\bar A}{\bar B}}(\rho^{AB})^{\otimes n}]&\geq&1-\exp{(-c\delta^2n)}\nn\\
{\rm Tr}[\Pi_{n,\delta}^{\bar B}(\rho^B)^{\otimes n}]&\geq&1-\exp{(-c\delta^2n)}\nn
\end{eqnarray}
for any $\delta>0$ and $n$\cite{ahlswede80}. Define
\begin{eqnarray}
{\tilde D}_{n,\delta}&:=&{\rm Tr}[\Pi_{n,\delta}^{{\bar B}}\Pi_{n,\delta}^{{\bar A}{\bar B}}(\rho^{AB})^{\otimes n}\Pi_{n,\delta}^{{\bar A}{\bar B}}\Pi_{n,\delta}^{{\bar B}}]\nn\\
{\tilde\rho}_{n,\delta}^{{\bar A}{\bar B}}&:=&\frac{\Pi_{n,\delta}^{{\bar B}}\Pi_{n,\delta}^{{\bar A}{\bar B}}(\rho^{AB})^{\otimes n}\Pi_{n,\delta}^{{\bar A}{\bar B}}\Pi_{n,\delta}^{{\bar B}}}{{\tilde D}_{n,\delta}}.\label{eq:mitsuo}
\end{eqnarray}
Due to Lemma \ref{lmm:gentlemeasurement}, we have
\begin{eqnarray}
\left\|(\rho^{AB})^{\otimes n}-{\tilde\rho}_{n,\delta}^{{\bar A}{\bar B}}\right\|_1\leq5\exp{\left(-\frac{c\delta^2n}{4}\right)}\label{eq:tadao}
\end{eqnarray}
in addition to
\begin{eqnarray}
{\tilde D}_{n,\delta}\geq1-2\exp{\left(-\frac{c\delta^2n}{2}\right)}.\label{eq:akio}
\end{eqnarray}
Let $\Pi_{n,\delta}'^{{\bar B}}$ be the projection onto the subspace of ${\mathcal H}_{n,\delta}^{\bar B}$ spanned by the eigenvectors of ${\tilde\rho}_{n,\delta}^{\bar B}$, corresponding to the eigenvalues not smaller than
\begin{align}
\lambda^*:=2^{-n(S(\rho^B)+\delta)}\cdot\exp{\left(-\frac{c\delta^2n}{2}\right)}\nn
\end{align}
Define
\begin{eqnarray}
D_{n,\delta}'&:=&{\rm Tr}[\Pi_{n,\delta}'^{\bar B}{\tilde\rho}_{n,\delta}^{{\bar A}{\bar B}}],\nn\\
{\hat\rho}_{n,\delta}^{{\bar A}{\bar B}}&:=&\frac{\Pi_{n,\delta}'^{\bar B}{\tilde\rho}_{n,\delta}^{{\bar A}{\bar B}}\Pi_{n,\delta}'^{\bar B}}{D_{n,\delta}'}.\label{eq:yasuo}
\end{eqnarray}
We then have
\begin{eqnarray}
D_{n,\delta}'&=&1-{\rm Tr}[(I^{\bar B}-\Pi_{n,\delta}'^{\bar B}){\tilde\rho}_{n,\delta}^{{\bar B}}]\nn\\
&\geq&1-\lambda^*\times{\rm rank}[{\tilde\rho}_{n,\delta}^{{\bar B}}]\nn\\
&\geq&1-2^{-nc\delta^2\log{e}/2}\nn\\
&\geq&1-\exp{\left(-\frac{c\delta^2n}{2}\right)},\label{eq:mariko}
\end{eqnarray}
where we used the fact that
\begin{eqnarray}
{\rm rank}[{\tilde\rho}_{n,\delta}^{{\bar B}}]\leq{\rm dim}{\mathcal H}_{n,\delta}^{\bar B}\leq2^{n(S(\rho^B)+\delta)}.\nn
\end{eqnarray}
Therefore, due to the gentle measurement lemma, we have
\begin{eqnarray}
\left\|{\tilde\rho}_{n,\delta}^{{\bar A}{\bar B}}-{\hat\rho}_{n,\delta}^{{\bar A}{\bar B}}\right\|_1\leq2\exp{\left(-\frac{c\delta^2n}{4}\right)}.\label{eq:kishou}
\end{eqnarray}
From (\ref{eq:tadao}), (\ref{eq:kishou}) and the triangle inequality, we obtain
\begin{eqnarray}
\left\|(\rho^{AB})^{\otimes n}-{\hat\rho}_{n,\delta}^{{\bar A}{\bar B}}\right\|_1\leq7\exp{\left(-\frac{c\delta^2n}{4}\right)}.\label{eq:rhorhohat}
\end{eqnarray}
From Definitions (\ref{eq:mitsuo}), (\ref{eq:yasuo}) and Inequalities (\ref{eq:akio}), (\ref{eq:mariko}), the maximum eigenvalue $\lambda^+$ of ${\hat\rho}_{n,\delta}^{{\bar A}{\bar B}}$ is bounded as
\begin{eqnarray}
\lambda^+&\leq&\frac{2^{-n(S(\rho^{AB})-\delta)}}{{\tilde D}_{n,\delta}D_{n,\delta}'}\nn\\&\leq&\frac{2^{-n(S(\rho^{AB})-\delta)}}{1-3\exp{\left(-\frac{c\delta^2n}{2}\right)}}\leq2^{-n(S(\rho^{AB})-2\delta)}\label{eq:azuma}
\end{eqnarray}
for sufficiently large $n$. By definition, we also have
\begin{eqnarray}
\frac{1}{\lambda^+}{\hat\rho}_{n,\delta}^{{\bar A}{\bar B}}(V)\leq (\Pi^A)^{\otimes n}\otimes\Pi_{n,\delta}'^{\bar B}.\label{eq:kikuta}
\end{eqnarray}

Let $\{p(dV),V\}$ be an ensemble of unitaries on $({\mathcal H}^A)^{\otimes n}$ that satisfies (\ref{eq:pdvcond}), and define
\begin{eqnarray}
{\hat\rho}_{n,\delta}^{{\bar A}{\bar B}}(V):=V^{\bar A}{\hat\rho}_{n,\delta}^{{\bar A}{\bar B}}V^{\dagger{\bar A}}.\label{eq:dfnrhoV}
\end{eqnarray}
As an ensemble average, we have
\begin{eqnarray}
{\bar\rho}_{n,\delta}^{{\bar A}{\bar B}}&:=&{\mathbb E}[{\hat\rho}_{n,\delta}^{{\bar A}{\bar B}}(V)]=(\pi^A)^{\otimes n}\otimes{\hat\rho}_{n,\delta}^{\bar B}.\label{eq:jetstream}
\end{eqnarray}
Inequality (\ref{eq:rhorhohat}) then implies
\begin{eqnarray}
&&\left\|{\bar\rho}_{n,\delta}^{{\bar A}{\bar B}}-(\pi^A)^{\otimes n}\otimes(\rho^B)^{\otimes n}\right\|_1
=\left\|{\hat\rho}_{n,\delta}^{\bar B}-(\rho^B)^{\otimes n}\right\|_1\nn\\
&&\leq\left\|{\hat\rho}_{n,\delta}^{{\bar A}{\bar B}}-(\rho^{AB})^{\otimes n}\right\|_1\leq7\exp{\left(-\frac{c\delta^2n}{4}\right)},\label{eq:hakoneonsen}
\end{eqnarray}
where the second line follows from the monotonicity of the trace distance. Due to (\ref{eq:yasuo}) and (\ref{eq:mariko}), the minimum nonzero eigenvalue $\lambda^-$ of (\ref{eq:jetstream}) is bounded as
\begin{eqnarray}
\lambda^-\geq\frac{\lambda^*}{d_A^nD_{n,\delta}'}\geq d_A^{-n}\cdot\lambda^*,\nn
\end{eqnarray}
which leads to
\begin{eqnarray}
\lambda&:=&\frac{\lambda^-}{\lambda^+}\nn\\
&\geq&2^{-n\left[\log{d_A}+S(\rho^B)-S(\rho^{AB})+3\delta\right]}\cdot\exp{\left(-\frac{c\delta^2n}{2}\right)}\nn\\
&=&2^{-n\left[I(A:B)_\rho+3\delta\right]}\cdot\exp{\left(-\frac{c\delta^2n}{2}\right)}.\nn
\end{eqnarray}

Suppose $V_1,\cdots,V_N$ are unitaries that are randomly and independently chosen from an ensemble $\{p(dV),V\}$. Due to (\ref{eq:kikuta}) and the operator Chernoff bound (Lemma 3 in \cite{berry08}), we have
\begin{eqnarray}
\!\!\!\!&&{\rm Pr}\left\{\frac{1}{N}\sum_{i=1}^N{\hat\rho}_{n,\delta}^{{\bar A}{\bar B}}(V_i)\notin\left[(1-\epsilon_{{}_1}){\bar\rho}_{n,\delta}^{{\bar A}{\bar B}},(1+\epsilon_{{}_1}){\bar\rho}_{n,\delta}^{{\bar A}{\bar B}}\right]\right\}\nonumber\\
\!\!\!\!&=&{\rm Pr}\left\{\frac{1}{N}\sum_{i=1}^N\frac{{\hat\rho}_{n,\delta}^{{\bar A}{\bar B}}(V_i)}{\lambda^+}\notin\left[(1-\epsilon_{{}_1})\frac{{\bar\rho}_{n,\delta}^{{\bar A}{\bar B}}}{\lambda^+},(1+\epsilon_{{}_1})\frac{{\bar\rho}_{n,\delta}^{{\bar A}{\bar B}}}{\lambda^+}\right]\right\}\nonumber\\
\!\!\!\!&\leq&2d_A^{n}d_B^{n}\exp{\left(-\frac{N\lambda\epsilon_{{}_1}^2}{2}\right)}\nonumber
\end{eqnarray}
for any $\epsilon_{{}_1}\in(0,1]$, which implies that
\begin{eqnarray}
&&{\rm Pr}\left\{\left\|\frac{1}{2^{nR}}\sum_{i=1}^{2^{nR}}{\hat\rho}_{n,\delta}^{{\bar A}{\bar B}}(V_i)-{\bar\rho}_{n,\delta}^{{\bar A}{\bar B}}\right\|_1\leq2\epsilon_{{}_1}\right\}\nonumber\\
&&\geq1-2d_A^{n}d_B^{n}\exp{\left(-\frac{2^{nR}\lambda\epsilon_{{}_1}^2}{2}\right)}\label{eq:chercher}
\end{eqnarray}
for an arbitrary $R>0$. Substituting $\epsilon_{{}_1}=2^{-1-n\delta}$, we obtain
\begin{eqnarray}
&&{\rm Pr}\left\{\left\|\frac{1}{2^{nR}}\sum_{i=1}^{2^{nR}}{\hat\rho}_{n,\delta}^{{\bar A}{\bar B}}(V_i)-{\bar\rho}_{n,\delta}^{{\bar A}{\bar B}}\right\|_1\leq2^{-n\delta}\right\}\nonumber\\
&&\geq1-2d_A^{n}d_B^{n}\exp{\left(-\frac{2^{n(R-2\delta)}\lambda}{8}\right)}.\nn
\end{eqnarray}
Therefore, if $R$ satisfies
\begin{eqnarray}
R>I(A:B)_\rho+5\delta+\frac{1}{2}c\delta^2\log{e},\label{eq:RgtI}
\end{eqnarray}
and if $n$ is sufficiently large so that Inequality (\ref{eq:azuma}) holds and the R.H.S. in (\ref{eq:chercher}) is greater than $0$, there exists a set of unitaries $\{V_i\}_{i=1}^{2^{nR}}$ such that
\begin{eqnarray}
\left\|\frac{1}{2^{nR}}\sum_{i=1}^{2^{nR}}{\hat\rho}_{n,\delta}^{{\bar A}{\bar B}}(V_i)-{\bar\rho}_{n,\delta}^{{\bar A}{\bar B}}\right\|_1\leq2^{-n\delta}.\label{eq:misonolove}
\end{eqnarray}
Using unitaries in the set, construct a random unitary operation ${\mathcal V}_n$ on ${\bar A}$ as (\ref{eq:defvn}).

The total error is evaluated as follows. From (\ref{eq:rhorhohat}), (\ref{eq:dfnrhoV}) and the monotonicity of the trace distance, we have
\begin{align}
\left\|{\mathcal V}_n((\rho^{AB})^{\otimes n})-\frac{1}{2^{nR}}\sum_{i=1}^{2^{nR}}{\hat\rho}_{n,\delta}^{{\bar A}{\bar B}}(V_i)\right\|_1\leq7\exp{\left(-\frac{c\delta^2n}{4}\right)}.\label{eq:ultramanG}
\end{align}
Due to (\ref{eq:hakoneonsen}), (\ref{eq:misonolove}), (\ref{eq:ultramanG}) and the triangle inequality, we obtain
\begin{align}
&&\left\|{\mathcal V}_n((\rho^{AB})^{\otimes n})-(\pi^A)^{\otimes n}\otimes(\rho^B)^{\otimes n}\right\|_1\nn\\
&&\leq2^{-n\delta}+14\exp{\left(-\frac{c\delta^2n}{4}\right)}\nn
\end{align}
for any $R>I(A:B)_\rho$, $\delta\in(0,1]$ satisfying (\ref{eq:RgtI}) and sufficiently large $n$. Thus we obtain (\ref{eq:speedy}).\hfill$\blacksquare$

\end{prf}

\subsection{Proof of Theorem \ref{thm:cliff}}\label{app:okaimonohakochira}

We prove Theorem \ref{thm:cliff} by showing that $M(U)\leq K(U)$ holds for any generalized Clifford operator $U$, which implies $M(U)=K(U)$ due to Lemma \ref{lmm:mgeqk}.  From Equality (\ref{eq:opsch2}), we have
\begin{eqnarray}
K(U)&=&S(AR_A)_{\Psi_U}\nn\\
&=&S(AR_A)_{\Psi_U}+\log{d}-\log{d}\nn\\
&=&S(AR_A)_{\Psi_U}+S(R_B)_{\Psi_U}-S(B)_{\Psi_U}\nn\\
&=&S(AR_A)_{\Psi_U}+S(R_B)_{\Psi_U}-S(AR_AR_B)_{\Psi_U}\nn\\
&=&I(AR_A:R_B)_{\Psi_U}.\nn
\end{eqnarray}
Thus it suffices to prove that any $R$ satisfying $R>I(AR_A:R_B)_{\Psi_U}$ also satisfies $R\geq M(U)$. 

 Fix an arbitrary $R>I(AR_A:R_B)_{\Psi_U}$ and choose sufficiently small $\delta$ and sufficiently large $n$. Define
\begin{eqnarray}
{\vec p}&:=&(p_1,\cdots,p_n)\in\{1,\cdots, d\}^n\nn\\
{\vec q}&:=&(q_1,\cdots,q_n)\in\{1,\cdots, d\}^n\nn\\
{\bm \sigma}_{{\vec p}{\vec q}}&:=&\sigma_{p_1q_1}\otimes\cdots\otimes\sigma_{p_nq_n},\nn
\end{eqnarray}
and consider the ensemble of unitaries
\begin{align}
\left\{1/d^{2n},\;{\bm \sigma}_{{\vec p}{\vec q}}\right\}_{{\vec p}{\vec q}\in\{1,\cdots, d\}^{2n}}\nn
\end{align}
on $R_B^{\otimes n}=\bar{R}_B$. Because of Schur's lemma, the ensemble satisfies
\begin{eqnarray}
&&\frac{1}{d^{2n}}\sum_{{\vec p}{\vec q}}{\bm \sigma}_{{\vec p}{\vec q}}^{\bar{R}_B}\proj{\phi}^{\bar{R}_B}{\bm \sigma}_{{\vec p}{\vec q}}^{\dagger\bar{R}_B}=(\pi_d^{R_B})^{\otimes n}\nonumber
\end{eqnarray}
for any $|\phi\rangle\in{\mathcal H}^{{\bar R}_B}$. Therefore, due to Lemma \ref{lmm:decoupling}, there exists a subset $\{{\vec p}_k{\vec q}_k\}_{k=1}^{2^{nR}}\subset\{1,\cdots, d\}^{2n}$ such that
\begin{align}
\left\|\frac{1}{2^{nR}}\sum_{k=1}^{2^{nR}}{\bm \sigma}_{{\vec p}_k{\vec q}_k}^{\bar{R}_B}(\Psi_U^{\otimes n})^{\bar{A}\bar{R}_A\bar{R}_B}{\bm \sigma}_{{\vec p}_k{\vec q}_k}^{\dagger\bar{R}_B}\right.\nn\\
\left.-(\Psi_U^{\otimes n})^{\bar{A}\bar{R}_A}\otimes(\pi_d^{R_B})^{\otimes n}\right\|_1\leq\epsilon_n,\label{eq:cliffdec}
\end{align}
where
\begin{align}
\epsilon_n:=2^{-n\delta}+16\exp{\left(-\frac{c\delta^2n}{4}\right)}.\label{eq:seses}
\end{align}

Without loss of generality, we assume that the basis $\{\ket{t}\}_{t=1}^d$, by which the generalized Pauli operators are defined as (\ref{eq:dfncliff}), is the Schmidt basis of $\ket{\Phi_d}^{AR_A}$ and $\ket{\Phi_d}^{BR_B}$.  That is, we assume that
\begin{eqnarray}
\ket{\Phi_d}^{AR_A}=\frac{1}{\sqrt{d}}\sum_{t=1}^d|t\rangle^A|t\rangle^{R_A},\;\ket{\Phi_d}^{BR_B}=\frac{1}{\sqrt{d}}\sum_{t=1}^d|t\rangle^B|t\rangle^{R_B}.\nn
\end{eqnarray}
A simple calculation then leads to
\begin{eqnarray}
\sigma_{pq}^{R_B}\ket{\Phi_d}^{BR_B}&=&(\sigma_{pq}^T)^B\ket{\Phi_d}^{BR_B}\nn\\
&=&\exp{(2\pi iqp/d)}\cdot\sigma_{-p,q}^B\ket{\Phi_d}^{BR_B},\nn
\end{eqnarray} 
where the superscript $T$ denotes the transposition with respect to the basis $\{\ket{t}\}_{t=1}^d$ defined by
\begin{eqnarray}
X^T=\sum_{t,t'=1}^d\langle t|X|t'\rangle\cdot|t'\rangle\!\langle t|\nn
\end{eqnarray}
for $X\in{\mathcal L}({\mathcal H}^B)$. Therefore, for some phase $\theta_{pq}'\in{\mathbb R}$ we have
\begin{eqnarray}
&&\sigma_{pq}^{R_B}|\Psi_U\rangle^{ABR_AR_B}\nn\\
&=&(U^{AB}\otimes\sigma_{pq}^{R_B})\ket{\Phi_d}^{AR_A}\ket{\Phi_d}^{BR_B}\nonumber\\
&=&\exp{(2\pi iqp/d)}\cdot U^{ AB}(I^A\otimes\sigma_{-p,q}^B)\ket{\Phi_d}^{AR_A}\ket{\Phi_d}^{BR_B}\nonumber\\
&=&e^{i\theta_{pq}'}(\sigma_{p'q'(pq)}^A\otimes\sigma_{r's'(pq)}^B)U^{AB}\ket{\Phi_d}^{AR_A}\ket{\Phi_d}^{BR_B}\nonumber\\
&=&e^{i\theta_{pq}'}(\sigma_{p'q'(pq)}^A\otimes\sigma_{r's'(pq)}^B)|\Psi_U\rangle^{ABR_AR_B}.\nonumber
\end{eqnarray}
Tracing out $B$, we obtain
\begin{eqnarray}
\sigma_{pq}^{R_B}\Psi_U^{AR_AR_B}\sigma_{pq}^{\dagger R_B}=\sigma_{p'q'(pq)}^A\Psi_U^{AR_AR_B}\sigma_{p'q'(pq)}^{\dagger A},\nonumber
\end{eqnarray}
which implies
\begin{eqnarray}
{\bm \sigma}_{{\vec p}_k{\vec q}_k}^{\bar{R}_B}(\Psi_U^{\otimes n})^{\bar{A}\bar{R}_A\bar{R}_B}{\bm \sigma}_{{\vec p}_k{\vec q}_k}^{\dagger\bar{R}_B}={\bm \sigma}_{{\vec p}_k'{\vec q}_k'}^{\bar{A}}(\Psi_U^{\otimes n})^{\bar{A}\bar{R}_A\bar{R}_B}{\bm \sigma}_{{\vec p}_k'{\vec q}_k'}^{\dagger\bar{A}}\nonumber
\end{eqnarray}
for a subset $\{{\vec p}_k'{\vec q}_k'\}_{k=1}^{2^{nR}}\subset\{1,\cdots, d\}^{2n}$. Using the subset, construct a random unitary operation ${\mathcal V}_n$ on ${\mathcal S}({\mathcal H}^{\bar A})$ as
\begin{eqnarray}
{\mathcal V}_n(\cdot)=\frac{1}{2^{nR}}\sum_{k=1}^{2^{nR}}{\bm \sigma}_{{\vec p}_k'{\vec q}_k'}^{\bar{A}}(\cdot){\bm \sigma}_{{\vec p}_k'{\vec q}_k'}^{\dagger\bar{A}}.\nn
\end{eqnarray}
We then have
\begin{align}
\left\|{\mathcal V}_n((\Psi_U^{\otimes n})^{\bar{A}\bar{R}_A\bar{R}_B})-(\Psi_U^{\otimes n})^{\bar{A}\bar{R}_A}\otimes(\Psi_U^{\otimes n})^{\bar{R}_B}\right\|_1\leq\epsilon_n\nn
\end{align}
from (\ref{eq:cliffdec}). Note that $\Psi_U^{R_B}=\pi_d^{R_B}$. Due to Lemma \ref{lmm:markovifdecouple} and the proof thereof (see Appendix \ref{app:prfmardec}), it follows that the state ${\mathcal V}_n((\Psi_U^{\otimes n})^{\bar{A}\bar{R}_A\bar{R}_B})$ is $\epsilon_n$-recoverable from $AR_A$. Therefore, from Definition 9 and Theorem 11 in \cite{waka15_rec}, we obtain $R\geq M(U)$. Note that the error vanishes exponentially to $n$ due to (\ref{eq:seses}).
\hfill$\blacksquare$

\section{Proof of Lemma \ref{lmm:horiemon} and \ref{lmm:ccpowerofu}}\label{app:frenchtoast}

\subsection{Proof of Lemma \ref{lmm:horiemon}}\label{sec:prfhoriemon}
The proof is based on an idea which is used in \cite{pawlowski09} to prove that information causality is satisfied in quantum mechanics.

For Inequality (\ref{eq:regain}), define $L^0={\hat K}^0=\emptyset$, and denote system $B$ before the first step by $B_0$. By the data processing inequality and the chain rule, we have
\begin{eqnarray}
&&I({\vec X}:B_\gamma,L^\gamma,{\hat K}^\gamma)\nonumber\\
&\leq&I({\vec X}:B_{\gamma-1},L^{\gamma-1},{\hat K}^\gamma)\nonumber\\
&=&I({\vec X}:B_{\gamma-1},L^{\gamma-1},{\hat K}^{\gamma-1})\nonumber\\
&&\;\;+I({\vec X}:{\hat K}_\gamma|B_{\gamma-1},L^{\gamma-1},{\hat K}^{\gamma-1})\nonumber\\
&=&I({\vec X}:B_{\gamma-1},L^{\gamma-1},{\hat K}^{\gamma-1})\nonumber\\
&&\;\;+I({\vec X},B_{\gamma-1},L^{\gamma-1},{\hat K}^{\gamma-1}:{\hat K}_\gamma)\nonumber\\
&&\;\;-I(B_{\gamma-1},L^{\gamma-1},{\hat K}^{\gamma-1}:{\hat K}_\gamma)\nonumber\\
&\leq&I({\vec X}:B_{\gamma-1},L^{\gamma-1},{\hat K}^{\gamma-1})+H({\hat K}_\gamma)\nonumber\\
&\leq&I({\vec X}:B_{\gamma-1},L^{\gamma-1},{\hat K}^{\gamma-1})+\log{|{\hat{\mathcal K}}_\gamma|}\nonumber
\end{eqnarray}
for $\gamma=1,\cdots,\Gamma$. Hence we obtain
\begin{eqnarray}
&&I({\vec X}:B_\Gamma,L^\Gamma,{\hat K}^\Gamma)\nonumber\\
&&\!\!\!\!\!\!\!\!\!=\sum_{\gamma=1}^\Gamma\left[I({\vec X}:B_\gamma,L^\gamma,{\hat K}^\gamma)-I({\vec X}:B_{\gamma-1},L^{\gamma-1},{\hat K}^{\gamma-1})\right]\!\!\nonumber\\
&&\!\!\!\!\!\!\!\!\!\leq\!\!\sum_{\gamma=1}^\Gamma\log{|{\hat{\mathcal K}}_\gamma|}\;=\;C_{\rm tot}.\nonumber
\end{eqnarray}

For Inequality (\ref{eq:regain2}), observe that
\begin{eqnarray}
&&H({\vec X})-H({\vec X}|{\vec X}')=I({\vec X}:{\vec X}')\leq I({\vec X}:B_\Gamma)\nonumber\\
&&\leq I({\vec X}:B_\Gamma,L^\Gamma,{\hat K}^\Gamma)\leq C_{\rm tot}\label{eq:enemyyy}
\end{eqnarray}
due to the data processing inequality. By definition, we have
\begin{align}
H({\vec X})=nR.\label{eq:enemy}
\end{align}
Fano's inequality\cite{cover05} implies
\begin{align}
H({\vec X}|{\vec X}')\leq h(P_e)+nRP_e,\label{eq:enemyy}
\end{align}
Substituting (\ref{eq:enemy}) and (\ref{eq:enemyy}) to (\ref{eq:enemyyy}), we obtain (\ref{eq:regain2}).\hfill$\blacksquare$

\subsection{Proof of Lemma \ref{lmm:ccpowerofu}}\label{app:drinkbar}

The proof of the first statement is based on a protocol proposed in \cite{dominic07}. Let $B_1$ and $B_2$ be $d$-dimensional quantum systems, and let $\{\sigma_i^A\}_{i=1}^{d^2}$ be the set of generalized Pauli operators on ${\mathcal H}^A$. Define states
\begin{eqnarray}
\ket{\Phi_{i}}^{AB_1}:=\sigma_i^A\ket{\Phi_d}^{AB_1}=(\sigma_i^T)^{B_1}\ket{\Phi_d}^{AB_1}\label{eq:specialtopping}
\end{eqnarray}
and
\begin{eqnarray}
&&\!\!\!\!\!\!\!\!\!\!\!\!\!\!\!\!\!\!\!\!\!\!\!\!\!\!\!\!\!\ket{\Psi_{U,i}}^{AB_1BB_2}:=U^{AB}\ket{\Phi_{i}}^{AB_1}\ket{\Phi_d}^{BB_2}\nn\\
&&\;\;\;\;\;\;=(\sigma_i^T)^{B_1}\ket{\Psi_{U}}^{AB_1BB_2}\label{eq:sigmaT}
\end{eqnarray}
for $i=1,\cdots,d^2$. Let us introduce notations
\begin{eqnarray}
{\vec i}\!\!&:=&\!\!i_1\cdots i_n\in\{1,\cdots,d^2\}^n,\nonumber\\
{\bm\sigma}_{\vec i}\!\!&:=&\!\!\sigma_{i_1}\otimes\cdots\otimes\sigma_{i_n}\nonumber\\
|{\bm\Phi}_{\vec i}\rangle^{{\bar A}{\bar B}_1}\!\!&:=&\!\!|\Phi_{i_1}\rangle^{AB_1}\otimes\cdots\otimes|\Phi_{i_n}\rangle^{AB_1},\nonumber\\
|{\bm\Psi}_{U,{\vec i}}\rangle^{{\bar A}{\bar B}_1{\bar B}{\bar B}_2}\!\!&:=&\!\!|\Psi_{U,i_1}\rangle^{AB_1BB_2}\otimes\cdots\otimes|\Psi_{U,i_n}\rangle^{AB_1BB_2},\nn
\end{eqnarray}
and
\begin{eqnarray}
\rho({\mathcal U}_n,{\vec i}):={\mathcal U}_n(|{\bm\Phi}_{\vec i}\rangle^{{\bar A}{\bar B}_1}|\Phi_{d}^{\otimes n}\rangle^{{\bar B}{\bar B}_2}).\nn
\end{eqnarray}
From (\ref{eq:specialtopping}) and (\ref{eq:sigmaT}), we have
\begin{eqnarray}
\rho({\mathcal U}_n,{\vec i})=({\bm\sigma}_{\vec i}^T)^{{\bar B}_1}\rho({\mathcal U}_n)({\bm\sigma}_{\vec i}^*)^{{\bar B}_1}\nn
\end{eqnarray}
and
\begin{eqnarray}
|{\bm\Psi}_{U,{\vec i}}\rangle^{{\bar A}{\bar B}_1{\bar B}{\bar B}_2}=({\bm\sigma}_{\vec i}^T)^{{\bar B}_1}|{\bm\Psi}_{U}\rangle^{{\bar A}{\bar B}_1{\bar B}{\bar B}_2},\nn
\end{eqnarray}
where we defined
\begin{eqnarray}
{\bm\sigma}_{\vec i}^*:=({\bm\sigma}_{\vec i}^T)^\dagger.\nn
\end{eqnarray}
Therefore, due to the unitary invariance of the fidelity (see Equality (\ref{eq:fuwatoro})), Condition (\ref{eq:mathcalu}) implies
\begin{align}
F\left(\rho({\mathcal U}_n,{\vec i}),|{\bm\Psi}_{U,{\vec i}}\rangle\right)\geq1-\epsilon,\nn
\end{align}
which leads to
\begin{align}
F\left(\rho({\mathcal U}_n,{\vec i})^{{\bar B}_1{\bar B}{\bar B}_2},{\bm\Psi}_{U,{\vec i}}^{{\bar B}_1{\bar B}{\bar B}_2}\right)\geq1-\epsilon\label{eq:yutori}
\end{align}
for any ${\vec i}$ by taking the partial trace.

Due to Schur's lemma, we have
\begin{eqnarray}
\frac{1}{d^2}\sum_{i=1}^{d^2}\sigma_i^A\ket{\Phi_d}\!\bra{\Phi_d}^{AB_1}\sigma_i^{\dagger A}=\pi_d^A\otimes\pi_d^{B_1}.\nn
\end{eqnarray}
Thus the average state of $\ket{\Psi_{U,i}}$ with respect to the uniform distribution $p_i=1/d^2\:(i=1,\cdots,d^2)$ is given by
\begin{eqnarray}
\bar{\Psi}_U^{AB_1BB_2}&:=&\frac{1}{d^2}\sum_{i=1}^{d^2}{\Psi_{U,i}}^{AB_1BB_2}\nonumber\\
&=&\pi_d^{B_1}\otimes U^{AB}(\pi_d^{A}\otimes{\Phi_d}^{BB_2})U^{\dagger AB}.\nonumber
\end{eqnarray}
The reduced state of $\bar{\Psi}_U$ on $B_1BB_2$ is
\begin{eqnarray}
\bar{\Psi}_U^{B_1BB_2}&=&\pi_d^{B_1}\otimes{\rm Tr}_A[U^{AB}(\pi_d^{A}\otimes{\Phi_d}^{BB_2})U^{\dagger AB}]\nonumber\\
&=&\pi_d^{B_1}\otimes{\rm Tr}_{AB_1}[U^{AB}({\Phi_d}^{AB_1}\otimes{\Phi_d}^{BB_2})U^{\dagger AB}]\nonumber\\
&=&\pi_d^{B_1}\otimes\sum_{s=0}^{d^2-1}c_s^2F_s^B|\Phi_d\rangle\!\langle\Phi_d|^{BB_2}F_s^{\dagger B},\nonumber
\end{eqnarray}
where $c_s$ and $F_s$ are defined by (\ref{eq:opsch}). The von Neumann entropy of states ${\Psi}_{U,i}^{B_1BB_2}$ and $\bar{\Psi}_U^{B_1BB_2}$ are then 
\begin{eqnarray}
S(B_1BB_2)_{\Psi_{U,i}}=S(A)_{\Psi_{U,i}}=\log{d}\nonumber
\end{eqnarray}
and
\begin{eqnarray}
S(B_1BB_2)_{{\bar\Psi}_U}=\log{d}+K(U),\nonumber
\end{eqnarray}
respectively, where the latter follows from (\ref{dfn:kofu}) and the orthonormality of $\{F_s|\Phi_d\rangle^{BB_2}\}_s$.
Thus the Holevo information (\cite{schumacher97,holevo98}), corresponding to the signal states $\{\Psi_i^{B_1BB_2}\}_{i=1}^{d^2}$ and the uniform distribution $p_i=1/d^2\:(i=1,\cdots,d^2)$, is given by
\begin{eqnarray}
\chi(\{p_i,\Psi_i^{B_1BB_2}\})\!\!&=&\!\!S(B_1BB_2)_{{\bar\Psi}_U}\!-\!\frac{1}{d^2}\sum_{i=1}^{d^2}S(B_1BB_2)_{\Psi_{U,i}}\nonumber\\
			&=&\!\!K(U).\nonumber
\end{eqnarray}

Due to the Holevo-Schumacher-Westmoreland theorem\cite{schumacher97,holevo98}, for any $\epsilon>0$ and sufficiently large $n$, there exists a subset ${\mathcal C}_n\subset\{1,\cdots,d^2\}^n$ of cardinality $n(K(U)-\epsilon)$, such that all elements in the set
\begin{align}
\{{\bm\Psi}_{U,{\vec i}}^{{\bar B}_1{\bar B}{\bar B}_2}\}_{{\vec i}\in {\mathcal C}_n}\nn
\end{align}
are distinguishable up to a small error $\epsilon$. That is, there exists a measurement on ${\bar B}_1{\bar B}{\bar B}_2$, described by a set of measurement operators $\{D_{\vec i}\}_{{\vec i}\in {\mathcal C}_n}$, such that we have
\begin{eqnarray}
P_e:=\frac{1}{|{\mathcal C}_n|}\sum_{{\vec i}\in{\mathcal C}_n}\left(1-{\rm Tr}\left[D_{\vec i}{\bm\Psi}_{U,{\vec i}}^{{\bar B}_1{\bar B}{\bar B}_2}\right]\right)\leq\epsilon\nn
\end{eqnarray}

Consider the following protocol in which Alice transmits $n(K(U)-\epsilon)$ bits of classical message to Bob by ${\mathcal U}_n$, assisted by shared entanglement:
\begin{enumerate}
\item Alice and Bob initially share $\ket{\Phi_d}^{AB_1}\ket{\Phi_d}^{BB_2}$, where $B_1$ and $B_2$ are additional quantum registers that Bob has.
\item To send a message $k\in\{1,\cdots,2^{nR}\}$ where $R=K(U)-\epsilon$, Alice chooses $k$-th element ${\vec i}^k=i_1^k\cdots i_n^k$ in ${\mathcal C}_n$, and applies ${\bm\sigma}_{{\vec i}^k}$ on ${\bar A}$.
\item Alice and Bob apply ${\mathcal U}_n$.
\item Bob performs a measurement on ${\bar B}_1{\bar B}{\bar B}_2$ described by $\{D_{\vec i}\}_{{\vec i}\in {\mathcal C}_n}$.
\end{enumerate}
The state after Step 3) is equal to $\rho({\mathcal U}_n,{\vec i}^k)^{{\bar A}{\bar B}_1{\bar B}{\bar B}_2}$ for each $k$, which satisfies
\begin{align}
\left\|{\bm\Psi}_{{\mathcal U}_n,{\vec i}^k}^{{\bar B}_1{\bar B}{\bar B}_2}-{\bm\Psi}_{U,{\vec i}^k}^{{\bar B}_1{\bar B}{\bar B}_2}\right\|_1\leq2\sqrt{\epsilon}\nn
\end{align}
due to (\ref{eq:yutori}) and (\ref{eq:tracefid}). Therefore, due to (\ref{eq:defTD2}), the average error in transmitting the message is bounded above as
\begin{eqnarray}
P_e'&:=&\frac{1}{|{\mathcal C}_n|}\sum_{{\vec i}\in{\mathcal C}_n}\left(1-{\rm Tr}\left[D_{\vec i}{\bm\Psi}_{{\mathcal U}_n,{\vec i}}^{{\bar B}_1{\bar B}{\bar B}_2}\right]\right)\nonumber\\
&=&\frac{1}{|{\mathcal C}_n|}\sum_{{\vec i}\in{\mathcal C}_n}\left(1-{\rm Tr}\left[D_{\vec i}{\bm\Psi}_{U,{\vec i}}^{{\bar B}_1{\bar B}{\bar B}_2}\right]\right)\nonumber\\
&&+\frac{1}{|{\mathcal C}_n|}\sum_{{\vec i}\in{\mathcal C}_n}{\rm Tr}\left[D_{\vec i}({\bm\Psi}_{U,{\vec i}}^{{\bar B}_1{\bar B}{\bar B}_2}-{\bm\Psi}_{{\mathcal U}_n,{\vec i}}^{{\bar B}_1{\bar B}{\bar B}_2})\right]\nonumber\\
&\leq&\epsilon+4\sqrt{\epsilon}\leq5\sqrt{\epsilon},\nn
\end{eqnarray}
which completes the proof.\hfill$\blacksquare$

\bibliographystyle{IEEEtran}
\bibliography{markov}

%\begin{IEEEbiographynophoto}{Eyuri Wakakuwa}
%received his doctorate from The University of Tokyo
%in 2015, and was a postdoctoral fellow at the University of Electro-Communications, Japan, until 2017. He is currently an assistant professor at The University of Tokyo. His research focuses on quantum information theory and the foundation of quantum mechanics.
%\end{IEEEbiographynophoto}
%
%\begin{IEEEbiographynophoto}{Akihito Soeda}
%received his doctorate from The University of Tokyo in 2011 and, from then to 2013, was a research fellow at the Centre for Quantum Technologies.  He is currently an assistant professor at The University of Tokyo.  His main research focus is on quantum information theory and its applications to other areas of physics.
%\end{IEEEbiographynophoto}
%
%\begin{IEEEbiographynophoto}{Mio Murao}
%received M.S. and Ph.D. at Ochanomizu University in Tokyo, Japan in 1993 and 1996, respectively.  She worked as a postdoctoral fellow at Harvard University (US), Imperial College, London (UK) and RIKEN (Japan).  She was appointed as Associate Professor in 2001 and Professor in 2015 in the Department of Physics, the School of Science, the University of Tokyo. Her research interests cover a wide range of theoretical topics in quantum information and quantum physics.  She currently focuses on investigating entanglement and other non-local properties of quantum mechanics and their applications for distributed quantum information processing.
%\end{IEEEbiographynophoto}

\end{document}